\documentclass{aa}

\usepackage{graphicx}
\usepackage{capt-of}
\usepackage[varg]{txfonts}
\usepackage{hyperref}
\hypersetup{
    hidelinks,
    colorlinks = false
}
\DeclareRobustCommand{\VAN}[3]{#2}
\let\VANthebibliography\thebibliography
\def\thebibliography{\DeclareRobustCommand{\VAN}[3]{##3}\VANthebibliography}

\bibpunct{(}{)}{;}{a}{}{,} 

\usepackage{graphicx}	
\usepackage{amsmath}	
\usepackage{amssymb}	
\usepackage{comment}
\usepackage{xcolor}
\usepackage{multirow}
\usepackage{multicol}

\newcommand{\RomanNumeralCaps}[1]{\MakeUppercase{\romannumeral #1}}

\newcommand\tZAMS{\ensuremath{t_\mathrm{ZAMS}}\xspace}
\newcommand\tformation{\ensuremath{t_\mathrm{frm}}\xspace}
\newcommand\tRGB{\ensuremath{t_\mathrm{RGB}}\xspace}
\newcommand\tAGB{\ensuremath{t_\mathrm{AGB}}\xspace}
\newcommand\tenvloss{\ensuremath{t_\mathrm{env}}\xspace}
\newcommand\tdepletion{\ensuremath{t_\mathrm{dpl}}\xspace}
\newcommand\tdecay{\ensuremath{t_\mathrm{dec}}\xspace}

\newcommand\verticalscale{\ensuremath{H}\xspace}
\newcommand\aspectratio{\ensuremath{h}\xspace}
\newcommand\angularmomentum{\ensuremath{l}\xspace}

\begin{document}



\title{Planetary migration in wind-fed non-stationary accretion disks in binary systems}


\author{A.~D.~Nekrasov
    \inst{1}
    \and
    V.~V.~Zhuravlev
    \inst{2}
    \and
    S.~B.~Popov
    \inst{2}
}

\institute{Dr.\ Remeis-Sternwarte~\&~ECAP, Univ. Erlangen-N\"urnberg, Sternwartstr. 7, 96049 Bamberg, Germany\\
    \email{alex.nekrasov@fau.de}
    \and
    Sternberg Astronomical Institute, Lomonosov Moscow State University, Universitetsky prospekt 13, 119234, Moscow, Russia
}

\date{
Received April 26, 2024; accepted December 11, 2024}


  \abstract
   {An accretion disk can be formed around a secondary star in a binary system when the primary companion leaves the Main sequence and starts to lose mass at an enhanced rate.}
   {We study the accretion disk evolution and planetary migration in wide binaries.}
   {We use a numerical model of a non-stationary alpha-disk with a variable mass inflow. We take into account that the low-mass disk has an extended region that is optically thin along the rotation axis. We consider irradiation by both the host star and the donor. The migration path of a planet in such a disk is determined by the migration rate varying during the disk evolution.}
   {Giant planets may open/close the gap several times over the disk lifetime. We identify the new type of migration specific to parts of the growing disk with a considerable radial gradient of an aspect ratio. Its rate is enclosed between the type \RomanNumeralCaps{2} and the fast type \RomanNumeralCaps{1} migration rates being determined by the ratio of time and radial derivatives of the disk aspect ratio. Rapid growth of the wind rate just before the envelope loss by the donor leads to the formation of a zone of decretion, which may lead to substantial outward migration. In binaries with an initial separation $a\lesssim 100$\,AU migration becomes most efficient for planets with 60--80 Earth masses. This results in approaching a short distance from the host star where tidal forces become non-negligible. Less massive Neptune-like planets at the initial orbits $r_\mathrm{p} \lesssim 2$\,AU can reach these internal parts in binaries with $a \lesssim 30$\,AU.}
   {In binaries, mass loss by the primary component at late evolutionary stages can significantly modify the structure of a planetary system around the secondary component resulting in mergers of relatively massive planets with a host star.}



\maketitle



\section{Introduction}
\label{sec:intro}

In the latest Decadal survey, one of the key themes is entitled ``Worlds and suns in context'' \citep{nationalacademiesofsciences2021}. 
In particular, this title indicates that at the next step of studies we have to advance a joint description of the properties and evolution of stars and planets. 
One of the elements of this broad context is the problem of star-planet interactions. 

There are many different possibilities when a star and a planet intensively interact with each other, e.g., via the tides, the magnetic field, the stellar wind, etc.~\citep[see][]{lanza2015}. 
However, all of them require a planet 
to be located close to the star. Such a situation is not expected in naive scenarios of planet formation as conditions at distances close to the star are not suitable for planet growth. Thus, an additional ingredient of a formation mechanism might be involved -- migration. 

Planetary migration is an actively studied topic, see, e.g., a brief review and references in \citet{papaloizou2021} and the recent review by \citet{paardekooper2023}. Usually, migration is studied in protoplanetary disks \citep{armitage2010}. 
This involves many interlinked physical processes resulting in numerous possibilities of a planetary orbit evolution. Planets can migrate both toward or away from
a star in different parts of a disk or/and at different times. Additionally, 
the trapping regions can appear.
The gravitational interaction of disk and planet might be even more complicated due to planet-planet interactions and interactions with planetesimals~\citep[see, e.g.,][]{raymond2022}. Most important,  migration is the main mechanism for the formation of ``hot'' and ``warm'' planets, especially gas and ice giants (hot jupiters, warm neptunes, etc.). Thus, migration can finally allow intensive star-planet interactions, especially tidal
interactions (see, e.g., \citealt{lazovik2021} and references therein). 

It is interesting and important to find and explore additional possibilities for planetary migration due to an interaction with a disk. In this paper, we advance the analysis initiated by \citet{kulikova2019}. The main idea is rather simple: in a binary system, the formation of an accretion disk around the secondary Main sequence star is possible during the late evolutionary stages of its primary companion. The disk is fed by the stellar wind from the primary. If a planetary system 
pre-exists around the secondary component, then a new episode of migration due to a planet-disk interaction might appear. We study the circumstellar disks, though the circumbinary disks at late stages of star evolution are also considered in the literature \citep{kluska2022}. So, a similar study is possible for planets in circumbinary disks.

About one-half of intermediate-mass stars are members of binary systems \citep{duchene2013, offner2023}. 
Now, it is well-established that planets in binaries, including close systems with semimajor axis $a\lesssim 100$\,AU, are quite frequent~\citep{bonavita2020}. Also see the catalogs of planets in binary and multiple stellar systems in \citet{schwarz2016} and \citet{thebault2015}\footnote{The online versions of these catalogs are available at \href{https://adg.univie.ac.at/schwarz/multiple.html}{https://adg.univie.ac.at/schwarz/multiple.html}, \href{https://exoplanet.eu/planets_binary/}{https://exoplanet.eu/planets\_binary/}.}. The majority of systems are too wide to allow the formation of significant accretion disk 
which can change the orbital distribution of planets. Nevertheless, this may not be the case in more than ten percent of binaries in the appropriate mass range. 
See also~\citet{hillen2016}, where the authors find that in systems with detected circumbinary disks, the secondary component shows the presence of a compact circumstellar accretion disk.

In \citet{kulikova2019} the authors considered the stationary disk only. However, due to the evolution of the primary, the rate of accretion is variable, especially during periods of enhanced mass loss. In this paper, we present a model of a non-stationary disk around a solar-like Main sequence star in pair with an evolved more massive companion (the donor, hereafter). The wind from the donor which is enhanced during the evolution at the red giant and asymptotic giant branches (RGB and AGB, respectively, hereafter) leads to the formation of a disk. Detailed calculations of the structure of such disks tracking accurately the donor evolution is the main part of this study. Our final results represent the migration of a single planet of a different mass in such a disk. 
 
Overall, we obtain that planets can experience significant migration in systems with $a\lesssim 100$\,AU, which leads to the appearance of ``hot'' and ``warm'' massive planets, which may experience intensive tidal interaction with the host star.

In the next Section we describe the scenario used in this study. It includes the binary evolution model, the two accretion disk models, and finally~--~the model of planetary migration. In Sect.~\ref{sec:numerical_method} the details about our numerical approach are given. Next, in  Sect.~\ref{sec:results} we present our results. These results and cautions on our modeling are discussed in Sect.~\ref{sec:discussion}. We draw the final conclusions in Sect.~\ref{sec:conclusions}. We show the additional Figures in the appendix~\ref{app:figures}. Details of calculations are presented in the appendices~\ref{app:temperature},~\ref{app:thin_temperature},~\ref{app:donor},~\ref{sec_app:numerical_method}, and~\ref{sec_app:decretion}.

\section{Model}
\label{sec:model}

We consider binary systems with major semi-axis $a \leq 100$\,AU. Primaries have initial masses, $M_1$, 
below $8$\,M$_{\sun}$, which guarantees the evolution of the star with smooth envelope loss at late stages (i.e, without a supernova explosion) and the formation of a white dwarf. The secondary component in each system under consideration is formed as a Solar-like star: $M_2=$~1\,M$_{\sun}$  and evolves slower than the primary ($M_2 < M_1$). 
We neglect any possible interactions in a multi-planetary system, i.e., only planet-disk interactions are analyzed. 
When the primary star evolves into a giant, the increasing stellar wind is captured by the secondary, and an accretion disk can be formed around it. The planet interacts with this disk which leads to a significant modification of the planetary orbit. 

The disk is assumed to be coplanar with the orbital plane of the binary and with the equatorial plane of the secondary star. The eccentricity of the binary system is assumed to be zero, $e = 0$. 

\subsection{Binary evolution}
\label{s:binary}

\begin{figure}
    \centering
    \begin{minipage}{0.49\columnwidth}
    \includegraphics[width=\columnwidth]{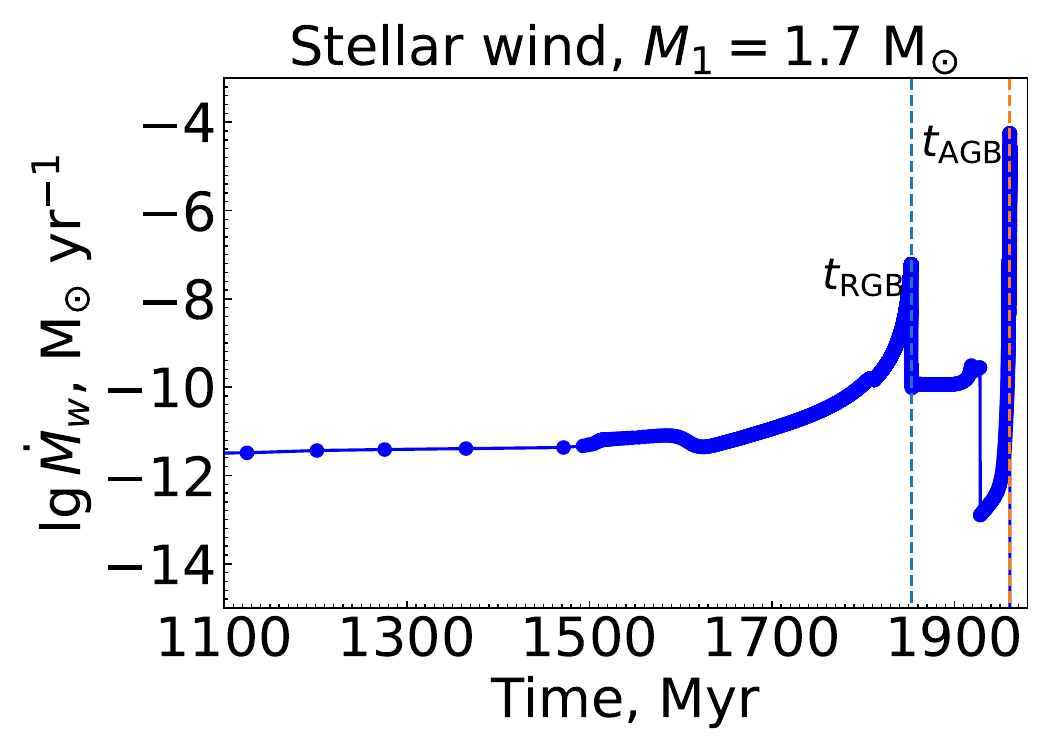}
    \end{minipage}
    \begin{minipage}{0.49\columnwidth}
    \includegraphics[width=\columnwidth]{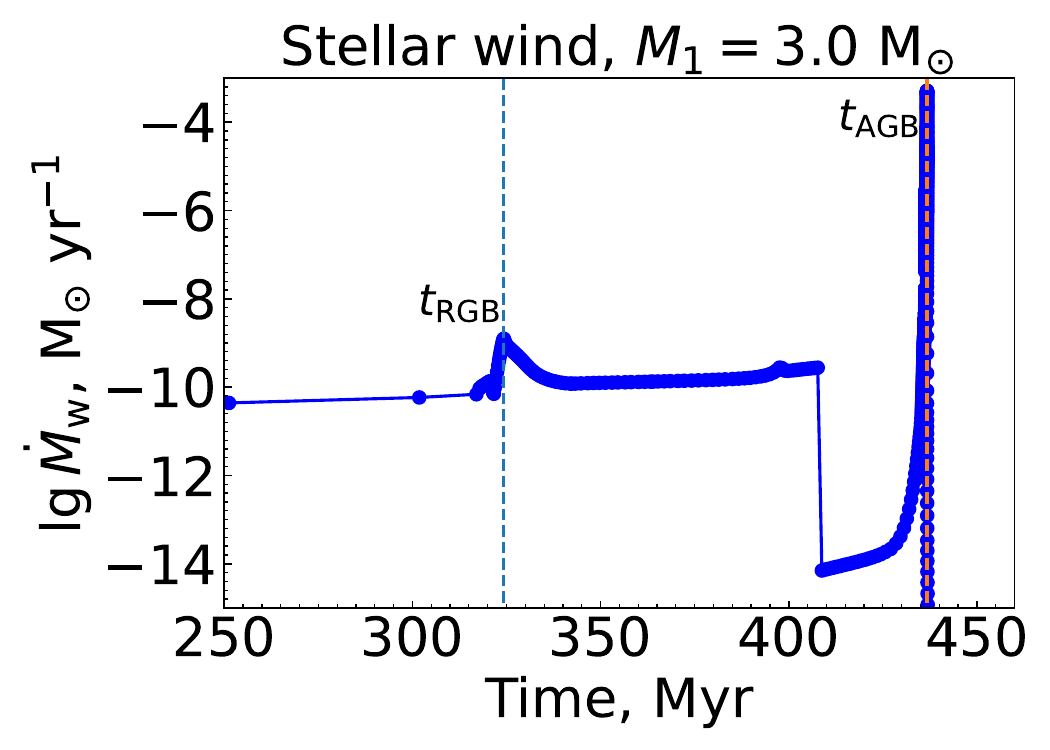}
    \end{minipage}    
    \caption{Stellar wind rates for low mass donors with initial masses $M_1 = 1.7$\,M$_{\sun}$ (left) and 3.0\,M$_{\sun}$ (right), calculated using MESA. The first peak of the stellar wind corresponds to the red giant branch (RGB) stage. The second peak associated with the envelope loss by the star corresponds to the stage of the asymptotic giant branch (AGB). The zero moment of time $t = t_{\mathrm{ZAMS}} = 0$ corresponds to the Zero Age Main Sequence (ZAMS).}
    \label{fig:low_mass_wind}
\end{figure}

\begin{figure}
    \centering
    \begin{minipage}{0.49\columnwidth}
    \includegraphics[width=\columnwidth]{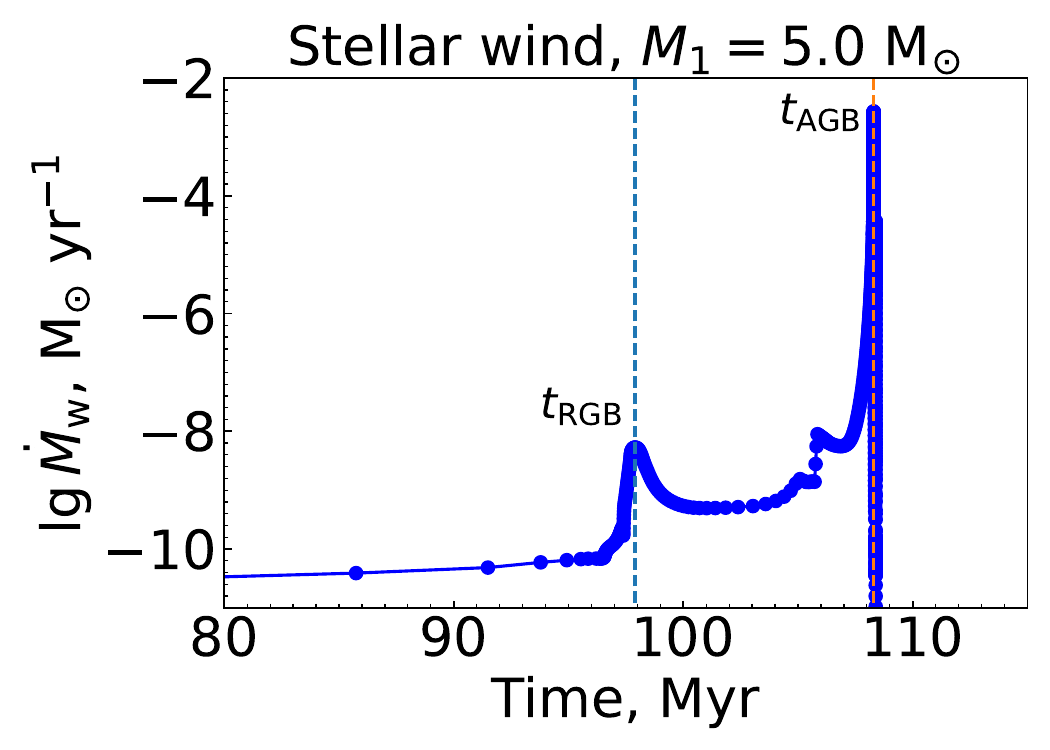}
    \end{minipage}
    \begin{minipage}{0.49\columnwidth}
    \includegraphics[width=\columnwidth]{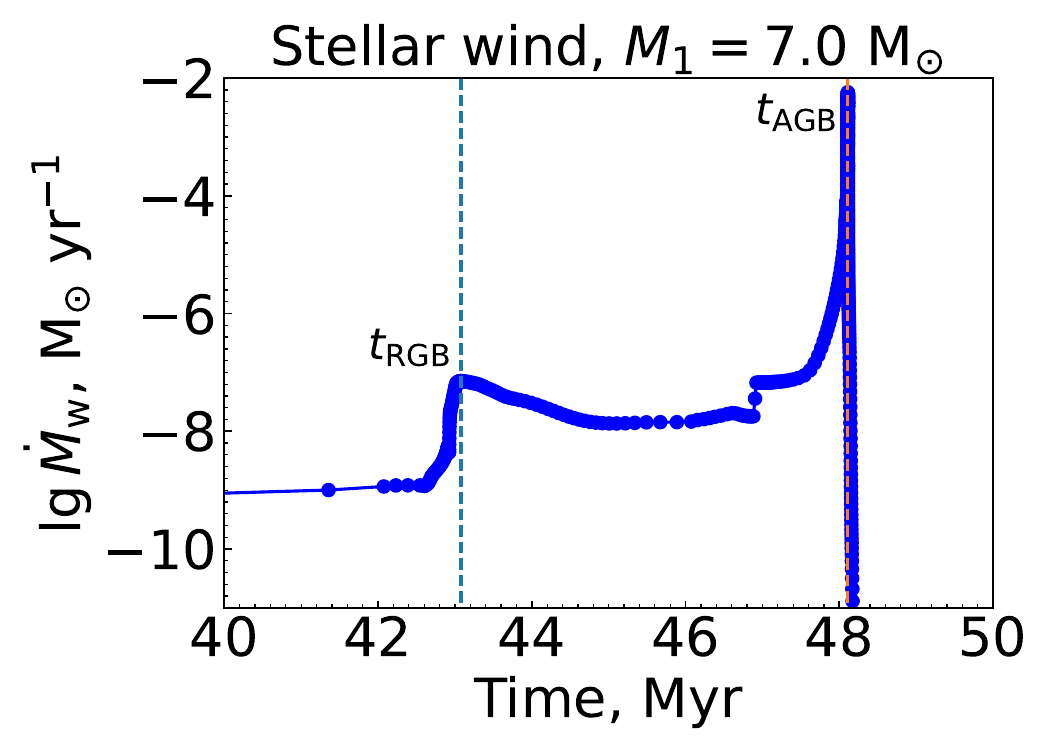}
    \end{minipage}    
    \caption{Stellar wind rates for high mass donors with initial masses $M_1 = 5.0$\,M$_{\sun}$ (left) and $7.0$\,M$_{\sun}$ (right), calculated using MESA. The peaks of the stellar wind have the same nature as in Fig.~\ref{fig:low_mass_wind}. The zero moment of time $t = t_{\mathrm{ZAMS}} = 0$ corresponds to ZAMS.}
    \label{fig:high_mass_wind}
\end{figure}

The evolution of a binary system is mainly determined by the evolution of the primary 
which is calculated using the code MESA~\citep{paxton2011}.\footnote{We use MESA tracks used for exoplanet population synthesis in~\citet{andryushin2021}. The tracks are obtained with MESA version 10398 for Solar metallicity.} Calculations are done for Zero Age Main Sequence (ZAMS) masses listed in Table~\ref{tab:model_free_parameters_binary}. 
The formation and structure of an accretion disk around the secondary are mainly determined by the stellar wind of the primary. The profile of the mass loss rate by the primary, $\dot M_\mathrm{w}$, at late stages of its evolution is shown in Figs.~\ref{fig:low_mass_wind},~\ref{fig:high_mass_wind}. 
It can be seen that the mass loss can be quite variable, e.g., the timescale of wind variations, $t_{\mathrm{w}} \equiv \dot M_{\mathrm{w}} / \ddot M_{\mathrm{w}}$, can decrease down to $10^3$\,years, while the characteristic viscous timescale 
for a disk with the size of several AU or more is typically higher than $10^5$\,years. That is why a non-stationary model of the disk is an essential element of this study. 
It is also important that the primary component can lose up to more than half 
of its mass due to the stellar wind.
Correspondingly, it is necessary to take into account the secular 
decrease of $M_1$, as this changes the orbital parameters, 
\begin{equation}
\label{eq:mass_loss}
M_1 (t) = M_\mathrm{1}(0) - \int\limits_{0}^{t} \dot M_\mathrm{w} (t^\prime) \mathrm{d} t^\prime.
\end{equation}

\begin{table}
\caption{The main free parameters of the binary used for modeling}
{\tiny
\centering
\begin{tabular}{|c|c|c|c|c|}
\hline
$M_2$ (M$_{\odot}$) & $M_{1, \rm init}$ (M$_{\odot}$) & $M_{1, \rm final}$ (M$_{\odot}$) & $a_{\rm init}$ (AU) & $a_{\rm final}$ (AU) \\
\hline
$1.0$ & $1.7$ & $0.55$ & $[10.0, 100.0]$ & $[17.1, 173.5]$ \\
$1.0$ & $3.0$ & $0.58$ & $[10.0, 100.0]$ & $[24.1, 248.7]$ \\
$1.0$ & $5.0$ & $0.82$ & $[10.0, 100.0]$ & $[30.4, 320.2]$ \\
$1.0$ & $7.0$ & $0.93$ & $[10.0, 100.0]$ & $[36.5, 387.5]$ \\
\hline
\end{tabular}
}
\tablefoot{The columns show: the secondary component mass $M_2$, the donor initial mass $M_{1, \rm init}$, the donor final mass $M_{1, \rm final}$, the binary initial major semi-axis range $a_{\rm init}$, the binary final major semi-axis range $a_{\rm final}$. Eccentricity of the system is assumed to be zero, $e=0$.}
\label{tab:model_free_parameters_binary}
\end{table}

The mass of a secondary component is considered to be constant.
For initial binary separations studied here $ a \in (10, 100)$\,AU, 
the evolving donor does not fill its Roche lobe. 
This allows us to apply a simplified treatment using the Bondi-Hoyle accretion regime  \citep{bondi1944}. The Bondi-Hoyle accretion rate is defined as
\begin{equation}
\label{eq:bondi_accretion_rate}
\dot{M}_{\mathrm{acc}} = \dot M_{\mathrm{w}} \frac{r_{\mathrm{a}}^2}{4 a^2}.
\end{equation}
Here $r_{\mathrm{a}}$ is the Bondi radius (aka the gravitational capture radius), which is 
\begin{equation}
\label{eq:bondi_accretion_radius}
r_{\mathrm{a}} = \frac{2 G M_2}{v_{\mathrm{rel}}^2 + c_{\mathrm{w}}^2},
\end{equation}
where $c_{\mathrm{w}}$ is the sound speed in the stellar wind, 
$v_{\mathrm{rel}}=\sqrt{v_{\mathrm{w}}^2 + v_{\mathrm{s}}^2}$ is the relative velocity of the wind and the companion star, $v_{\mathrm{w}}$ is the wind velocity at the secondary star location, 
$v_{\mathrm{s}}= \sqrt{ G M_1^2 / (M_1 + M_2) / a}$ is the orbital velocity of the secondary, and $G$ is the gravitational constant. In this work, we choose $c_{\mathrm{w}} = 10$\,km\,s$^{-1}$
 and $v_{\mathrm{w}} = 20$\,km\,s$^{-1}$ \citep{habing1996}.

We make a rather simplified assumption about the wind velocity. However, this is justified by the fact that the formation of a sufficiently large disk as well as significant migration of planets are possible only at the RGB and AGB stages. These stages are characterized by the low velocity of the stellar wind, see, e.g., \citet{soker2000, bennett2010} and references therein. We discuss the impact of the wind velocity on the disk evolution in Sect.~\ref{sec:wind_props}. 

Binary separation changes with time due to the primary mass decrease and angular momentum loss by the system. Following the standard approach~\citep[e.g., see][]{postnov2014} we obtain an equation for the binary separation $a = a(t)$ evolution, 
\begin{equation}
\label{eq:separation}
\frac{\dot a}{a} = \frac{\dot M_{\mathrm{w}}}{M_1} \left(1 + \frac{r_{\mathrm{a}}^2}{4 a^2} - 2 \frac{r_{\mathrm{a}}^2}{4 a^2} \frac{1}{q} - \left(1 - \frac{r_{\mathrm{a}}^2}{4 a^2}\right) \frac{q}{1 + q}\right),
\end{equation}
where $q \equiv M_2/M_1$ is the stellar mass ratio, and $r_{\mathrm{a}}$ is defined from  Eq.~\eqref{eq:bondi_accretion_radius}. $\dot M_{\mathrm{w}}$ is defined positive. Note that $M_1 (t)$, $\dot M_{\mathrm{w}} (t)$, $r_{\mathrm{a}} (t)$, and $q(t)$ are all functions of time and that  Eq.~\eqref{eq:separation} assumes that there is no stage with a common envelope~\citep[see, e.g.,][]{iben1993} in the system.See the initial and final binary separation values in Table~\ref{tab:model_free_parameters_binary}. We note that we obtain Eq.~\eqref{eq:separation} under the assumption of zero eccentricity of the system, $e=0$. However, this may not be the case in the observed binaries, see, e.g.,~\citet{wu2024}. We 
relegate the study of non-stationary wind-fed disk 
in eccentric binaries for future work.

\subsection{Accretion disk}

An axisymmetric circumstellar disk is formed inside the Roche lobe of the planet-hosting Main sequence secondary, see~\citet{deval-borro2009}. The outer boundary of the disk is limited by the tidal truncation radius. A stable disk cannot expand beyond this radius due to the tidal torque influence of the primary. The radius is estimated using Table~1 from~\citet{paczynski1977} as
\begin{equation}
\label{eq:r_tidal}
r_{\mathrm{out}} = r_{\mathrm{tidal}} \approx 0.896 \left( \frac{1}{1 + q}\right)^{0.0467} r_{\mathrm{l}}.
\end{equation}
Here $r_{\mathrm{l}}$ is the Roche lobe radius~\citep{eggleton1983}, 
\begin{equation}
\label{eq:r_lobe}
\frac{r_{\mathrm{l}}}{a} = \frac{0.49 q^{2/3}}{0.6 q^{2/3} + \ln\left(1 + q^{1/3}\right)}.
\end{equation}

We choose the value of the inner disc boundary as $r_{\mathrm{in}} = R_2$. Actually, it
can be defined, e.g., by truncation by the magnetic field or/and wind of the secondary~\citep[see, e.g.,][]{armitage2017}. However, the precise value of $r_{\mathrm{in}}$ has little effect on the overall evolution of the disk.

The disk is assumed to be geometrically thin. 
Since the disk mass is expected to be comparatively small, we neglect its self-gravity. 
The rotation is taken to be Keplerian: $\Omega = \Omega_{\mathrm{K}}$, where 
$\Omega_{\mathrm{K}} \equiv \sqrt{GM_2/r^3}$ with 
$r$ being the distance from the host star.

The viscous torque, $F = F (r, t)$, is selected as an independent function describing the radial evolution of the disk, see, e.g.,~\citet{lipunova2000} or~\citet{lipunova2015}. It is defined as 
\begin{equation}
\label{eq:torque}
F = -2 \pi r \nu \Sigma r \frac{\mathrm{d} \Omega}{\mathrm{d} r} r = 3\pi r^2 \Omega \nu \Sigma,
\end{equation}
where $\nu$ is a kinematic viscosity in the disk, and $\Sigma$ is the disk surface density. 
The second equality is valid for the Keplerian rotation. 

We introduce the specific angular momentum as 
\begin{equation}
\label{eq:angular_momentum}
\angularmomentum = r^2 \Omega = \sqrt{G M_2 r}.
\end{equation}
Here on the right-hand side, it is assumed that the disk is Keplerian. This quantity simplifies the treatment of equations. 
We define the values $\angularmomentum=\angularmomentum_{\rm in}$ and
$\angularmomentum=\angularmomentum_{\rm out}$ which correspond to $r=r_{\rm in}$, $r=r_{\rm out}$, 
respectively. Both of these values are used below. 

With the notations at hand, the disk evolution is determined by the nonlinear diffusion equation 
\begin{equation}
\label{eq:diffusion_equation}
\frac{\partial \Sigma}{\partial t} = \frac{(G M_2)^2}{4 \pi \angularmomentum^3} \frac{\partial^2 F}{\partial \angularmomentum^2} + \left(\frac{\partial \Sigma}{\partial t}\right)_{\mathrm{ext}},
\end{equation}
obtained by~\citet{lyubarskij1987} 
with an additional source function from~\citet{perets2013}.

The initial condition for Eq.~\eqref{eq:diffusion_equation} is limited to the sufficiently small absolute value  
so that the initial spurious amount of matter does not affect further evolution, see Sect.~\ref{sec:numerical_method}.
Boundary conditions necessary to solve Eq.~\eqref{eq:diffusion_equation} are specified as follows, e.g.,~\citet{armitage2017},
\begin{equation}
\label{eq:boundary}
\left. F \right|_{\angularmomentum_{\mathrm{in}}} = 0, \quad \left. \frac{\partial F}{\partial \angularmomentum} \right|_{\angularmomentum_{\mathrm{out}}} = 0,
\end{equation}
where the inner boundary condition corresponds to the zero viscous stress at  $r_{\mathrm{in}}$, while the outer one corresponds to the zero mass outflow rate at the outer boundary, $r_{\mathrm{out}}$. This corresponds to 
the situation when the angular momentum is removed from the outer edge of the disk by the tidal action of the primary. 

A different situation takes place when the disk does not expand up to $r_{\mathrm{out}}$ due to low optical thickness. In this case, the disk is optically thick 
in its own plane within 
$r = r_{\tau} < r_{\mathrm{out}}$. The edge $r=r_{\tau}$
can be estimated using the optical thickness in the radial direction introduced by Eq.~\eqref{eq:optical_thickness_radial} below, also see the introduction to the App.~\ref{app:donor_conical}. Since the standard disk theory employed in this work 
is based on the local energy balance with respect to the whole disk plane, it becomes hardly reliable at $r>r_{\tau}$. Therefore, 
instead of solving Eq.~\eqref{eq:diffusion_equation}
beyond $r_{\tau}$ we assume that the ambient optically thin flow decrets due to the angular momentum coming from the disk with the velocity 
\begin{equation}
\label{eq:boundary_fl}
\left.  v_r\right|_{\angularmomentum_{\tau} < \angularmomentum < \angularmomentum_{\mathrm{out}}} = \frac{3}{2} \frac{\nu}{r},
\end{equation}
where $v_r$ is the radial velocity of matter in a disk
and $\angularmomentum_{\tau}$ is evaluated at $r_{\tau}$. Equation~\eqref{eq:boundary_fl} restated in terms of $F$ reads 
\begin{equation}
\label{eq:boundary_fl_F}
\left.  \left ( \frac{F}{\angularmomentum} + 
\frac{\partial F}{\partial \angularmomentum} \right ) \right|_{\angularmomentum_{\tau} < \angularmomentum < \angularmomentum_{\mathrm{out}}} = 0.
\end{equation}
Numerical tests show that such a boundary condition imposed at the 
stage of disk depletion is physically plausible since it is 
equivalent to a weakly evaporating tenuous flow outside $r_{\tau}$.
Hereafter, we refer to Eq.~\eqref{eq:boundary_fl_F} as the floating boundary condition.

The source of matter from the stellar wind is calculated similarly to~\citet{perets2013} as
\begin{equation}
\label{eq:mass_source}
\left(\frac{\partial \Sigma}{\partial t}\right)_{\mathrm{ext}} = \frac{\dot{M}_{\mathrm{acc}}}{2 \pi r r_{\mathrm{a}}} = \frac{\dot{M}_{\mathrm{w}} r_{\mathrm{a}}}{8 \pi r a^2} \theta \left(r_{\mathrm{a}} - r \right),
\end{equation}
where $\theta \left(r_{\mathrm{a}} - r \right)$ is the Heaviside function.

\subsubsection{Quasi-stationary limit}

One can obtain the stationary solution of Eq.~\eqref{eq:diffusion_equation} assuming~$\partial \Sigma/\partial t = 0$, as 
\begin{equation}
\label{eq:nu_sigma}
\nu \Sigma = \frac{ \dot M_{\mathrm{acc}} (t)}{3 \pi} f ( \angularmomentum, \angularmomentum_{\mathrm{a}} (t)).
\end{equation}
Equation~\eqref{eq:nu_sigma} has been obtained 
by~\citet{kulikova2019} using the zero torque condition at the inner boundary along with the explicit 
form of the source of matter, see Eq.~\eqref{eq:mass_source}. 
They showed that 
the additional factor $f$ can be written as\footnote{Note that in~\citet{kulikova2019} a similar eq.~(20) contains a misprint.} 
\begin{equation}
\label{eq:geometric_factor}
f (\angularmomentum, \angularmomentum_{\mathrm{a}}) = \left\{
\begin{aligned} & 1 - \frac{\angularmomentum_{\mathrm{in}}}{\angularmomentum} - \frac{\angularmomentum^2}{3 \angularmomentum_{\mathrm{a}}^2} + \frac{\angularmomentum_{\mathrm{in}}^3}{3 \angularmomentum_{\mathrm{a}}^2 \angularmomentum}, & \angularmomentum \leq \angularmomentum_{\mathrm{a}}, \\
& \frac{2 \angularmomentum_{\mathrm{a}}}{3 \angularmomentum} - \frac{\angularmomentum_{\mathrm{in}}}{\angularmomentum} + \frac{\angularmomentum_{\mathrm{in}}^3}{3 \angularmomentum_{\mathrm{a}}^2 \angularmomentum}, & \angularmomentum > \angularmomentum_{\mathrm{a}}, \\ \end{aligned} \right.,
\end{equation}
where $\angularmomentum_{\mathrm{in}}$ and  $\angularmomentum_{\mathrm{a}}$ are values of the specific angular momentum 
at $r_{\mathrm{in}}$ and $r_{\mathrm{a}}$, respectively, see Eq.~\eqref{eq:angular_momentum}. 
Here, we additionally assume that the local accretion rate through the disk vanishes beyond $r_{\mathrm{a}}$. 


\subsubsection{Radial velocity of gas in a disk}
\label{sec:drift}

Viscous torque, Eq.~\eqref{eq:torque}, is related to the local accretion rate in a disk 
according to the azimuthal projection of the Euler equation for the accreting matter, 
\begin{equation}
\label{eq:continuity_equation}
\dot M (\angularmomentum) = \frac{\mathrm{d} F}{\mathrm{d} \angularmomentum},
\end{equation}
where by definition
\begin{equation}
\dot M (\angularmomentum) \equiv - 2 \pi r (\angularmomentum) \Sigma (F, \angularmomentum) v_{\mathrm{r}} > 0.
\end{equation}
Using Eq.~\eqref{eq:torque} we obtain 
\begin{equation}
\label{eq:radial_drift}
v_{\mathrm{r}} = - \frac{1}{2 \pi r (\angularmomentum) \Sigma (F, \angularmomentum)} \dot M (\angularmomentum) = -\nu \frac{3}{F} \frac{\mathrm{d} F}{\mathrm{d} r}.
\end{equation}
Equation~\eqref{eq:radial_drift} is used below to get the rate of the type \RomanNumeralCaps{2}
planetary migration, see Sect.~\ref{s:migration}, as well as to consider the  disk structure in Sect.~\ref{sec:results_accretion}.

\subsection{Energy balance}
\label{sec:temperature}

Equation~\eqref{eq:diffusion_equation} must be solved taking into account details of internal heating and cooling of the disk. Geometrically thin disks are known to have a local energy balance: thermal energy extracted from the mean shear motion of matter, e.g., via turbulent cascade, at some $r$ is emitted away at the same distance from the host star. A number of further simplifying assumptions on the turbulent dissipation of energy as well as on the transfer of thermal energy across the disk leads us to the energy balance equation, which determines kinematic viscosity standing in Eq.~\eqref{eq:torque}. Thus, these assumptions allow us to establish the missing relationship between $F$ and $\Sigma$, which is necessary to solve Eq.~\eqref{eq:diffusion_equation}.


Following~\citet{shakura1972} and~\citet{shakura1973} we assume that the turbulent viscosity 
is expressed through the dimensionless parameter $\alpha$,
\begin{equation}
\label{eq:alpha_viscosity}
\nu = \alpha c_{\mathrm{s}} \verticalscale = \alpha c_{\mathrm{s}}^2 \Omega^{-1} = \alpha \frac{\gamma k_{\mathrm{B}}}{\mu m_{\mathrm{H}}} T_{\mathrm{c}} \frac{r^{3/2}}{\sqrt{G M_2}},
\end{equation}
where $\verticalscale$ is the disk vertical scale height and $\verticalscale \approx c_{\mathrm{s}} / \Omega$ due to the vertical hydrostatic equilibrium. The sound speed, $c_{\mathrm{s}}$, is related to the temperature according to the ideal gas equation of state, 
$c_{\mathrm{s}}^2 = \frac{\gamma k_{\mathrm{B}}}{\mu m_{\mathrm{H}}} T_{\mathrm{c}}$, where $\gamma$ is a specific heat ratio, $k_{\mathrm{B}}$ is the Boltzmann constant, $\mu$ is the mean molecular weight in the disk matter, $m_{\mathrm{H}}$ is a hydrogen atom mass, and $T_{\mathrm{c}}$ is the disk temperature at its midplane. In our simulations we set $\gamma = 1.4$, $\mu = 2.3$, which corresponds to a standard hydrogen-helium composition of the gas.

We set $\alpha = 0.01$ for all our calculations unless otherwise stated. This is a conservative estimate of dimensionless viscosity based on spectroscopic and interferometric observations of protoplanetary disks combined with each other within the alpha-disk paradigm, see~\citet{hartmann1998}. The more recent observations indicate that $\alpha$ varies from disk to disk or/and may be different in the inner and the outer parts of the disk approximately in the range $10^{-4}$--$10^{-2}$, see, e.g.,~\citet{rosotti2020}, \citet{trapman2020}, \citet{flaherty2024} and the review by \citet{rosotti2023}. For this reason, we also perform additional calculations taking the lower $\alpha=0.001$ discussed in Sect.~\ref{sec:discussion_disk_turbulence}.

While considering the radiative energy transfer in a disk we do not restrict the model with 
the optically thick case only. 
We estimate the Rosseland optical thickness in the $z$-direction as
\begin{equation}
\label{eq:optical_thickness_vertical}
\tau_{\mathrm{z}, \mathrm{R}} (r, t) = \frac{\kappa_{\mathrm{R}} (r, t) \Sigma (r, t)}{2},
\end{equation}
where $\kappa_{\mathrm{R}} (r,t)$ is the Rosseland mean opacity taken from~\citet{semenov2003}\footnote{See the web link to the results of numerical calculations of opacity for different sets of parameters in the cited paper -- \href{https://www2.mpia-hd.mpg.de/home/henning/Dust_opacities/Opacities/opacities.html}{https://www2.mpia-hd.mpg.de/home/henning/Dust\_opacities/Opacities/opacities.html}}. Everywhere below it is assumed that $\kappa_{\mathrm{R}} (r,t)$ is taken at the disk midplane. Following \citet{nakamoto1994} we also consider disks which are optically thin in the $z$-direction, i.e., $\tau_{\mathrm{z}, \mathrm{R}} < 1$.

Additionally, we introduce an estimate of optical thickness in the radial direction, 
\begin{equation}
\label{eq:optical_thickness_radial}
\tau_{\mathrm{r}, {\mathrm{R}}} (r, t) = \int\limits_{r}^{r_{\mathrm{out}}} \frac{\tau_{\mathrm{z}, {\mathrm{R}}} (r, t)}{\verticalscale (r, t)} \mathrm{d } r.
\end{equation}
We note that the integration limits chosen in the definition~\eqref{eq:optical_thickness_radial} 
yield the lower estimate of optical thickness in the radial direction at the given 
$r$ as far as we assume the normal behavior of surface density decreasing outwards.

The local energy balance is valid in regions with $\tau_{\mathrm{r }, {\mathrm{R}}} > 1$ only.
We assume that the condition $\tau_{\mathrm{r }, {\mathrm{R}}} > 1$
provides a reasonable restriction for the validity of our 
model to low-mass disks that are optically thin in the $z$-direction.

The midplane temperature which determines the viscosity of a disk is obtained from the energy balance equation, for the details of its derivation see Appendix~\ref{app:temperature},
\begin{multline}
\label{eq:energy_equation}
T_{\mathrm{c}}^4 = \frac{ 9 \nu \left(T_{\mathrm{c}}\right) \Sigma \Omega_{\mathrm{K}}^2}{8 \sigma_{\mathrm{B}}} \left( 1 + \frac{3 \kappa_{\mathrm{R}}\left(T_{\mathrm{c}}\right) \Sigma}{8} + \frac{1}{2 \kappa_{\mathrm{P}}\left(T_{\mathrm{c}}\right) \Sigma} \right) + \\
+ \left(T^4_{\mathrm{I}, \mathrm{A}} + T^4_{\mathrm{I},\mathrm{D}} + T^4_{\mathrm{w}}\right) \left(1 + \frac{1}{2\kappa_{\mathrm{P}}\left(T_{\mathrm{c}}\right) \Sigma} \right),
\end{multline}
where $\sigma_{\mathrm{B}}$ is the Stefan-Boltzmann constant, $\kappa_{\mathrm{P}}$ is the Planck mean opacity essential for optically thin regions where $\tau_{\mathrm{z}, \mathrm{R}} < 1$ (but $\tau_{\mathrm{r}, \mathrm{R}} > 1$, see Appendix~\ref{app:thin_temperature}) obtained by~\citet{semenov2003}. 
Note that Eq.~\eqref{eq:energy_equation} contains terms due to
 irradiation flux from the host star, $F_{\mathrm{I}, \mathrm{A}} \sim \sigma_{\mathrm{B}} T_{\mathrm{I},\mathrm{A}}^4$, the irradiation flux from the donor, $F_{\mathrm{I}, \mathrm{D}} \sim \sigma_{\mathrm{B}} T_{\mathrm{I},\mathrm{D}}^4$ (see Appendix~\ref{app:donor} for the details), 
and additional heating of the disk due to dissipation of the kinetic energy of settling matter, $F_{\mathrm{w}} \sim \sigma_{\mathrm{B}} T_{\mathrm{w}}^4$ and the rest ``viscous'' flux is labeled as $F_{\mathrm{v}}$.

\subsection{Planetary migration}
\label{s:migration}
In this work, we consider a planet on a circular orbit with size $r_{\mathrm{p}}$. 
The eccentricity of a planetary orbit is assumed to be zero, $e = 0$. 
The orbit inclination with respect to the disk is equal to zero as well.
The largest dynamically stable circular planetary orbit can be estimated as 
\begin{equation}
\label{eq:max_planet_orbit}
\frac{r_{\mathrm{p}}^{\mathrm{max}}}{a} = 0.464 - 0.380 \frac{M_1}{M_1 + M_2},
\end{equation}
see \citet{holman1999}. 
Normally, a planet transfers its angular momentum to the disk, 
so that it migrates inwards. 
The planet continues to migrate until reaching the orbit where tidal interactions with the central star become significant. We estimate this critical distance, $r_{\mathrm{p}}^\mathrm{fin}$, as the radius from which a planet can migrate due to tidal forces down to the stellar surface 
by $t_\mathrm{life}$. For calculations, we use the rate of tidal migration (TM hereafter) from~\citet{jackson2008} defined as 
\begin{equation}
\label{eq:complete_migration_orbit}
\frac{r_{\mathrm{p}}^{\mathrm{fin}} (t_{\mathrm{life}})}{R_2} = \left(\frac{117}{4 Q} \left( \frac{m_{\mathrm{p}}}{M_2}\right) \sqrt{\frac{G M_2}{R_2^3} } t_{\mathrm{life}} \right)^{2/13}.
\end{equation}
Here $Q$ is the stellar tidal dissipation parameter. Its expected values for solar-like stars are $Q$ $\sim$ $10^5$--$10^6$ (we select $Q = 10^6$ for the estimation), but higher values of $Q$ up to $10^8$ are also possible. In Eq.~\eqref{eq:complete_migration_orbit}, $m_{\mathrm{p}}$ is the planetary mass which is assumed to be constant (i.e., we neglect mass accretion by the planets), and $t_{\mathrm{life}}$ is the disk characteristic lifetime, or, equally, 
total migration time of the planet. A numerical estimate of the distance~\eqref{eq:complete_migration_orbit} for the given values of parameters is: $r_{\mathrm{p}}^{\mathrm{fin}} (t_{\mathrm{life}}) / R_2 \approx 2.6 \left( m_{\mathrm{p}} / 300 \mathrm{\,m}_{\oplus}\right)^{2/13} \left(t_{\mathrm{life}} / \mathrm{Myr} \right)^{2/13}$. 
Obviously, this value is usually close to unity. For example, the largest value in our simulations is achieved for a planet with $m_{\mathrm{p}} = 300$\,m$_{\oplus}$ in a system with  $t_{\mathrm{life}} \sim 120$\,Myr:  $r_{\mathrm{p}}^{\mathrm{fin}} (t_{\mathrm{life}}) \approx 5.4$~$R_2$. 

We consider two types of migration: 
type \RomanNumeralCaps{1} migration for less massive planets
and type \RomanNumeralCaps{2} migration for more massive planets which open a gap in the disk.

Type \RomanNumeralCaps{1} migration rate is estimated according to~\citet{tanaka2002} as
\begin{equation}
\label{eq:migration_type1}
\frac{\mathrm{d} r_{\mathrm{p}}}{\mathrm{d} t} = - \left(2.72 + 1.08 \beta\right) \frac{m_{\mathrm{p}}}{M_2^2} \aspectratio^{-2} \Sigma r_{\mathrm{p}}^3 \Omega \left(r_{\mathrm{p}}\right),
\end{equation}
where $\aspectratio = \verticalscale / r$ is the aspect ratio of the disk at the planetary orbit, 
$\beta$ is the power law index of the surface density profile 
$\Sigma \propto r^{-\beta}$.

The transition from the type~\RomanNumeralCaps{1} migration to the type~\RomanNumeralCaps{2} migration is determined by the planet-to-star mass ratio,  $q \equiv m_{\mathrm{p}} / M_2$, 
see, e.g.,~\citet{baruteau2014}. Its critical value is
\begin{equation}
\label{eq:critical_mass_planet}
q_{\mathrm{crit}} = \frac{100}{\mathcal{R}} \left( \!(X+1)^{1/3} - (X - 1)^{1/3} \right)^{-3},
 \end{equation}
where $X = \sqrt{1 + \frac{3}{800}  \aspectratio^3 \mathcal{R}}$ is the numerical coefficient and $\mathcal{R} = r_{\mathrm{p}}^2 \Omega_{\mathrm{p}}/\nu$ is the Reynolds number defined by the viscosity of the disk.
If the planet-to-star mass ratio $q$ is less than the critical value, 
then the type \RomanNumeralCaps{1} migration takes place. The type \RomanNumeralCaps{2} migration 
occurs in the other case.

We compare $q$ with the critical mass ratio all the time the planet migrates embedded in the evolving disk. 
So,  the planet can change the type of migration in the course of the disk evolution. 
The transition from type \RomanNumeralCaps{2} to type \RomanNumeralCaps{1} happens not instantaneously. We assume that the gap can be closed on a characteristic timescale which we estimate according to~\citet{armitage2005} as
\begin{equation}
\label{eq:tau_close}
\Delta t_{\mathrm{close}} \sim \frac{\verticalscale^2}{\nu} = \frac{1}{\alpha \Omega_{\mathrm{p}}}.
\end{equation}
For example, if a planet at first evolves according to type \RomanNumeralCaps{2} migration, and later occurs in circumstances that correspond to $q<q_{\mathrm{crit}}$, then 
the transition to the type \RomanNumeralCaps{1} happens only after $\Delta t_{\mathrm{close}} $.

The type \RomanNumeralCaps{2} migration rate is estimated according to 
\begin{equation}
\label{eq:migration_type2}
\frac{\mathrm{d} r_{\mathrm{p}}}{\mathrm{d} t} =  v_{\mathrm{r}} \frac{M_{\mathrm{d}}}{M_{\mathrm{d}} + m_{\mathrm{p}}} = \frac32 \alpha \aspectratio^2 r_{\mathrm{p}} \Omega_{\mathrm{p}} \frac{2 r_{\mathrm{p}}}{F} \frac{\mathrm{d} F}{\mathrm{d} r} \frac{M_{\mathrm{d}}}{M_{\mathrm{d}} + m_{\mathrm{p}}},
\end{equation}
where $M_{\mathrm{d}} = 2 \pi r_{\mathrm{p}}^2 \Sigma_{\mathrm{p}}$ is the
characteristic mass of the disk at the planetary orbit, 
$v_{\mathrm{r}}$ is the radial velocity of the accreting matter~\eqref{eq:radial_drift}, 
and $F$ is the viscous torque taken from Eq.~\eqref{eq:torque}. 
Note that at a given $r$ the value of $v_\mathrm{r }$ can be both 
positive or negative. Therefore, the outward type \RomanNumeralCaps{2} migration is possible. Equation~\eqref{eq:migration_type2} takes into account the decrease 
of the migration rate in the case when $m_{\mathrm{p}} > M_{\mathrm{d}}$, see~\citet{ivanov1999} and~\citet{scardoni2020}.

In order to improve a qualitative presentation of our results, we also define a characteristic migration timescale 
\begin{equation}
\label{eq:migration_timescale}
\hat t_{\mathrm{p}} = \frac{r_{\mathrm{p}}}{\left|\mathrm{d} r_{\mathrm{p}}/\mathrm{d} t\right|}. 
\end{equation}
For the type \RomanNumeralCaps{2} migration, the timescale given by Eq.~\eqref{eq:migration_timescale} corresponds to the local viscous time in a disk. The shorter the timescale, the faster the migration of the planet, and vice versa.

In this work, we would like to avoid a bulk of complexity associated with the description of migration developed recently, see~\citet{paardekooper2023}. Instead, our objective is to carry out a comprehensive study of simple migration in a complex disk in order to obtain a reliable basis for future work. This simplified 
approach allows us to study a parameterized 
problem of migration. We discuss the complexity of migration in 
Sect.~\ref{sec:discussion_migration}.

\section{Numerical setup of the model}
\label{sec:numerical_method}

Equations governing the evolution of the accretion disk have been described in the previous Section. These are Eq.~\eqref{eq:diffusion_equation} which determines $\Sigma = \Sigma (F, \angularmomentum, t)$ and Eq.~\eqref{eq:energy_equation} which determines  $T_{\mathrm{c}} = T_{\mathrm{c}} (\Sigma, \angularmomentum, t)$. These equations are related to each other through the constraint given by Eq.~\eqref{eq:torque}, which defines $F = F (\Sigma, T_{\mathrm{c}}, \angularmomentum, t)$. We use Eq.~\eqref{eq:torque} to express $\Sigma$ via $F$ and $T_{\mathrm{c}}$. The set of Eqs.~\eqref{eq:torque},~\eqref{eq:diffusion_equation},~\eqref{eq:energy_equation} allows us to determine $F$, $\Sigma$ and $T_{\mathrm{c}}$ as functions of the radial coordinate ($r$ and \angularmomentum are interchangeable variables) and time. The scheme employed to solve this non-linear problem numerically is described in detail in App.~\ref{sec_app:numerical_method}.

We additionally introduce the semi-analytical quasi-stationary disk model by replacing Eq.~\eqref{eq:diffusion_equation} with Eq.~\eqref{eq:nu_sigma}. The corresponding solution has
been obtained in~\citet{kulikova2019}. We compare our results with this solution. It is denoted 
as QSD (quasi-stationary disk), while the full non-stationary model described by Eq.~\eqref{eq:diffusion_equation} is denoted as NSD (non-stationary disk) hereafter. The non-stationary disk is referred to as the NS disk hereafter.

The numerical grid for disk evolution is defined as follows. We set a uniform spatial grid in terms of the specific angular momentum, \angularmomentum: $\{ \angularmomentum_n = \angularmomentum_{\mathrm{in}} + n ~ \Delta, \Delta = \left(\angularmomentum_{\mathrm{out}} - \angularmomentum_{\mathrm{in}}\right)/N, n = \overline{0, N}\}$. For all simulations, we set $N = 100$ spatial cells distributed uniformly in the range from $\angularmomentum_{\rm in}$ to $\angularmomentum_{\rm out}$. 
We set the non-uniform numerical grid over time $\{ t_{k}: t_0 < t_1 < \dots < t_{K}, k = \overline{0, K}\}$, where the slices $t_{k}$ are taken from the MESA track grid with optional additional grid thickening performed using a linear interpolation of MESA parameters. The MESA time grid that we use for post-MS stars concentrates cells to the RGB and AGB peaks and has from roughly $3000$ time cells for the most massive donor to more than $11000$ time cells for the least massive donor, see Figs.~\ref{fig:low_mass_wind} and~\ref{fig:high_mass_wind} for the time cells used. We also add $K^{\prime} = 500$ uniformly distributed time cells to have a more dense coverage of evolution stages far from RGB and AGB peaks. The implicit unconditionally stable numerical scheme, see App.~\ref{sec_app:accretion_numerical} for details, allows us to use an arbitrary ratio of grid steps by time and coordinate.

Besides the mathematical stability of the scheme, we ensure the physical convergence of the numerical solution. Regarding the time resolution, our time grid has a time step of about or less than $1$ year close to the RGB and AGB peaks. This is at least two orders of magnitude less than the characteristic wind variability timescale, which has its minima at RGB and AGB peaks. 
At the same time, the viscous timescalemuch larger. 

The sharp gradients of the opacity may occur due to evaporation of dust species, 
which leads to significant gradients of temperature and surface density (or, interchangeably, viscous torque). This implies the limitation of the size of the spatial grid. For $N = 100$ and typical (varying) outer radius, we get the grid cell size of $\sim 0.05$~AU at the inner edge of the disk. The grid cell size increases with radius $\propto \sqrt{r}$. We do not expect sharp gradients of the disk variables at its outer partshere the size of the grid cell gradually attains values up to one AU or to a few AU depending on the location of the outer boundary 
which can approach $100$~AU. We ensure that the numerical solution does not depend on the number of the spatial cells if $N \gtrsim 100$. 

The spurious initial value of the disk viscous torque is set to $F_{\rm init} = 10^{20} \sin\left(\frac{\pi}{2}\frac{\angularmomentum - \angularmomentum_{\rm in}}{\angularmomentum_{\rm out} - \angularmomentum_{\rm in}}\right)$~dyn\,cm, which satisfies the boundary conditions. 
However, we note that the exact choice of the initial condition is of small significance for the following solution, provided that the absolute value of this condition is sufficiently small. The disk loses its initial state over a viscous timescale, which is always much smaller than the time between ZAMS and RGB/AGB. The stability of the numerical solution has been tested for varying amplitude of the initial condition. It has been varied from $10^{20}$~dyn\,cm to zero. So, the zero initial condition is also acceptable in our scheme.

At the initial slice and for all subsequent slices the surface density $\Sigma$ is expressed through $F$ and $T_{\mathrm{c}}$ using Eq.~\eqref{eq:torque}. 
The temperature $T_{\mathrm{c}}$ is obtained as the solution of Eq.~\eqref{eq:energy_equation}. 
This is performed rearranging this equation to the form $f(T_{\mathrm{c}}) = 0$ and looking for 
the roots by the Brent algorithm~\citep{brent1973}, for which we use the eponymous method \textsc{brentq} from \textsc{python} library \textsc{scipy.optimize}.

The boundary condition is imposed according to a numerical form of Eq.~\eqref{eq:boundary_fl_F}. Note that the outer boundary of a disk defined in Eq.~\eqref{eq:r_tidal} depends on time: $\angularmomentum_{\mathrm{out}} = \angularmomentum_{\mathrm{out}} (a) = \angularmomentum_{\mathrm{out}} (t_k)$. Thus, the spatial step $\Delta = \Delta (t_k) = \Delta_k$ and spatial nodes $\angularmomentum_n = \angularmomentum_n (t_k) = \angularmomentum_{k, n}$ also depend on time. They must be updated at every slice accounting for the binary orbital evolution, see Eq.~\eqref{eq:separation} which is also integrated numerically. For the 
numerical simulations, we select the range of initial separations $a$ from $10$~AU to $100$~AU with the step of $10$~AU for all donor masses,see Table~\ref{tab:model_free_parameters_binary}.


After the calculation of the accretion disk evolution, we calculate the planet migration. The details of the scheme are described in App.~\ref{sec_app:migration}. We interpolate 
the numerical variables describing the accretion disk evolution to a better spatially and time-resolved grid. 
The spatial resolution of the new grid 
becomes less than $0.02$~AU. 
The time resolution of the new grid is about $1$ year during the rapidly varying wind epochs (RGB and AGB) and not worse than $100$ years during the whole evolution. 
Using this grid, we simulate planet migration according to equations described in Sect.~\ref{s:migration}. We apply the described scheme to a range of planet masses from $1$~m$_{\oplus}$ to $3000$~m$_{\oplus}$ and a range of initial planet orbit separations from $0.25$~AU to the largest dynamically stable circular planetary orbit, defined by Eq.~\eqref{eq:max_planet_orbit}. For the visual representation of results over the wide range of free parameters, we use bilinear interpolation to plot the contours of migration as described in App.~\ref{sec_app:contours}.

\section{Results}
\label{sec:results}

\subsection{Accretion disk}
\label{sec:results_accretion}

For further analysis, we define several time moments that are the same for all disks with the specified donor:
\begin{itemize}
\item $\tRGB$ is the moment of the maximum wind rate at the red giant stage, see the first peak in Figs.~\ref{fig:low_mass_wind},~\ref{fig:high_mass_wind};
\item $\tAGB$ is the moment of the maximum wind rate at the stage of the asymptotic giant branch, see the second (main) peak in Figs.~\ref{fig:low_mass_wind},~\ref{fig:high_mass_wind};
\item \tenvloss is the moment of the envelope loss by the donor when the disk is not supplied with matter anymore.
\end{itemize}

\begin{figure*}
    \centering
    \includegraphics[width=0.246\textwidth]{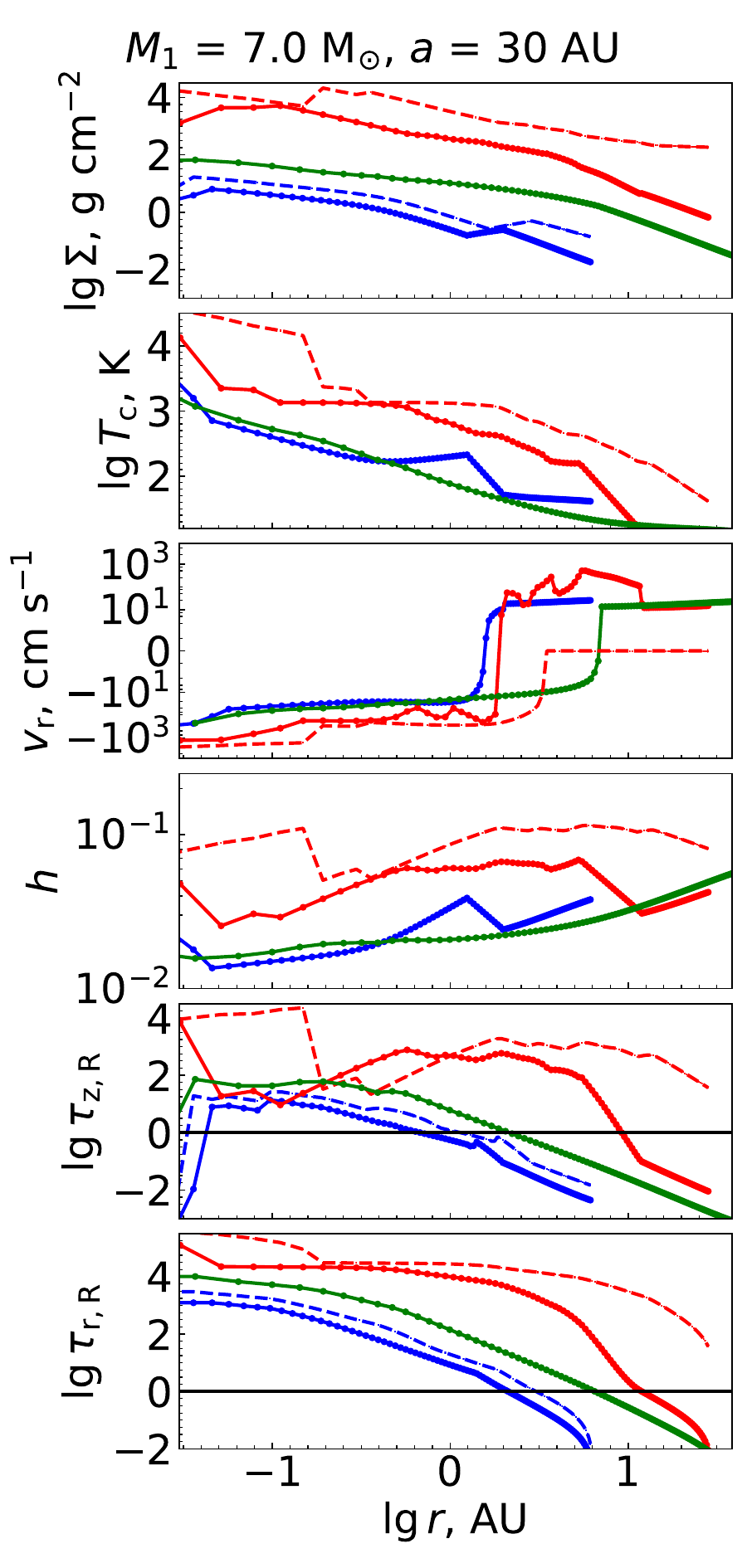}
    \includegraphics[width=0.246\textwidth]{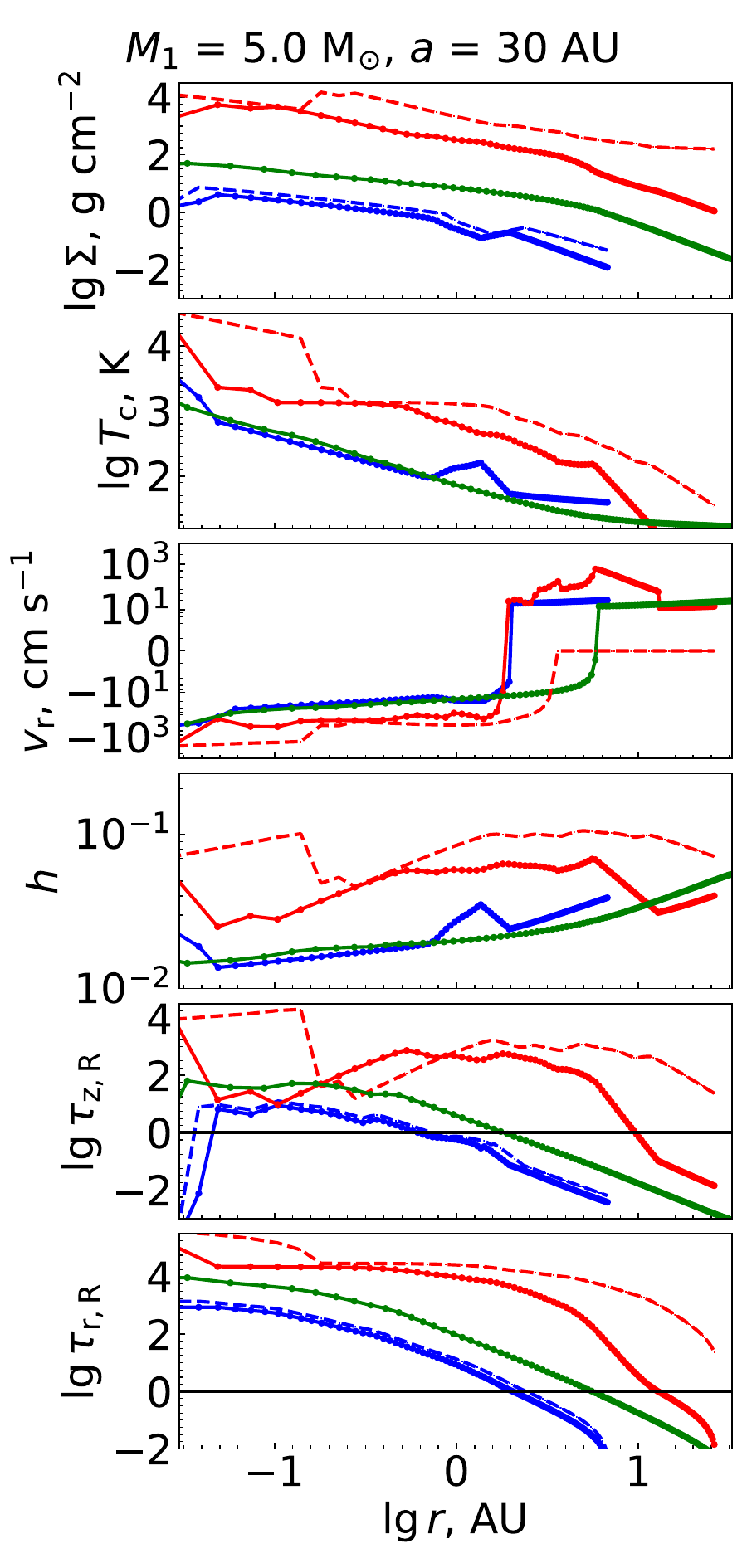}
    \includegraphics[width=0.246\textwidth]{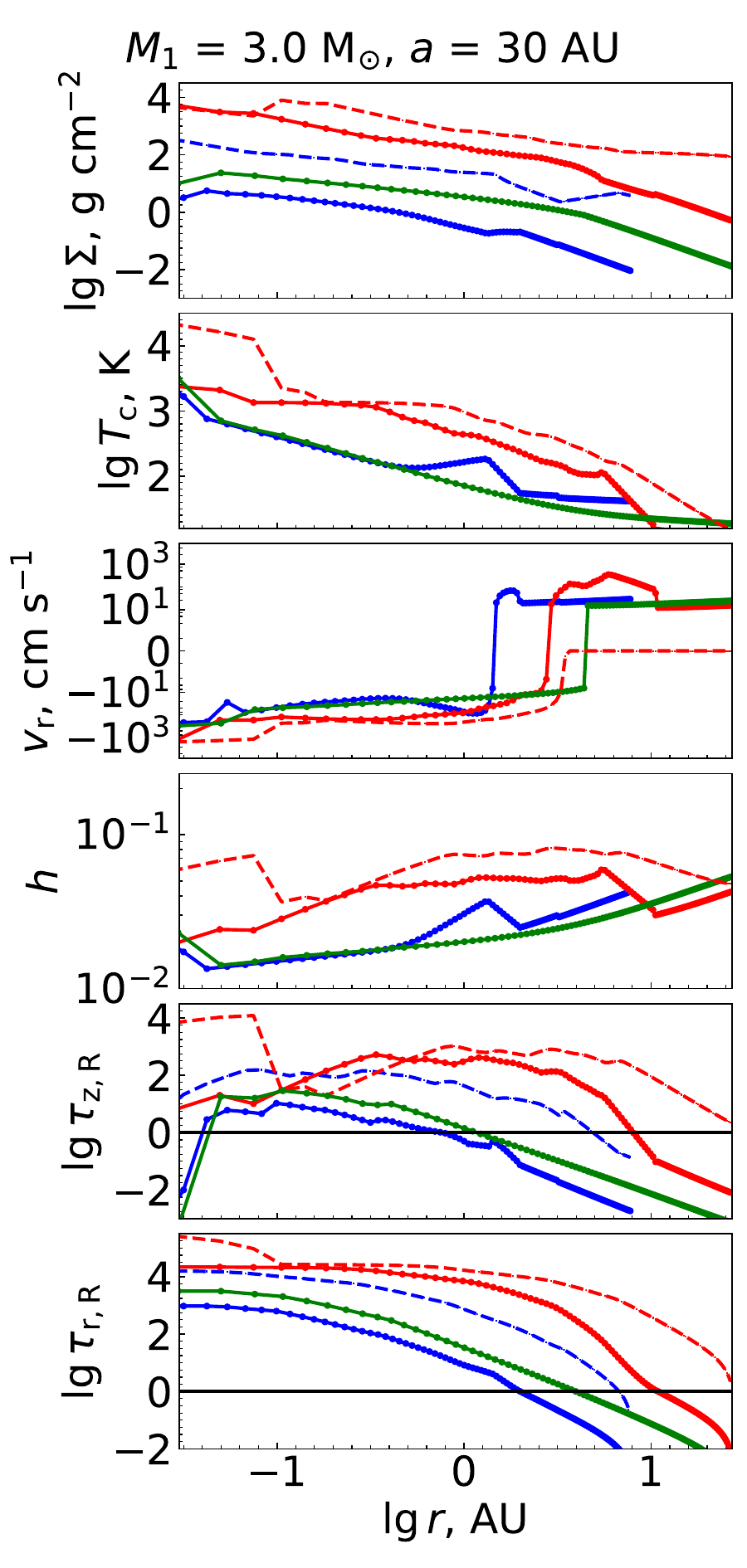}
    \includegraphics[width=0.246\textwidth]{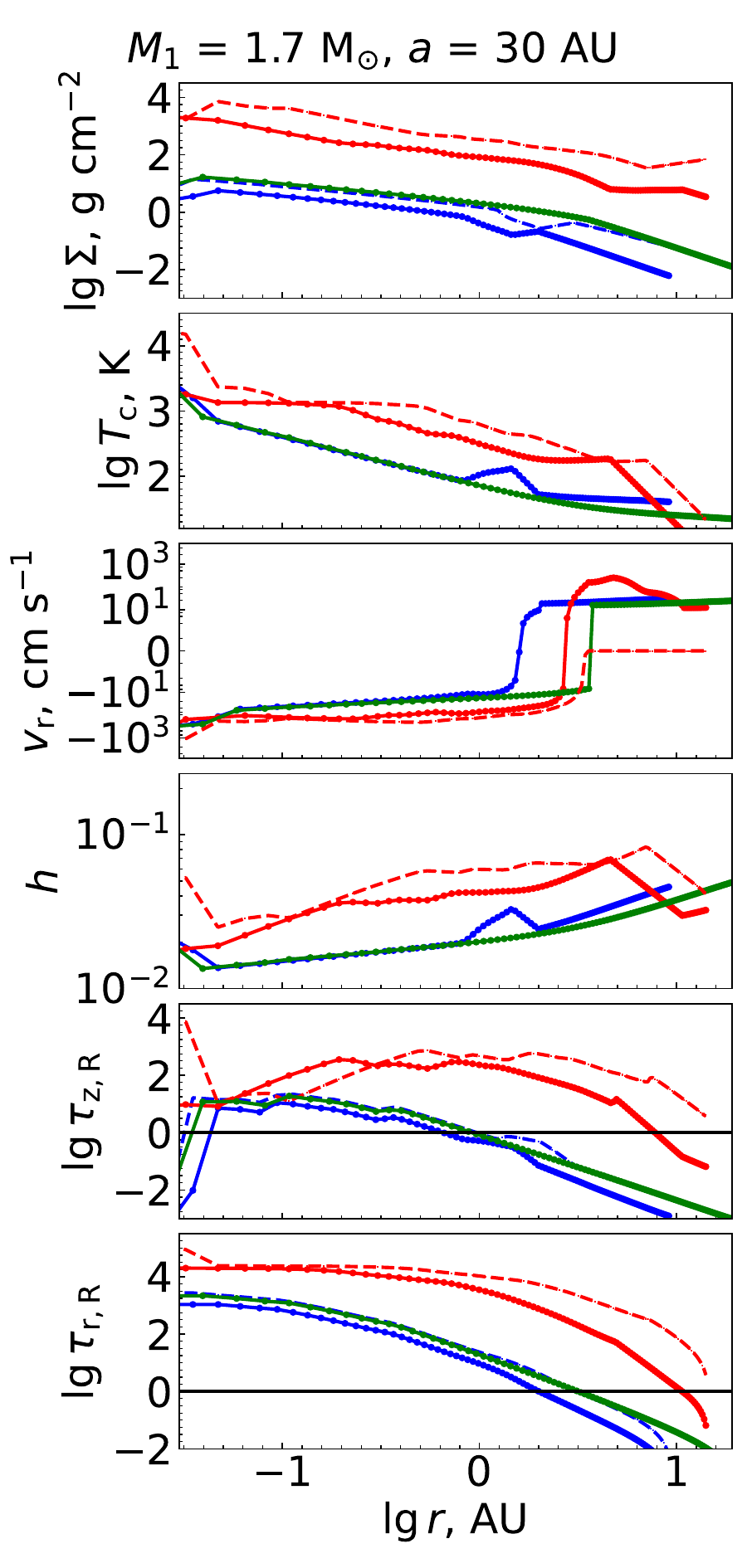}
    \caption{Each column corresponds to a different initial donor mass, $M_1$, of our sample. The initial binary separation is $a = 30$\,AU for all columns. The rows from 1 to 6 show the disk profiles of surface density $\Sigma$, midplane temperature $T_{\mathrm{c}}$, radial velocity $v_{\mathrm{r}}$, aspect ratio $\aspectratio = \verticalscale / r$, optical thickness in vertical and radial directions $\tau_{\mathrm{z}, \mathrm{R}}$, $\tau_{\mathrm{r}, \mathrm{R}}$, respectively. 
    Blue, red, and green curves represent a disk at the time $t=\tformation, \tAGB, \tdepletion$, respectively. For the values of the time moments see Table~\ref{tab:times}. Solid and dashed curves show, respectively, NSD and QSD. The markers of the solid curves correspond to the spatial grid of the solution. QSD curves are presented only on plots with $\Sigma$, $\tau_{\mathrm{z}, \mathrm{R}}$ and $\tau_{\mathrm{r}, \mathrm{R}}$, otherwise, for the moment $\tAGB$ only. See Table~\ref{tab:snow_bondi_radii} for the values of the Bondi radius and the snow line radius. }
    \label{fig:accretion_total}
\end{figure*}

Additionally, we introduce the time moments that specify the disk in the system with
particular initial separation:
\begin{itemize}
\item $\tformation$ is the moment of disk formation when the accretion rate becomes high enough to produce a disk with considerable size. 
We formally take $\tformation$ corresponding to the disk with size $2$~AU when the accretion rate for the first time attains $\dot M_{\mathrm{acc}} \simeq \dot M_{\mathrm{acc}}^{\mathrm{lim}}$ introduced below in Sect.~\ref{sec:analytics_general_disk_features}.
\item $\tdepletion$ is the moment after the envelope loss by the donor, which is in 
the midst of disk depletion during its closed evolution;
\item \tdecay is the moment of the disk decay at the end of its closed evolution when $r_{\tau} \to r_{\mathrm{in}}$. 
\end{itemize}
The time moments $\tformation$, $\tAGB$ and $\tdepletion$ are used to show the disk 
profiles in Fig.~\ref{fig:accretion_total}.

The universal sequence of the time moments is $\tZAMS < \tformation^* < \tRGB < \tformation^* < \tAGB < \tenvloss < \tdepletion < \tdecay$, where the asterisk at \tformation means, that this time moment can be either before \tRGB or after \tRGB depending on the initial system separation and initial donor mass. See also Table~\ref{tab:times}, where these time moments are presented for different donors in a system with initial binary separation $a=30$~AU.

Finally, we select several spatial regions in the disk. The regions are related to the optical thickness determined by the wind rate or, alternatively, the accretion rate. The optical thickness substantially varies across the disk.
 
\begin{itemize}

    \item Region 1 (R1): standard optically thick region, where  $\tau_{\mathrm{z}, \mathrm{R}} > 1$, i.e., $\tau_{\mathrm{r}, \mathrm{R}} \gg 1$. The disk contains an extended R1 at high accretion rates. Typically, this corresponds to  $\dot M_{\mathrm{acc}} \gtrsim 10^{-8}$~M$_{\sun}$\,yr$^{-1}$. This region is the most important for planetary migration because it is characterized by the largest disk mass up to $\sim 0.01$~M$_{\sun}$, the largest surface density up to $\sim 10^4$~g\,cm$^{-2}$ and the largest viscous torques. Below, in Sections \ref{sec:accretion_general} and \ref{sec:accretion_other_donors}, the disk with R1 extending up to a considerable value which is not less than 1~AU is represented by the curves taken at $t = \tAGB$ for all donors and by the curves at \tdepletion only for donors $M_1 = 7.0, 5.0$\,M$_{\sun}$, see Fig.~\ref{fig:accretion_total}.
    
    \item Region 2 (R2): transitional region where $\tau_{\mathrm{z}, \mathrm{R}} \lesssim 1$ but $\tau_{\mathrm{r }, \mathrm{R}} > 1$. Usually, a disk with an outer R2 having a size $\sim 2$~AU forms in epochs when $\dot M_{\mathrm{acc}} \gtrsim 10^{-11}$~M$_{\sun}$\,yr$^{-1}$, see the derivation of the corresponding estimate, Eq.~\eqref{eq:tau_r_estimated}, in Section~\ref{sec:analytics_general_disk_features}\footnote{To compare accretion rate and wind rate in different systems, one can use the estimate from Eq.~\eqref{eq:bondi_accretion_rate}: $\dot M_{\mathrm{acc}} / \dot M_{\mathrm{w}} = r_{\mathrm{a}}^2 / 4 a^2 \approx 0.04 (M_2 / \mathrm{M}_{\sun})^2 (a / 10 \mathrm{\,AU})^{-2} [(v_{\mathrm{w}} / 20 \mathrm{\,km\,s}^{-1})^2 + (c_{\mathrm{w}} / 10 \mathrm{\,km\,s}^{-1})^2]^{-2}$.}. In this case, a limited R1 appears in the inner part of the disk. As far as $\dot M_{\mathrm{acc}} \lesssim 10^{-11}$~M$_{\sun}$\,yr$^{-1}$, disk becomes fully optically thin in the vertical direction. Planet migration in R2 becomes slower in general. Nevertheless, it can still be non-negligible for high-mass planets experiencing type~\RomanNumeralCaps{1} migration in comparison with the migration of low-mass planets in R1. In Sections~\ref{sec:accretion_general} and~\ref{sec:accretion_other_donors}, the disk with R2 extending enough is represented by the curves at $\tformation$ for all donors and the curves at $\tdepletion$ only for donors $M_1 = 3.0, 1.7$\,M$_{\sun}$, see Fig~\ref{fig:accretion_total}.

    \item Region 3 (R3): optically thin region. In this region, the disk becomes optically thin in the radial direction: $\tau_{\mathrm{r }, \mathrm{R}} < 1$. An assumption of the local energy balance is far from being valid and thus, our model predictions must be inaccurate in this region. Therefore, we use a floating bound in our calculations, see Eq.~\eqref{eq:boundary_fl_F}. It separates R2 from R3. Also, we apply the condition of decretion of R3, see Eq.~\eqref{eq:boundary_fl_F}. The low value of optical thickness in the radial direction specific to R3 prevails at low accretion rates: disk usually becomes fully optically thin for $\dot M_{\mathrm{acc}} \ll 10^{-11}$~M$_{\sun}$\,yr$^{-1}$. The disk has extremely low mass if R3 is dominating, see Sections~\ref{sec:accretion_general} and~\ref{sec:accretion_other_donors} for details.
    The corresponding planet migration is also negligible, for details see Sect.~\ref{sec:migration_general}. 
    Note that these results were also checked within the model with the outer boundary imposed at the tidal truncation radius only.

\end{itemize}

The evolution of an accretion disk starts from the formation of R2 close to the host star. As the increasing stellar wind rate approaches the value corresponding to the accretion rate $\dot M_{\mathrm{acc}} \simeq 10^{-11}$~M$_{\sun}$\,yr$^{-1}$, 
R2 expands into the surrounding R3 where the accreted matter stays to be optically thin. The further increase of the wind rate leads to the formation of R1 close to the host star. Subsequently, R1 and R2 both spread outwards. Once the wind rate starts to decrease, the reverse process takes place. In the case when the wind ceases after the envelope loss by the donor, the reverse process continues until the disk decays. By the disk decay we mean the moment when both R1 and R2 vanish. In the systems with low-mass donors, the wind rate may decrease so much that the disk decays temporarily between $\tRGB$ and $\tAGB$. 

In Sect.~\ref{sec:accretion_general}, we present a detailed description of disk evolution in the system with the most massive donor $M_1 = 7.0$~M$_{\sun}$. After that, in Sect.~\ref{sec:accretion_other_donors}, we highlight the variations of disk evolution in the case of less massive donors, $M_1 = 5.0, 3.0$ and $1.7$~M$_{\sun}$. We start by demonstrating the importance of the non-stationary approach and then describe the heating of the disk by the donor. 

We also remind that the abbreviations NSD and QSD mean non-stationary and quasi-stationary disk models, respectively. When referring to a physical non-stationary disk, we use the notation ``NS disk''.

\subsubsection{Relevance of the non-stationary evolution model}
\label{sec:qsd_vs_nsd}

\begin{figure}
    \centering
    \begin{minipage}{\columnwidth}
    \includegraphics[width=\columnwidth]{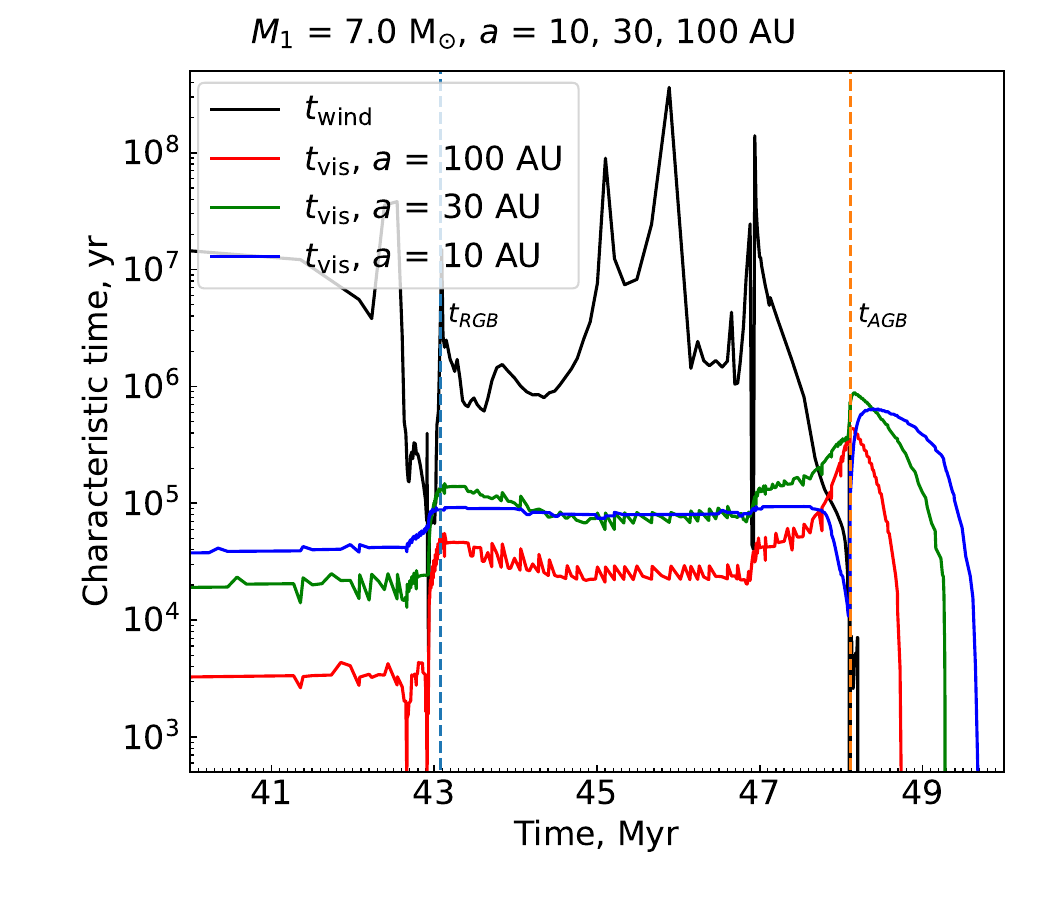}
    \includegraphics[width=\columnwidth]{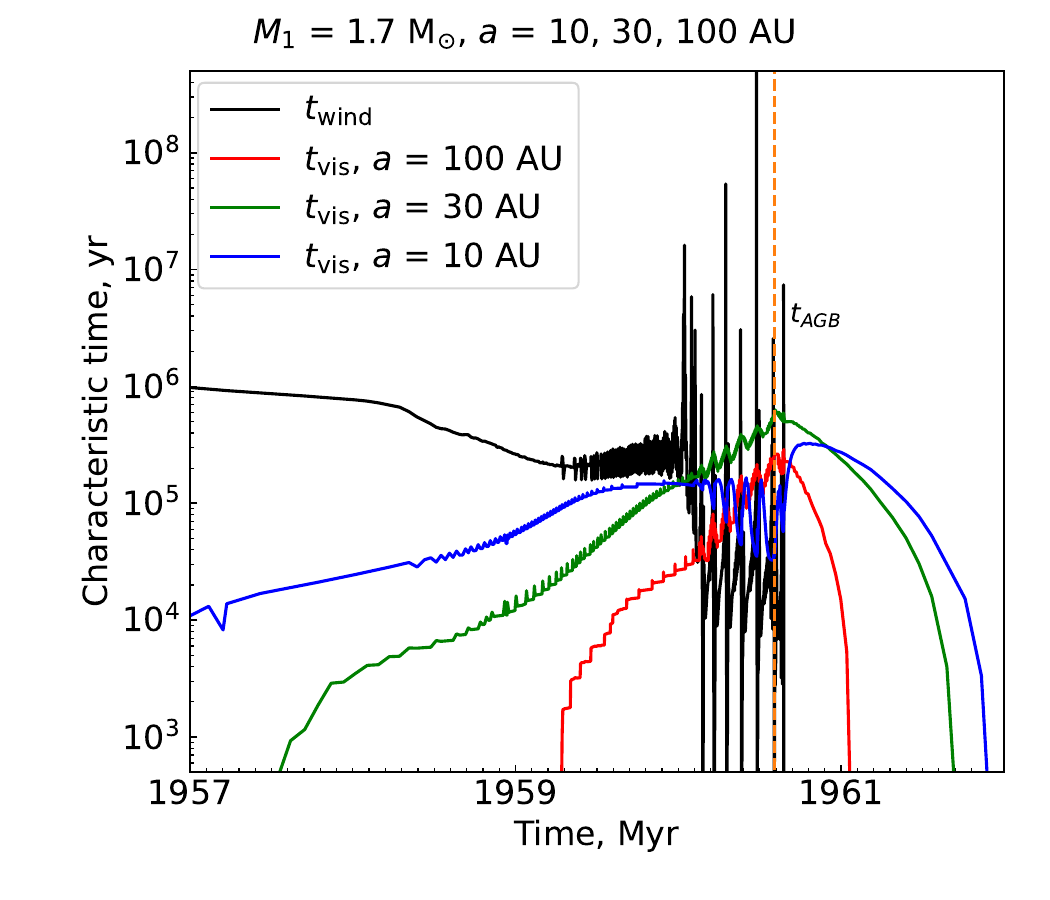}
    \end{minipage}
    \caption{Black curve: characteristic timescale of wind variability, $t_{\mathrm{w}} = \dot M_{\mathrm{w}} / \ddot M_{\mathrm{w}}$. Blue, green, and red curves: characteristic viscous timescale of the disk, $t_{\nu} = r^2 / \nu$, taken at the outer boundary of NS disk in the system with initial separations $a= 10, 30, 100$~AU, respectively. The initial donor mass $M_1 = 7.0$~M$_{\sun}$ and $M_1 = 1.7$~M$_{\sun}$ on the top and the bottom panels, respectively.}
    \label{fig:viscous_times}
\end{figure}

The results of~\citet{kulikova2019} are based on QSD. That model is satisfactory while the characteristic timescale of wind rate variation, $t_{\mathrm{w}}$, 
is much longer than the timescale of the disk non-stationary evolution. The latter scale is usually approximated by the viscous timescale of a disk, 
\begin{equation}
\label{t_visc}
t_{\nu} = 1/(\alpha \aspectratio^2 \Omega) \ll t_{\mathrm{w}}, 
\end{equation}
where $\Omega$ and $\aspectratio$ are taken at the outer boundary of the disk. Employing NSD, we compare $t_\nu$ with $t_{\mathrm{w}}$ in the systems with 
the lowest mass donor and the most massive one for several initial separations,
see Fig.~\ref{fig:viscous_times}. 

This comparison suggests that QSD should describe well the evolution of a disk supplied from the wind in epochs far from RGB and AGB peaks. Moreover, $t_{\nu}$ decreases in wider systems since the smaller accretion rate leads to the formation of smaller disks. This picture is complicated by the truncation of the disk at the tidal radius in the closest systems, see the curve for $a = 10$\,AU in Fig.~\ref{fig:viscous_times}. As one can see, the residual disk truncated at the tidal radius lives longer than the disk with the floating outer boundary represented by the curves for $a = 30, 100$\,AU, despite that the latter is initially larger and, thus, it has initially larger $t_{\nu}$. Such a peculiarity reflects the underestimation of the timescale of the truncated disk non-stationary evolution in comparison with its viscous timescale. In general, we find that an external long-term source of matter allows the disk to exist for a much longer time than $t_{\nu}$, which stays within $\lesssim 10^5$~years, except the time around AGB peak, where it attains $\sim 10^6$~years. In the latter case, the viscous timescale corresponds to the disk with the size $\sim 10$~AU and $\aspectratio \sim 0.05$, which is in accordance with the residual disk profiles in Fig.~\ref{fig:accretion_total}, see below for more details.

However, the condition of quasi-stationarity, Eq.~\eqref{t_visc}, is not valid during the whole period of the system evolution. It can be seen that $t_{\mathrm{w}}$ becomes comparable or even less than $t_{\nu}$ around $\tAGB$ for all donors and initial separations. This is also the case around $\tRGB$ for initially close systems with high mass donors. The predictions of QSD are inaccurate at these times. This is demonstrated below in Fig.~\ref{fig:accretion_total} by the large difference between the disk profiles of QSD and NSD at $t=\tAGB$. 
Additionally, the disk profiles of QSD and NSD significantly differ from each other at $t=\tformation$ for the donor with $M_1=$~3.0 $\mathrm{ M}_{\sun}$, see column 3 of Fig.~\ref{fig:accretion_total}, because in this case the sufficiently large disk with the size $\sim 2$ AU is accumulated right by the beginning of prominent oscillations of the wind which is typical for AGB phase of the lower mass donors. These wind oscillations can be seen on the bottom panel in Fig.~\ref{fig:viscous_times}.

We find that in all these cases NSD provides an order of magnitude smaller surface density and somewhat smaller aspect ratio of the disk than is predicted by QSD. Looking at the disk profiles at the moment of the increasing accretion rate, it is clear that NSD always lags behind QSD. Thus, disks in NSD are found to be more (less) massive than in QSD at the same moments of time during the decrease (increase) of the wind rate. In other words, the NSD solution tends to the QSD solution on a rather long timescale corresponding to a few viscous timescales of the disk, which has been checked in our test calculations. The lag between NSD and QSD is negligible as long as the condition $t_{\nu} \ll t_{\mathrm{w}}$ is valid, but the lag grows along with the ratio $t_{\nu}/ t_{\mathrm{w}}$.

We also note that, unlike QSD, NSD enabled us to study the residual disk left after the envelope loss by the donor. NSD provides accurate initial disk profiles that determine the isolated evolution of the residual disk without the matter inflow. Note that these initial data are the outcome of essentially non-stationary dynamics of wind-fed disks during the preceding AGB phase. Also, note that the radial velocity of the residual disk acquires a smooth profile and becomes negative throughout the disk up to its floating boundary. This is in accordance with its quasi-stationary state, see the corresponding row 3 in Fig.~\ref{fig:accretion_total}.

It is important to emphasize the peculiar profile of the radial velocity in the essentially NS disk.
NSD demonstrates an extended zone of decretion, $v_{\mathrm{r}}>0$, in the outer parts of the disk at $t=\tAGB$. The profile of the radial velocity significantly deviates from the known self-similar solution first obtained by \citep{lynden-bell1974}.
The zone of decretion starts from $\gtrsim 2$~AU ($\lesssim 2$)~AU for the least (most) massive donor. Thus, it goes far inside the Bondi radius, which is within $r_\mathrm{a} = 3\text{--} 4$ AU for all systems around $t=\tAGB$, see also Table~\ref{tab:snow_bondi_radii} for the exact values. 
It is also instructive to compare the zone of decretion 
at $t=\tAGB$ with that at $t=\tdepletion$ when the disk is close to a quasi-stationary state. 
Despite the latter case, the disk is even smaller than its progenitor at $t=\tAGB$, and its own zone of decretion starts considerably farther from the host star.

Decretion takes place due to the rapid accumulation of matter in the middle part of the disk. This part is limited by the Bondi radius from the outside and the radius where the local viscous timescale becomes longer than the timescale of wind variation from the inside. The corresponding abnormal distribution of surface density causes the gradient of the viscous stress to change sign. This leads to the inverse flux of an angular momentum through the middle part of the disk. Note that the absolute value of $v_{\mathrm{r}}$ in the middle of the zone of decretion can exceed the standard quasi-stationary value inferred from the instant disk temperature using Eq.~\eqref{eq:boundary_fl} up to several times when taken at the same location in a disk and 
up to an order of magnitude when taken at the outer boundary. 
This is visible, e.g., by the profile of $v_{\mathrm{r}}$ shown at $t=\tformation$ for the donor with $M_1=3.0\mathrm{~M}_{\sun}$, see the range $1.5\lesssim r \lesssim 2.0$ AU in Fig.~\ref{fig:accretion_total}, column 3.
We check the validity of an enhanced decretion in the essentially NS disk
performing an additional numerical test with a simple disk and wind models, see App.~\ref{sec_app:decretion} for details.

For the lower mass donors, $M_1=1.7\mathrm{~M}_{\sun}$ and $M_1=3.0\mathrm{~M}_{\sun}$, the zone of decretion exists during the whole AGB phase. That manifests itself in the form of prominent growing oscillations of the wind rate. For example, it lasts up to $\sim 0.5$ Myr in systems with the donor with $M_1=3.0\mathrm{~M}_{\sun}$. The zone of decretion shrinks outwards for a while, which occurs right after the sharp wind rate minimum. Just before that, for a short period $\sim 10^3 \text{--} 10^4$ yrs, the zone may split into two parts. Thus, an additional layer of accretion may emerge for a while in the range $2\text{--}4$ AU. We check that this feature occurs due to the heating of the outer parts of the disk by the donor. 
The higher mass donors, $M_1=5.0\mathrm{~M}_{\sun}$ and $M_1=7.0\mathrm{~M}_{\sun}$ have a single peak of the wind rate which occurs right at $t=\tAGB$, by definition. The corresponding zone of decretion arises already $\sim 1$~Myr before $t=\tAGB$ being attached to R2. However, it becomes significant only $\sim 0.1$~Myr before $t=\tAGB$ and vanishes shortly after the peak of the wind rate. Note that an additional layer of accretion appears for $\sim 5$~kyr shortly after the wind rate attains its maximum. However, this layer is attached to R2, which is adjacent to the floating boundary of the disk. At the same time, R2 widens due to a prominent expansion of the system as the donor intensively loses its mass. Similar to the systems with the lower mass donors, an additional layer of accretion emerges due to the heating of R2 by the donor. As the binary becomes wider, the zone of decretion becomes less extended for all donors. For the binary initial separation as large as $100$ AU, it stays attached to R2 adjacent to the floating boundary of the disk. 

\subsubsection{Donor impact on the disk structure}
\label{sec:donor_impact}

\begin{figure}
    \centering
    \includegraphics[width=\columnwidth]{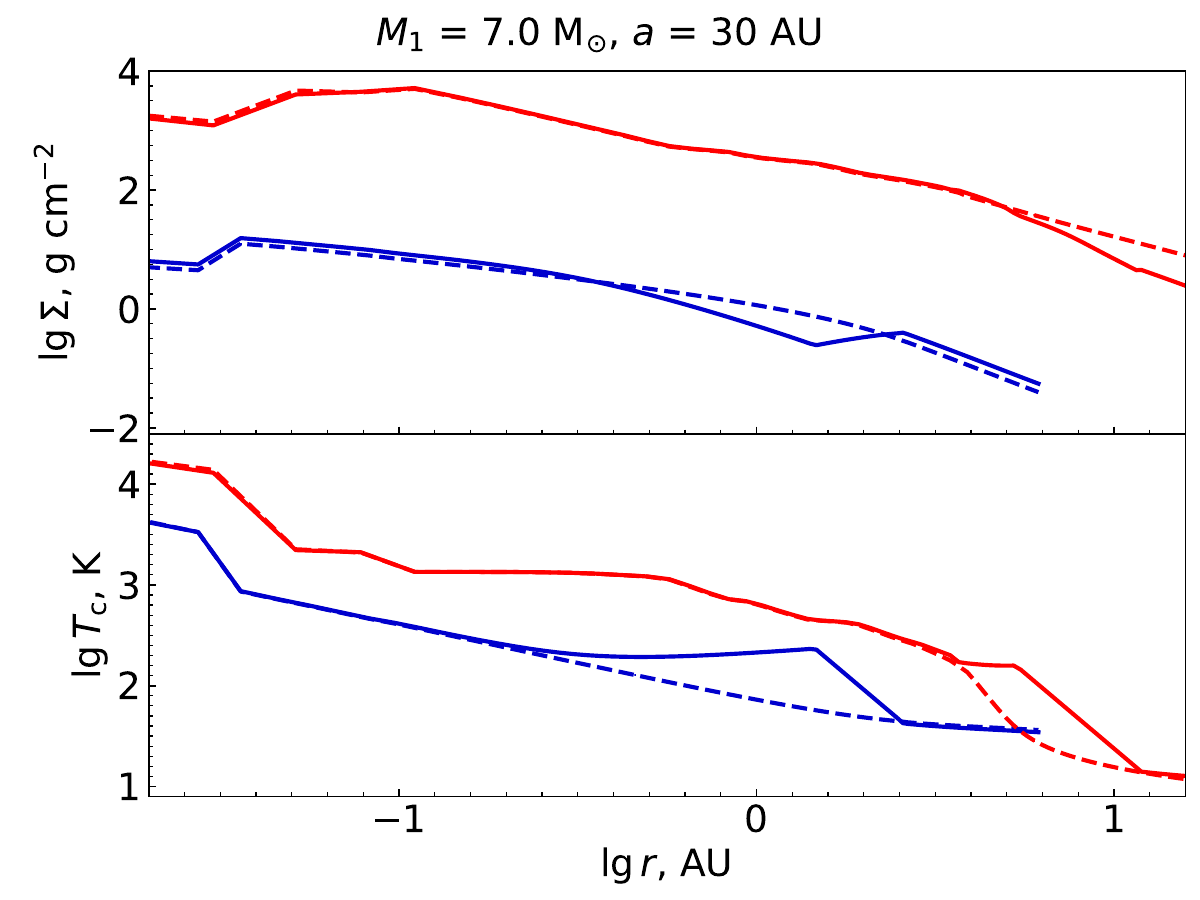}
    \caption{The donor heating of a disk sometime after the disk formation at $t=\tformation=42.966$\,Myr, see Table~\ref{tab:times}, (blue curve) and at $t=\tAGB=48.111$\,Myr (red curve). NSD profiles of surface density (top panel) and midplane temperature (bottom panel) are presented. The solid lines show the results with donor heating taken into account. The dashed lines show the results without accounting for the donor heating, i.e., for $T_{\mathrm{I, D}}^4$ set to zero in Eq.~\eqref{eq:energy_equation}.}
    \label{fig:donor_influence}
\end{figure}

The luminosity of the donor changes significantly in the course of donor evolution after the Main Sequence. The dramatic increase in the donor size leads to essential heating of the outer parts of the disk. This kind of heating requires the following condition, obtained for the conical disk geometry assumption, see Fig.~\ref{fig:donor_appendix} for a visual representation of the calculation:
\begin{equation}
\label{eq:estimate_maximum_size}
R_1 > \verticalscale_{\tau} \approx 1 \mathrm{\,AU} \left(\frac{r_{\tau}}{10 \mathrm{\,AU}}\right) \left(\frac{\aspectratio_{\tau}}{0.1}\right),
\end{equation}
where $\verticalscale_{\tau}$ is the vertical scale height of the disk at $r = r_{\tau}$. $\verticalscale_{\tau}$ approximates the largest vertical scale height in a disk. Let us remind that $r_{\tau}$ is the radial coordinate corresponding to the radial optical thickness $\tau_{\mathrm{r, R}} = 1$, while $\aspectratio_{\tau} = \verticalscale_{\tau} / r_{\tau}$.

The condition~\eqref{eq:estimate_maximum_size} suggests that the disk heating by the donor takes place for the donor size above one to several AU depending on the wind rate and the binary separation. Such a value is typical for the considered donors at RGB and AGB stages, $\tRGB \lesssim t \leq \tAGB$. Our calculations confirm this expectation, see the typical case in Fig.~\ref{fig:donor_influence} where a ``hump'' of $T_{\mathrm{c}}$ near the outer edge of the disk is caused by the donor impact. The temperature hump is seen on the solid curves, which are compared to the dashed curves obtained in the absence of the donor heating. Numerical tests show that the temperature hump exists in a disk for the whole range of the initial orbital separation studied here, $a \in (10, 100)$~AU, as well as for all considered initial donor masses. Note that this feature can be found also in Fig.~\ref{fig:accretion_total}. 

The rise of disk temperature produced by donor heating is usually several times larger than that produced solely by the host star. It can exceed the temperature of the disk obtained in the absence of the donor heating up to an order of magnitude for the most extreme combination we studied here: the high mass donor with $M_1 = 7.0$~M$_{\sun}$ and the closest binary with initial separation $a = 10$~AU. At the same time, the donor heating becomes weaker as the initial separation increases and/or we consider the less massive donor. The radial extent of the temperature hump caused by donor heating usually attains several AU in the outer parts of the disk, while the inner parts of the disk within $\sim 1$~AU remain unaffected due to the disk self-shadowing. The temperature hump also leads to the corresponding decrease of surface density, since $T_{\mathrm{c}}$ and $\Sigma$ are approximately related through the condition $F \sim \nu \Sigma \sim T_{\mathrm{c}} \Sigma \sim const$ valid at least in the quasi-stationary regime. As a consequence, an ordinary decrease of surface density toward the disk outskirts is replaced by an approximately flat radial profile $\Sigma (r) \approx const$, or in some cases, even by the inverse radial profile $\Sigma (r) \propto r^{\beta}, ~ \beta > 0$. Such a radial profile of surface density is a distinctive feature of a disk heated by a luminous donor. Moreover, it can be accompanied by the inverse radial profile of temperature, see the curves for 
a cooler disk at \tformation in Fig.~\ref{fig:donor_influence}.

The detailed analytical derivation of donor heating is relegated to the Appendix~\ref{app:donor}.

\subsubsection{The condition of disk formation}
\label{sec:analytics_general_disk_features}

Our simulations apply to disks that are optically thick in the radial direction. 
This corresponds to some lower limit on~$\dot M_{\mathrm{acc}}$. To obtain the 
condition of the optically thick disk formation, we 
put rough limits on the optical thickness in both $z$- and $r$-directions. We derive an estimate for $\tau_{\mathrm{z}, \mathrm{R}}$ from Eqs.~\eqref{eq:optical_thickness_vertical} and~\eqref{eq:energy_equation} assuming that $T_{\mathrm{c}} \approx T_{\mathrm{I}, \mathrm{A}}$ which is true for a low-mass disk, and also assuming that the disk can be described within the QSD approach. Thus, employing Eq.~\eqref{eq:nu_sigma} we get 
\begin{equation}
\label{eq:tau_estimated}
\tau_{\mathrm{z}, \mathrm{R}} \approx 0.03 \left( \frac{\kappa_0}{3.0\mathrm{~cm}^2 \mathrm{\,g}^{-1}}\right) \left( \frac{\alpha}{0.01}\right)^{-1} \left( \frac{r}{\mathrm{2~AU}}\right)^{-1} \left( \frac{\dot M_{\mathrm{acc}}}{10^{-11} \mathrm{~M}_{\sun} \mathrm{\,yr}^{-1}}\right),
\end{equation}
where $\kappa_0$ is the value of Rosseland opacity for a given temperature: at low temperatures, it varies within the range $\sim 1 \text{--} 5$~cm$^2$\,g$^{-1}$. We note that the normalization of distance at $r=2$~AU is a matter of convention. We chose this value as it is close to the location of the snow line in the preceding protoplanetary disk. The corresponding part of the protoplanetary disk has been a nursery of planets which later appeared to be embedded in the secondary disk studied in this work. Thus, such a choice of the secondary disk size can be justified by the general idea that the region of the snow line should be most populated by planets.

A similar rough estimate is derived for $\tau_{\mathrm{r}, \mathrm{R}}$ from Eqs.~\eqref{eq:optical_thickness_radial},~\eqref{eq:energy_equation}, and~\eqref{eq:tau_estimated} using the assumptions already made in order to obtain Eq.~\eqref{eq:tau_estimated}. In addition, it is necessary to 
employ the radial power-law scalings 
of the optical thickness in the vertical direction as well as the vertical scale height which are derived within QSD. We obtain the following dependencies: $\tau_{\mathrm{z}, \mathrm{R}} \sim r^{-1}$ and $\verticalscale \sim r^{5/4}$. 
Finally, we find 
\begin{multline}
\label{eq:tau_r_estimated}
\tau_{\mathrm{r}, \mathrm{R}} = \int\limits_{r}^{r_{\mathrm{out}}}  \frac{\tau_{\mathrm{z}, \mathrm{R}}}{\verticalscale} \mathrm{d} r^{\prime} \approx \frac{\tau_{\mathrm{z}, \mathrm{R}} (r)}{\verticalscale (r)} \int\limits_{r}^{r_{\mathrm{out}}} \left( \frac{r^{\prime}}{r}\right)^{-9/4} \mathrm{d} r^{\prime} = \tau_{\mathrm{z}, \mathrm{R}} (r) \frac{r}{\verticalscale (r)} \times \\
\times \frac45 \left[ 1 - \left(\frac{r}{r_{\mathrm{out}}} \right)^{5/4}\right] \approx \frac{\tau_{\mathrm{z}, \mathrm{R}} (r)}{\aspectratio} \frac{r_{\mathrm{out}} - r}{r_{\mathrm{out}}} \approx 1.0 \left( \frac{r_{\mathrm{out}} - r}{r_{\mathrm{out}}}\right) \times \\
\times \left( \frac{\aspectratio}{0.05}\right)^{-1} \left( \frac{\kappa_0}{3.0}\right) \left( \frac{\alpha}{0.01}\right)^{-1} \left( \frac{r}{\mathrm{2~AU}}\right)^{-1} \left( \frac{\dot M_{\mathrm{acc}}}{\dot M_{\mathrm{acc}}^{\mathrm{lim}}}\right),
\end{multline}
where $\dot M_{\mathrm{acc}}^{\mathrm{lim}}$ is the previously mentioned threshold value for the formation of a disk with a radial size of $2$~AU, which is optically thick in the radial direction. For a given normalization, this value is $\dot M_{\mathrm{acc}}^{\mathrm{lim}} \approx 1.7 \times 10^{-11}$~M$_{\sun}$\,yr$^{-1}$. 

Equation~\eqref{eq:tau_r_estimated} implies that disk of size $r \sim 2$~AU sufficient to cover the substantial part of pre-existing planetary system is formed provided that an accretion rate $\dot M_{\mathrm{acc}} \gtrsim \dot M_{\mathrm{acc}}^{\mathrm{lim}}$.
Clearly, if we considered the formation of a larger/smaller disk, the value of $\dot M_{\mathrm{acc}}^{\mathrm{lim}}$ would become larger/smaller. 
By our calculations of NSD, we check that the estimates~\eqref{eq:tau_estimated},~\eqref{eq:tau_r_estimated} are in reasonable accordance with an accurate value of optical thickness, so that, indeed, $\tau_{\mathrm{r}, \mathrm{R}} (r) \gtrsim 1$ in the significant part of the area where the disk can exist, $r \in (r_{\mathrm{in}}, r_{\mathrm{out}})$, as soon as the accretion rate exceeds its threshold value. Equation~\eqref{eq:tau_r_estimated} also shows that $\tau_{\mathrm{r}, \mathrm{R}}$ exceeds $\tau_{\mathrm{z}, \mathrm{R}}$ by a factor of $\sim 1 / \aspectratio$ in a geometrically thin disk.

The value of $\dot M_{\mathrm{acc}}^{\mathrm{lim}}$ estimated above is used
to define the moment of disk formation, $\tformation$. The corresponding 
disk profiles are shown in Fig.~\ref{fig:accretion_total}.

\subsubsection{Disk evolution for the most massive donor}
\label{sec:accretion_general}

\begin{table}
\caption{Notable moments of time (Myr) of donor and disk evolution}
{\tiny
\centering
\begin{tabular}
{|c|c|c|c|c|c|c|c|}
\hline
$M_1$ & \tformation & $\tRGB$ & $\tAGB$ & \tenvloss & \tdepletion & \tdecay \\
\hline
$1.7$ & $1850.04$ & $1852.55$ & $1960.59$ & $1960.65$ & $1961.18$ & $1961.76$ \\
\hline
$3.0$ & $436.32$ & $ 324.230 $ & $436.751$ & $436.779$ & $ 437.486$ & $438.236$ \\
\hline
$5.0$ & $105.91$ & $97.895$ & $108.295$ & $108.393$ & $ 108.963$ & $109.559$ \\
\hline
$7.0$ & $42.966$ & $43.079$ & $48.111$ & $48.191$ & $ 48.725$ & $49.279$ \\
\hline
\end{tabular}
}
\tablefoot{We show the moments of time for the systems shown in Fig.~\ref{fig:accretion_total}. The moments of time: the RGB accretion peak, $\tRGB$; the AGB accretion peak, $\tAGB$; the donor envelope loss, which causes the termination of wind, \tenvloss; the first formation of the sufficiently large disk, $\tformation$; formation of the most massive disk, $\tAGB$, which is practically equal to $\tAGB$; a depleted disk after the termination of wind, $\tdepletion$, which is in between $\tAGB$ and \tdecay, where \tdecay corresponds to the disk decay. Note that the values of $\tRGB$, $\tAGB$, and \tenvloss are universal for all systems with a given donor, while the values \tformation, \tdepletion and \tdecay are unique for the particular binary initial separation. Additionally, a disk in the system with donor $M_1 = 1.7$\,M$_{\sun}$ is smaller than $\sim 2$\,AU or vanishes at all during the interval from $t=1852.593$ Myr up to $t=1959.618$ Myr.}
\label{tab:times}
\end{table}

\begin{table}
\caption{Notable values for a disk taken at $\tRGB$ and $\tAGB$}
{\tiny
\centering
\begin{tabular}{|c|c|c|c|c|c|}
\hline
$M_1$ & stage & $M_{\mathrm{d}}$ (M$_{\odot}$) & $a_{\mathrm{d}}$ (AU) & $\Sigma$ (g\,cm$^{-2}$) & $T_{\mathrm{c}}$ (K) \\
\hline
\multirow{2}{*}{$1.7$} & RGB & $6.2 E{-6}$ & $3.61$ & $2.32$ &$110.3$ \\
\cline{2-6}
 & AGB & $3.9 E{-4} $ & $10.81$ & $91.7$ & $345.4$ \\
\hline
\multirow{2}{*}{$3.0$} & RGB & $7.2 E{-8}$ & $1.11$ & $0.05$ & $108.5$ \\
\cline{2-6}
 & AGB & $7.0 E{-4}$ & $10.80$ & $199.4$ & $445.1$ \\
\hline
\multirow{2}{*}{$5.0$} & RGB & $2.3 E{-7}$ & $1.54$ & $0.08$ & $210.7$ \\
\cline{2-6}
 & AGB & $1.3 E{-3}$ & $12.97$ & $351.6$ & $665.0$ \\
\hline
\multirow{2}{*}{$7.0$} & RGB & $3.4 E{-6}$ & $3.32$ & $0.94$ & $216.5$ \\
\cline{2-6}
 & AGB & $1.3 E{-3}$ & $11.53$ & $407.8$ & $698.1$ \\
\hline
\end{tabular}
}
\tablefoot{We show the values for the systems shown in Fig.~\ref{fig:accretion_total}. The values: the disk mass, $M_{\mathrm{d}}$, expressed in M$_{\sun}$; the disk size, $a_{\mathrm{d}}$ according to the location of the R2-R3 edge expressed in AU; the surface density, $\Sigma$, at $r = 1$\,AU; the midplane temperature, $T_{\mathrm{c}}$ at $r = 1$\,AU. Each pair of rows corresponds to a different donor mass, $M_1$, and stage, RGB or AGB.}
\label{tab:values_30}
\end{table}

\begin{table}
\caption{Bondi radius, $r_{\rm a}$, and snow line radius, $r_{\rm sl}$, taken at the time moments \tformation, \tAGB, \tdepletion}
{\tiny
\centering
\begin{tabular}{|c|c|c|c|c|c|c|}
\hline
$M_1$ & $r_{\rm sl} (\tformation)$ & $r_{\rm sl} (\tAGB)$ & $r_{\rm sl} (\tdepletion)$ & $r_{\rm a} (\tformation)$ & $r_{\rm a} (\tAGB)$ & $r_{\rm a} (\tdepletion)$ \\
\hline
$1.7$ & $0.366$ & $4.957$ & $0.401$ & $3.353$ & $3.483$ & $3.523$ \\
\hline
$3.0$ & $0.415$ & $3.291$ & $0.409$ & $3.138$ & $3.528$ & $3.529$ \\
\hline
$5.0$ & $0.396$ & $5.803$ & $0.484$ & $2.849$ & $3.502$ & $3.523$ \\
\hline
$7.0$ & $1.411$ & $5.505$ & $0.562$ & $2.608$ & $3.493$ & $3.523$ \\
\hline
\end{tabular}
}
\tablefoot{We show the radii for the systems shown in Fig.~\ref{fig:accretion_total}. All radii are expressed in AU. The snow line is defined as the radius of the disk temperature profile transition corresponding to $T_{\mathrm{c}}=150$~K.}
\label{tab:snow_bondi_radii}
\end{table}

Here we describe general features of NSD obtained in our simulations. Fig.~\ref{fig:accretion_total}, column 1 demonstrates the disk radial structure in the binary with the donor initial mass $M_1 = 7.0$~M$_{\sun}$ and intermediate initial separation $a=30$~AU. The selected time moments are given in Table~\ref{tab:times}, while some informative values describing the disk at $t = \tRGB$ and $t=\tAGB$ are given in Table~\ref{tab:values_30}. Note that these Tables provide results for all donors, whereas we relegate the description of disks produced by other donors to the next Section.

In Fig.~\ref{fig:accretion_total}, the curves taken at \tformation show the appearance of the disk when the accretion rate for the first time attains $\dot M_{\mathrm{acc}} \simeq \dot M_{\mathrm{acc}}^{\mathrm{lim}}$ due to the growing wind at the beginning of the donor RGB stage. Note that $\tformation$ is sufficiently close to $\tRGB$. Thus, we do not consider the structure of the disk at earlier stages when our crude assumption about the velocity of the wind is not valid. The curves generated at $t=\tAGB$ represent the most massive disk ever accumulated during the evolution of the binary. The curves at \tdepletion show the depleting disk $\sim 0.5$ Myr after the envelope loss by the donor.

We find that a sufficiently large disk is formed already at the RGB stage. Despite a decrease of the wind rate by order of magnitude between $\tRGB$ and $\tAGB$, the disk persists for $\sim 5$~Myr until the envelope loss by the donor because $\dot M_{\mathrm{acc}} \gtrsim \dot M_{\mathrm{acc}}^{\mathrm{lim}}$ stays all the time. Generally, the AGB stage of the donor is much shorter than its RGB stage. At the same time, the disk at the RGB stage is much smaller than its follower at the AGB stage, check the disk size values in Table~\ref{tab:values_30}. This is obviously explained by a much stronger wind at the AGB stage. The ratio of the peak wind rates at $\tAGB$ and $\tRGB$ is of order $\sim 10^5$. The peak intensity wind at $t=\tAGB$ gives birth to the largest disk with a size exceeding $\sim 11$ AU. Such a value substantially exceeds the corresponding Bondi radius, i.e., the distance from the host star limiting the wind accretion, which we find to be always within the range $2\text{--}4$~AU, with an exact value weakly depending on the initial binary separation. At the same time, the disk never reaches the tidal truncation radius, which takes the instant value $r_{\mathrm{out}} = 28.23$ AU at $t=\tAGB$ in the system with $a = 30$~AU. Shortly after the wind attains the peak intensity, the donor loses the envelope and the wind ceases: the disk starts its closed evolution characterized by an accretion timescale. An exact way of this closed evolution is determined by the structure of disk at $t=t_\mathrm{env}$, which we finally obtain simulating NS disk during the whole binary evolution. It takes approximately $1.1$ Myr for the disk matter to accrete onto the host star, see the difference between $t_\mathrm{dec}$ and $t_\mathrm{env}$ in Table~\ref{tab:times}. In principle, a new generation of planets can start to grow in long-lived massive disks, but analysis of this possibility is beyond the scope of our study. We check that the lifetime of the residual disk remaining after the envelope loss by the donor is in reasonable accordance with the viscous timescale estimated at the outer edge of the disk at $t=t_{\mathrm{env}}$. 

The low-mass disk accumulated by $t=\tformation$ has the surface density varying from $\Sigma \sim 1 \text{--} 10$~g\,cm$^{-2}$ in the inner parts down to $\Sigma \sim 10^{-1}$~g\,cm$^{-2}$ at the outskirts. The midplane temperature is found in the range $T_{\mathrm{c}}$~$\sim 10^2\text{--} 10^3$~K corresponding to a cool disk heated mostly by the host star and having a small aspect ratio approaching $\aspectratio \sim 0.01$. We note that temperature does not decrease even lower due to the donor heating: the profiles of $T_\mathrm{c}$ and $\aspectratio$ have a distinctive hump at the outskirts of the disk, see curve at \tformation in column 1 of Fig.~\ref{fig:accretion_total} and see Sect.~\ref{sec:donor_impact} for the general description of the donor influence. The donor heating is particularly efficient in R2 since the disk radiative cooling is suppressed by a factor of low vertical optical thickness. At the same time, dust starts to evaporate in the innermost part of the disk $\sim 0.1$~AU, as the temperature rises. This results in the transition of the disk to the optically thin state. Thus, the temperature rises further and exceeds $\sim 10^3$~K. In this case, the innermost part of the disk which is mostly heated by the host star, becomes puffed for a reason similar to that described just above for the disk outskirts: its radiative cooling is suppressed as compared to the optically thick middle part of the disk. We find that the low-mass disk described above has a mass as small as $\sim 10^{-6}$~M$_{\sun}$.

The largest disk accumulated by $t = \tAGB$ has a surface density up to $\Sigma \sim 10^3 \text{--} 10^4$~g\,cm$^{-2}$ in the inner parts down to $\Sigma \sim 10$~g\,cm$^{-2}$ at the outskirts. The midplane temperature of the largest disk attains $\sim 10^3$~K much farther away from the host star as compared with the low-mass disk, compare the corresponding curves at \tformation and \tAGB in Fig.~\ref{fig:accretion_total}. The snow line shifts beyond $r= 5$~AU. At the same time, the inner parts of the disk where dust is evaporated approach the size of $\sim 0.7$~AU. The corresponding optical thickness substantially decreases, however, the disk remains optically thick. Hence, the enhanced radiative cooling makes the inner part of the disk dense and geometrically thin, note the decrease of $\aspectratio$ by a few times as changing from the outskirts to the inner parts of the disk shown in column 1 Fig.~\ref{fig:accretion_total}. This feature becomes more distinct due to the donor heating, which makes the disk outskirts geometrically thicker similar to that of the low-mass disk. The disk acquires a distinctly flaring shape. We find that the innermost part of the largest disk is heated up to $T_\mathrm{c} \gtrsim 10^4$~K. The pure gas opacity gets sharply higher at such temperature due to ionization processes that inflate the disk close to its inner boundary, see the corresponding hump on the profile of $\aspectratio$ at $r \lesssim 0.05$~AU in Fig.~\ref{fig:accretion_total}, column 1. In this case, the surface density is found to have decreased considerably. We also note that the disk accumulates mass up to $0.0013\mathrm{~M}_{\sun}$ at $\tAGB$.

The residual disk left after the envelope loss by the donor is shown at half of its lifetime, see the \tdepletion curves in column 1 of Fig.~\ref{fig:accretion_total}. It is seen that despite the considerable time $\sim 0.5$~Myr has passed since the envelope loss, the disk still holds a substantial amount of matter. Its mass and size are $\sim 10^{-4}\mathrm{~M}_{\sun}$ and $\sim 6$~AU, respectively. We note that the disk radial structure becomes quite smooth in comparison with the epoch of wind accretion. The surface density profile is close to a power law, which indicates that the disk evolution proceeds well within a quasi-stationary regime \citep{lynden-bell1974, lyubarskij1987}. The disk is distinctly flaring with an aspect ratio substantially decreased as compared to the moment $t=\tAGB$. An aspect ratio varies from $\sim 0.03$ at the outskirts of the disk down to $\sim 0.015$ in its innermost part. One can see that, similar to the low-mass disk produced at the wind rate corresponding to $\dot M_{\mathrm{acc}} \simeq \dot M_{\mathrm{acc}}^{\mathrm{lim}}$, such a disk is already light enough to become optically thin due to evaporation of dust in its innermost part. The residual disk has a temperature and aspect ratio rather close to that of the low-mass disk, however, it is much larger and heavier than the low-mass disk generated by the weak wind from the donor. The results concerning the residual disk presented here allow us to predict the appearance of extended cool accretion disks around MS stars in wide binaries with young white dwarfs.

As the initial separation of the binary becomes larger, the amount of wind matter captured by the disk decreases. This shortens its lifetime and decreases its mass and size. As above in this Section, we describe here general results with an emphasis on the sufficiently large disk with $\dot M_{\mathrm{acc}} \gtrsim \dot M_{\mathrm{acc}}^{\mathrm{lim}}$, or equivalently, the disk size not less than $\sim 2$~AU. For $a$ increasing up to $100$~AU, such a disk does not exist far away from $t=\tRGB$ anymore. Nevertheless, for the widest system with $a=100$~AU, disk approaches the size (the mass) $r \sim 2$~AU ($10^{-6}\mathrm{~M}_{\sun}$) right at $t=\tRGB$. At the same time, the existence of a disk with a size not less than $\sim 2$~AU before $t=\tAGB$ shortens up to $0.5$~Myr. However, disk holds the size (mass) around $\sim 1$~AU ($10^{-7}\mathrm{~M}_{\sun}$) for a considerably longer time $\sim 5$~Myr. For the initial binary separation increasing from $30$~AU to $100$~AU, the mass of the largest disk accumulated by $t=\tAGB$ decreases by half an order of magnitude taking the values down to $0.0003 \mathrm{~M}_{\sun}$. The corresponding disk size takes the values down to $8$~AU (compare this value with the corresponding value in Table~\ref{tab:values_30}). After the envelope loss by the donor, the disk decays approximately twice as fast as the binary initial separation increases from $30$~AU up to $100$~AU. Thus, the residual disk lifetime in the system with initial $a=100$~AU is around $\sim 0.5$~Myr.

In the binaries closer than initial $a=30$~AU, the disk becomes larger as compared to the case demonstrated in Fig.~\ref{fig:accretion_total} and Table~\ref{tab:values_30}. For example, the system with the initial separation $a=20$~AU gives birth to the largest disk we meet in calculations. It attains the size $\sim 18$~AU about $ 50$~kyr after $t=\tAGB$. However, the disk produced in close binaries with initial separation approaching $10$~AU extends up to the tidal truncation radius around $t=\tRGB$ as well as $t=\tAGB$. In this case, we obtain the most massive but more compact disk. At $t=\tRGB$, its size stays within $\sim 2$~AU. Despite that, the disk contains much more matter as compared to a wider binary: the disk mass attains $3\times 10^{-5}\mathrm{~M}_{\sun}$, which should be compared to an order of magnitude smaller value given in Table~\ref{tab:values_30}. Similarly, at $t=\tAGB$, we obtain the disk a few times more massive as compared to the binary with initial $a=30$~AU. Its mass and size are, respectively, $0.0071\mathrm{~M}_{\sun}$ and $12.7$~AU.

\subsubsection{Disk evolution for less massive donors}
\label{sec:accretion_other_donors}

The results outlined in the previous Section basically apply to less massive donors. The disk radial structure is shown for donors with initial masses $M_1 = 5.0$, $3.0$, $1.7$~M$_{\sun}$ in Fig.~\ref{fig:accretion_total}, columns 2--4, respectively. As in the previous Section, we first analyze the disk properties in binaries with the initial separation $a=30$ AU and then explain the changes corresponding to the changes of the initial binary separation. Being generated at their own $t=\tformation, \tAGB$ and $\tdepletion$, the corresponding radial profiles of disk look qualitatively similar to those for the most massive donor, compare columns 2--4 to column 1 in Fig.~\ref{fig:accretion_total}. At the same time, the number of significant differences can be identified by the values contained in Tables \ref{tab:times} and \ref{tab:values_30}. Both qualitative and quantitative differences between the disks produced by different donors are determined by variations of the wind rate profile depending on the donor mass, see Figs.~\ref{fig:low_mass_wind}--\ref{fig:high_mass_wind}. Below we highlight the main deviations from the disk formation and evolution obtained for donor $M_1 = 7.0$~M$_{\sun}$ while proceeding with less massive donors. Additionally, we point out some important features of the disk that are common to all donors.

Contrary to the case with the most massive donor, sufficiently large disk with $\dot M_{\mathrm{acc}} \gtrsim \dot M_{\mathrm{acc}}^{\mathrm{lim}}$ is not formed at the RGB stage of intermediate-mass donors with $M_1=$3.0$\mathrm{~M}_{\sun}$ and $5.0\mathrm{~M}_{\sun}$. However, such disk is formed in the case of low mass donor with $M_1=1.7\mathrm{~M}_{\sun}$. In this case, the disk exists for $2.5$ Myr before $t=t_{\mathrm{RGB}}$ and attains the size of $3.5$~AU which is even larger than the corresponding disk produced by the most massive donor, see Table~\ref{tab:values_30}. Note that its mass is twice that of the disk produced in the binary with $M_1=7.0\mathrm{~M}_{\sun}$ and is greater by almost two orders of magnitude than the disk produced in the binary with $M_1=3.0\mathrm{~M}_{\sun}$. Such a situation takes place due to the peculiarities of stellar evolution. Indeed, the lowest mass red giant demonstrates a more powerful and long-standing wind than heavier donors, comparing the plots in Figs.~\ref{fig:low_mass_wind},~\ref{fig:high_mass_wind}. As soon as the wind from the low-mass red giant abruptly weakens, the disk depletes by $\sim 0.1$~Myr after $t=t_\mathrm{RGB}$ and shrinks back to a small size. The intermediate-mass donors with $M_1=$3.0$\mathrm{~M}_{\sun}$ and 5.0$\mathrm{~M}_{\sun}$ produce a sufficiently large disk for the first time, respectively, $\sim 0.4$~Myr and $\sim 2.4$~Myr before the peak intensity wind at $t=\tAGB$. Whereas the lowest mass donor does the same once again $\sim 1$~Myr before $t=\tAGB$, see Table~\ref{tab:times}.

Since the ratio of the peak wind rates at $t_{\mathrm{AGB}}$ and $t_{\mathrm{RGB}}$ stays in the range $10^4$--$10^6$ for less massive donors, their AGB stage also produces the largest and the most massive disk throughout the whole evolution of the system, see Table \ref{tab:values_30}. The largest disk size weakly depends on the donor mass and attains $\sim 13$ AU for the intermediate mass donor with $M_1 =5.0\mathrm{~M}_{\sun}$. As in the case of the most massive donor, the largest disk size substantially exceeds the corresponding Bondi radius for less massive donors. We check that for $a=30$~AU no donor produces the disk at the AGB stage expanding up to the tidal truncation radius. The latter takes values from $r_{\mathrm{out}} = 26.07$ down to $14.08$ AU at $t=\tAGB$ while changing from $M_1 =5.0\mathrm{~M}_{\sun}$ down to $M_1 =1.7\mathrm{~M}_{\sun}$, respectively. Note that despite the same initial separation of the binary, $r_{\mathrm{out}}$ increases by $t=\tAGB$ more for the heavier donors, which is explained by the growing expansion of the binary due to the more intensive mass loss by the heavier donors. We also check that the closed evolution of the residual disk produced by less massive donors lasts not less than in the case of the most massive donor. The residual disk exists even longer for the intermediate-mass donors decaying $1.1$--$1.5$ Myr after they lose their envelopes, check Table \ref{tab:times}.

We note that the low-mass disks shown in Fig.~\ref{fig:accretion_total} for $t=\tformation$ are nearly identical to each other having the same mass $\sim 10^{-6}$~M$_{\sun}$ for all donors. This is not the case at $t=\tAGB$ when the surface density gradually increases by a factor of a few in the middle of the disk while changing from the lowest mass donor to the most massive one. Accordingly, the largest disk accumulated by $t=\tAGB$ is more massive in binaries with heavier donors, see Table \ref{tab:values_30}. Contrary to the cases of the most massive donor and the donor with $5.0\mathrm{~M}_{\sun}$, the innermost part of this disk produced by the less massive donors, $1.7\mathrm{~M}_{\sun}$ and $3.0\mathrm{~M}_{\sun}$, is not puffed, see the corresponding values at \tAGB in Fig.~\ref{fig:accretion_total}. It takes the smallest value of an aspect ratio slightly above $0.01$, while its temperature stays well below $10^4$~K. Additionally, the snow line typically shifts closer to the host star in a disk produced by the less massive donors. For example, it is located at $r=3.3$~AU and $5.8$~AU for donors $3.0\mathrm{~M}_{\sun}$ and $5.0\mathrm{~M}_{\sun}$, respectively, see Table~\ref{tab:snow_bondi_radii}.

It is notable that the peak values of the donor wind rate at $t=\tAGB$ for the lowest mass donor, $1.7\mathrm{~M}_{\sun}$, and the most massive one, $7.0\mathrm{~M}_{\sun}$, show a considerably larger difference than masses accumulated by disk in these two cases, see Figs.~\ref{fig:low_mass_wind},~\ref{fig:high_mass_wind} and Table~\ref{tab:values_30}. This is explained by the variation of an expansion rate of the binary due to the decrease of the donor mass in the systems with different donors. Indeed, the different expansion rates of the binaries with different donors change the ratio of the cross sections for the gravitational capture of wind in systems with the same initial separation, and finally, the ratio of accretion rates in the course of binary evolution. Using Eq.~\eqref{eq:separation}, we make sure that by $t=\tAGB$, the binaries with initial separation $a=30$~AU expand by a factor of $\sim 2$ and $\sim 4$ for the donors with $M_1=1.7\mathrm{~M}_{\sun}$ and $7.0\mathrm{~M}_{\sun}$, respectively. Since the corresponding values of Bondi radius in these two systems are almost the same, see Table~\ref{tab:snow_bondi_radii}, we find that the growth of the disk in the systems with heavier donors is inhibited mostly by the relatively faster expansion of the binary. For the same reason, the largest disk size weakly depends on the donor mass (this has already been pointed out above in this Section). Moreover, the size of such a disk in the system with donor $M_1=5.0~\mathrm{M}_{\sun}$ slightly exceeds that of the disk in the system with the most massive donor, $M_1=7.0~\mathrm{M}_{\sun}$, see Table~\ref{tab:values_30}.

The residual disk left after the envelope loss by the donor evolves similarly for all donors. As we change to the lowest mass donor, its mass and size by $t=\tdepletion$ decreases down to $\sim 10^{-5}\mathrm{~M}_{\sun}$ and $\sim 3$~AU, respectively. Note that the humps of temperature and aspect ratio arising in the optically thin innermost part of the disk become more significant for the less massive donor, compare the corresponding curves at \tdepletion shown in Fig.~\ref{fig:accretion_total}.

The less massive donors in wider systems with initial separation increasing up to $100$~AU do not produce a sufficiently large disk at $t=\tRGB$. In this case, disk size (mass) rises up to $\sim 1$~AU ($10^{-7}\mathrm{~M}_{\sun}$) for the lowest mass donor. As $a$ increases up to $100$~AU, the existence of a disk with a size not less than $\sim 2$~AU before $t=t_{\mathrm{AGB}}$ shortens up to $\sim 0.3$, and $0.4$~Myr for donors with $M_1=1.7$, and $5.0\mathrm{~M}_{\sun}$, respectively, while it remains almost unchanged for donor with $M_1=3.0\mathrm{~M}_{\sun}$. Nevertheless, the disk holds size (mass) around $\sim 1$~AU ($10^{-7}\mathrm{~M}_{\sun}$) for a considerably longer time $\sim 0.7$ and $2.4$~Myr for donors with $M_1=1.7$ and $5.0\mathrm{~M}_{\sun}$, respectively. A different trend in the formation of the disk in wider systems with donor $M_1 = 3.0\mathrm{~M}_{\sun}$ is explained by the too-low wind rate right up to the beginning of the AGB stage. We check that for the initial binary separation increasing from $30$~AU up to $100$~AU, the mass of the largest disk accumulated by $t=\tAGB$ decreases by half an order of magnitude for all less massive donors. For example, the size (mass) of such disk becomes as small as $6$~AU ($7\times 10^{-5} \mathrm{~M}_{\sun}$) for $a=100$~AU in the binary with $M_1=1.7\mathrm{~M}_{\sun}$.

In the closer binaries with $a$ down to $10$~AU, the disk attains the tidal truncation radius around $t=\tRGB$ and $t=\tAGB$ for all less massive donors. In this case, each given donor produces the most massive but compact disk. At $t=\tRGB$, the disk size stays within $\sim 3$~AU for the lowest mass donor with $M_1=1.7\mathrm{~M}_{\sun}$ and even smaller for the intermediate mass donors. The corresponding disk mass takes values $3\times 10^{-5}$, $5\times 10^{-7}$ and $8\times 10^{-7} \mathrm{~M}_{\sun}$ for the donors $M_1=1.7$, $3.0$ and $5.0\mathrm{~M}_{\sun}$, respectively. These values should be compared with the corresponding data in Table~\ref{tab:values_30}. Similarly to the case of the most massive donor, the less massive donors produce at $t=\tAGB$ a few times more massive but more compact disk as compared to the corresponding binary with initial $a = 30$~AU. The disk mass (size) is $0.0011$, $0.0036$, $0.0058$~M$_\sun$ ($6.4$, $8.9$, $10.5$ AU) for donors $1.7$, $3.0$, $5.0$~M$_\sun$, respectively.

\subsection{Planetary migration}
\label{sec:results_migration}

\subsubsection{Analytical estimates concerning planetary migration}
\label{sec:results_migration_analytics}

We consider migration of a single constant mass planet embedded in the NS disk. The type \RomanNumeralCaps{1} migration timescale can be estimated analytically in the low-mass disk within QSD under the assumption $T_{\rm c} \approx T_{\rm I, A}$. The same assumptions were used to obtain Eq.~\eqref{eq:tau_estimated}. With these assumptions and the given values of parameters from Eq.~\eqref{eq:tau_estimated}, i.e., $\dot M_{\mathrm{acc}} = \dot M_{\mathrm{acc}}^{\mathrm{lim}}$, $r = 2$\,AU, we obtain the following upper limit on the migration timescale,
\begin{multline}
\label{eq:timescale_estimate}
\left. \hat{t_{\mathrm{p}}} \right|_{\rm\RomanNumeralCaps{1}} \approx 2.8 \times 10^{8} \mathrm{~yrs} \left( \frac{2.72 + 1.08 \beta}{4.34}\right)^{-1} \left( \frac{\alpha}{0.01}\right) \left( \frac{m_{\mathrm{p}}}{\mathrm{m}_{\oplus}}\right)^{-1} \times \\
\times \left( \frac{r}{\mathrm{2~AU}}\right)^2 \left( \frac{\dot M_{\mathrm{acc}}}{\dot M_{\mathrm{acc}}^{\mathrm{lim}}}\right)^{-1}.
\end{multline}
Equation~\eqref{eq:timescale_estimate} is an estimate of the migration rate taking place in a low-mass disk, which we conventionally defined in Sect.~\ref{sec:analytics_general_disk_features} as the one with $\dot M_{\rm acc} \simeq \dot M_{\rm acc}^{\rm lim}$. Therefore, the migration of the Earth-like planets is expected to be negligible in such a disk. At the same time, the Earth- to the Neptune-like planets on the close orbits within $1$~AU are expected to migrate weakly given the long-term existence of a low-mass disk with $\dot M_{\rm acc} \simeq \dot M_{\rm acc}^{\rm lim}$ before $t=\tRGB$. According to Eq.~\eqref{eq:timescale_estimate}, the type \RomanNumeralCaps{1} migration should be significant for giant planets provided that the low-mass disk exists at least for $\sim 1$ Myr. We further expect that giant planets embedded in the largest disk around $t=\tAGB$ should reach the distance of TM, $r_{\rm p}^{\rm fin}$. On the other hand, the Earth- to the Neptune-like planets can migrate far from their original orbits at the AGB stage since the accretion rate is enhanced by a few orders of magnitude for hundreds of kyr. These conclusions on planetary migration at the AGB stage should be the case for planets that pre-existed on orbits within $\sim 10$ AU which is constrained by the disk size, see Table~\ref{tab:values_30}.

Giant planets can open the gap and undergo the type \RomanNumeralCaps{2} migration. We do not consider the other types of planet influence on the disk. As soon as the viscous timescale of a disk, $t_\nu$, is significantly less than its lifetime, see Fig.~\ref{fig:viscous_times}, we expect that type \RomanNumeralCaps{2} migration alone can shift the planets up to $r_{\rm p}^{\rm fin}$. It is instructive to derive the approximate equation for $q_{\rm crit}$ valid for values of disk viscosity and aspect ratio met in this work. As far as $X$ entering Eq.~\eqref{eq:critical_mass_planet} is close to unity even for $\aspectratio$ highly exceeding $\alpha$, we have 
\begin{equation*}
q_{\mathrm{crit}} \approx \frac{100}{\mathcal{R}}\left (2^{1/3} - 
\frac{1}{8} \left ( \frac{\aspectratio}{\alpha}\right )^{1/3}\right )^{-3},
\end{equation*}
where the second term in brackets can also be safely omitted for our case $\alpha=0.01$ and $\aspectratio \lesssim 0.1$. We finally obtain
\begin{equation}
\label{q_cr_simple}
q_{\mathrm{crit}} \approx 1.25 \times 10^{-3} 
\left ( \frac{\alpha}{0.01} \right ) 
\left ( \frac{\aspectratio}{0.05} \right )^2.
\end{equation}
According to this estimate, we expect that the opening gap mass of the planet varies in the range $m_{\mathrm{p}}^{\mathrm{crit}} = 15\text{--}1500~\mathrm{m}_\oplus$ as we find that the thickness of the NS disk corresponds to $\aspectratio \sim 0.01$ close to the time of the disk formation or decay and attains $\aspectratio \sim 0.1$ at $t=\tAGB$, see Sect.~\ref{sec:accretion_general} and~\ref{sec:accretion_other_donors}. As soon as the opening gap mass 
strongly depends on the aspect ratio, $m_{\mathrm{p}}^{\mathrm{crit}} \propto \aspectratio^2$, we expect that the outcome of migration of the giants embedded in NS disk is determined by a sequence of type~\RomanNumeralCaps{1} and type~\RomanNumeralCaps{2} migration periods changing each other.
The change between the migration types can be caused by either the radial gradient or the time evolution of the disk aspect ratio.

Let us estimate the ratio between the type \RomanNumeralCaps{1} and \RomanNumeralCaps{2} migration rates for the gap opening mass of the planet, $q=q_{\rm crit}$. For that, we compare the velocities~\eqref{eq:migration_type1} and~\eqref{eq:migration_type2} in order to obtain an estimate
\begin{multline}
\label{eq:migration_velocities}
\frac{\left.\mathrm{d} r_{\mathrm{p}} / \mathrm{d} t\right|_2}{\left.\mathrm{d} r_{\mathrm{p}}/ \mathrm{d} t \right|_1} = \frac{v_2}{v_1} = \frac{1.5 \alpha \aspectratio^2 f^{-1} M_{\mathrm{d}} / (M_{\mathrm{d}} + m_{\mathrm{p}})}{(2.72 + 1.08 \beta) \aspectratio^{-2} m_{\mathrm{p}} M_{\mathrm{d}} / (2 \pi M_2^2)} = \\
= \frac{3 {\pi } \alpha \aspectratio^4}{(2.72 + 1.08 \beta) f} \frac{M_2^2}{m_{\mathrm{p}}^2} \frac{1}{1 + M_{\mathrm{d}} / m_{\mathrm{p}}} = 0.98 \left( \frac{\alpha}{0.01} \right) \left( \frac{\aspectratio}{0.08} \right)^4 \times \\
\times \left( \frac{f}{1} \right)^{-1} \left( \frac{2.72 + 1.08 \beta}{4.34} \right)^{-1} \left( \frac{m_{\mathrm{p}}}{300\,\mathrm{m}_{\oplus}} \right)^{-2} \left( \frac{M_2}{\mathrm{M}_{\sun}} \right)^{2} \frac{1}{1 + M_{\mathrm{d}} / m_{\mathrm{p}}},
\end{multline}
where the notations from Eqs.~\eqref{eq:migration_type1},~\eqref{eq:migration_type2} are used. It is also assumed that migration occurs in a quasi-stationary disk, which allows us to rewrite the term $(2 r / F) (\mathrm{d}F / \mathrm{d}r) = 1 / f$, where $f$ is the factor from Eq.~\eqref{eq:nu_sigma}. Equation~\eqref{eq:migration_velocities} reproduces the known result that the type \RomanNumeralCaps{1} migration proceeds faster than the type \RomanNumeralCaps{2} migration for sufficiently massive planets in cold and low-mass disks ($\aspectratio \lesssim 0.08$ and $M_{\mathrm{d}} \lesssim m_{\mathrm{p}}$), see \citet{ward1997}. Note the strong dependence of this feature on the disk aspect ratio. 

Therefore, the fast type \RomanNumeralCaps{1} migration takes place as soon as the ratio given by Eq.~\eqref{eq:migration_velocities} becomes less than unity for $m_\mathrm{p}$ given by Eq.~\eqref{q_cr_simple}. Explicitly,
\begin{multline}
\label{eq:migration_velocities_add}
\frac{v_2}{v_1} = 
0.078 \left( \frac{\alpha}{0.01} \right)^{-1} \left( \frac{f}{1} \right)^{-1} \left( \frac{2.72 + 1.08 \beta}{4.34} \right)^{-1} \frac{1}{1 + M_{\mathrm{d}} / m_{\mathrm{p}}}.
\end{multline}
This estimate shows that the fast type \RomanNumeralCaps{1} migration is expected in any disk
with sufficient viscosity. Note that this is not the case in \citet{kulikova2019}, where the type \RomanNumeralCaps{2} migration is considered in the limit of a heavy disk only, while the disk characteristic mass is actually smaller than the opening gap mass of the planet. The latter takes place because \citet{kulikova2019} do not consider the relatively dense part of the disk beyond the snow line. 

\subsubsection{General description of migration}
\label{sec:migration_general}

\begin{figure}
    \centering
    \includegraphics[width=\columnwidth]{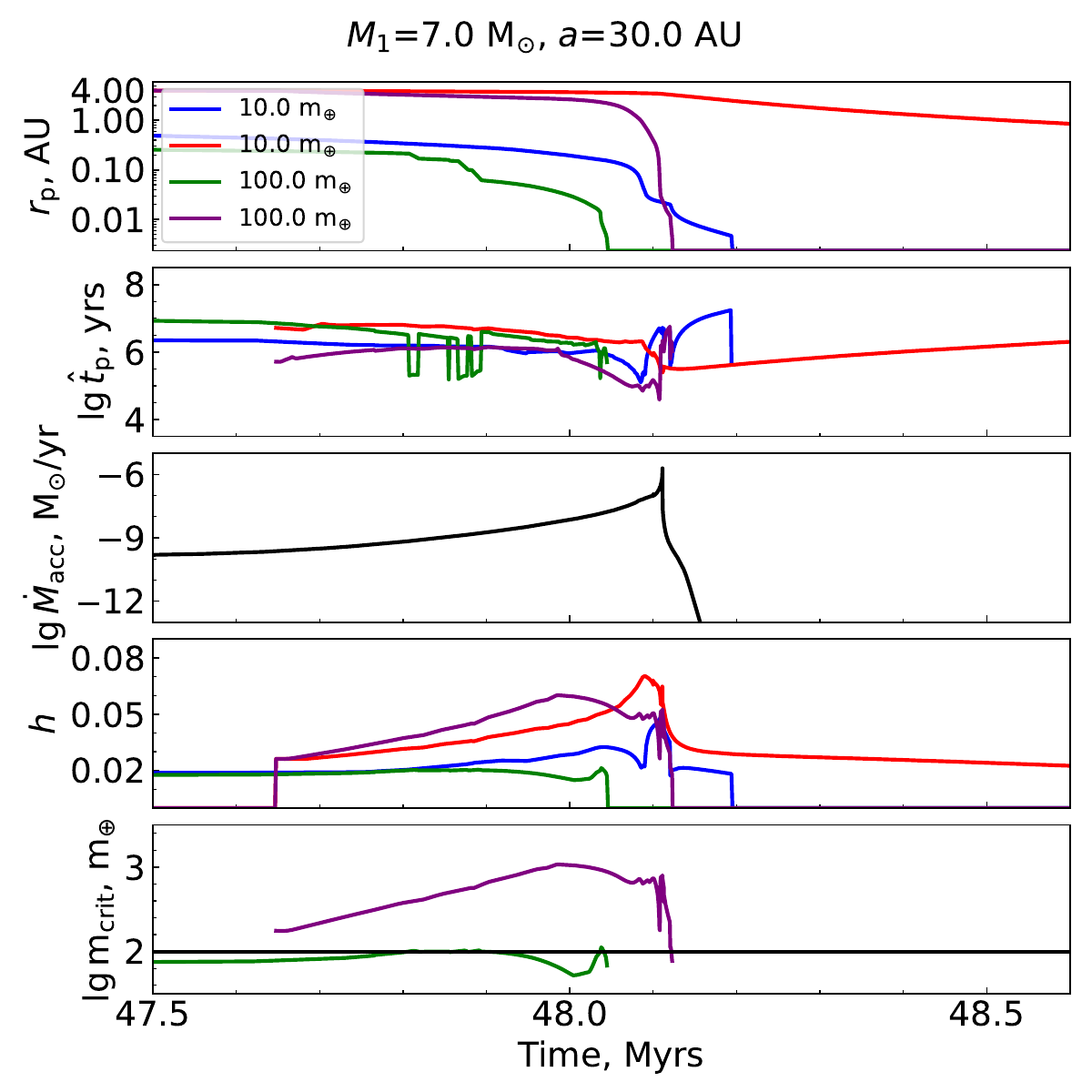}
    \caption{Orbital evolution (top panel) and the corresponding characteristic migration time (second panel from top) for planets with masses $m_{\mathrm{p}} = 10$~m$_{\oplus}$ and $100$~m$_{\oplus}$ at initial orbits $r_{\mathrm{p}} = 1$~AU and $4$~AU. Additional panels from top to bottom show the total accretion rate as well as the disk aspect ratio and the critical planet mass of the gap formation~\eqref{eq:critical_mass_planet} in the vicinity of the planet for the system with initial $M_1 = 7.0$~M$_{\sun}$, $a = 30$~AU. Note that the critical planetary mass of the gap formation is shown only for giants. 
    }
    \label{fig:migration_general_1}
\end{figure}

The examples of migration tracks are shown in Fig.~\ref{fig:migration_general_1} and in Fig.~\ref{fig:migration_different_donor_masses} for planets $m_{\mathrm{p}}=10~\mathrm{m}_\oplus$ and $100~\mathrm{m}_\oplus$ and for all donors in the system with intermediate initial separation. The dependence of the planetary orbit on time is accompanied by the migration timescale as well as the disk aspect ratio and the critical mass of the gap formation in the vicinity of the migrating planet. These additional quantities help to follow the condition of gap formation around the giants. The initial planetary orbits are set to $1$ and $4$~AU, with the latter being close to the largest stable orbit, $r_{\rm p}^{\rm max}$, in the particular case of high-mass donors in the systems with initial $a=30$ AU. We show migration around the AGB stage only. However, it can be seen that the planet with $m_\mathrm{p}=100~ \mathrm{m}_\oplus$ in the system with a donor with $M_1=$~$7.0~\mathrm{M}_\sun$ and planets $m_{\mathrm{p}}=10, 100~\mathrm{m}_\oplus$ in the system with a donor with $M_1=$~$1.7~\mathrm{M}_\sun$ have already considerably changed their initial $1$~AU orbit by this time. This is explained by migration in a disk accumulated at the RGB stage. In turn, this makes a lighter planet reach the distance of TM before the donor loses its envelope. The latter is not the case in systems with donors with $M_1=3.0~\mathrm{M}_\sun$ and $5.0~\mathrm{M}_\sun$. At the same time, a heavier planet migrates up to $r_{\rm p}^{\rm fin}$ irrespective of the donor mass and initial orbit size.

It is seen that distant planets start migrating not long before $t=\tAGB$ when the disk spreads beyond their initial orbit. Also, a distant giant with $m_{\mathrm{p}}=100~\mathrm{m}_\oplus$ always starts migrating according to the type \RomanNumeralCaps{1}. Then, in systems with low-mass donors, migration can switch to the type \RomanNumeralCaps{2} in the innermost part of a disk due to oscillations of the accretion rate and the disk parameters.

Once a low-orbit giant appears in a disk, it usually opens the gap for the following reasons. First, $m_{\rm p}^{\rm crit}$ depends mainly on the disk aspect ratio according to Eq.~\eqref{eq:critical_mass_planet} or~\eqref{q_cr_simple}. Second, the disk aspect ratio is smaller closer to the host star and, additionally, at an earlier time compared to the AGB peak. If it were a stationary system, the planet would undergo the type \RomanNumeralCaps{2} migration further on. However, the growth of the accretion rate on the way to the AGB peak leads to the growth of the disk aspect ratio, which closes the gap and migration changes to the type \RomanNumeralCaps{1}. We find that this always leads to an increase in the migration rate in our model. As far as the temporal growth rate of $\aspectratio$ exceeds the rate of its decrease around the planet migrating inwards in a flaring disk, the type \RomanNumeralCaps{1} migration holds further on. On the contrary, as soon as the temporal growth of $\aspectratio$ slows down or ceases, the decrease of $\aspectratio$ around the planet migrating inwards in a flaring disk opens the gap back again. Numerical analysis reveals an additional type of migration which may last for a substantial time. In our simplified model of migration, this new type of migration exhibits jumps between the migration of the types \RomanNumeralCaps{1} and \RomanNumeralCaps{2}, see Figs.~\ref{fig:migration_general_1} and \ref{fig:migration_different_donor_masses} (left plot). This situation takes place provided that $\aspectratio$ of a disk grows up faster (slower) than it decreases around the planet changing its orbit with the type \RomanNumeralCaps{2} (\RomanNumeralCaps{1}) migration rate, see Sect.~\ref{sec:rolling_migration} for details.

The profile of $\aspectratio$ measured in the vicinity of a planet migrating in NS disk may be quite complex already in the case of a simple shape of the accretion rate at the AGB peak and without the change between different types of migration (see the migration track of the low-mass planet starting from $1$~AU in the systems with high-mass donors). The prominent peak of such a $\aspectratio$ closely follows the peak of the accretion rate in the neighboring case of a distant low-mass planet initially situated at $4$ AU. However, in the former case of the smaller initial orbit, such a peak of $\aspectratio$ is diminished by the substantial spatial gradient of $\aspectratio$ in the NS disk down from $0.7$ AU. This behavior is observed in the model close to $t=\tAGB$, see Fig.~\ref{fig:accretion_total}. The same effect is seen for the giant planet, where it is 
additionally complicated by the change between the types of migration in the low-orbit case, see Fig.~\ref{fig:migration_general_1}. The shaping of $\aspectratio$ measured in the vicinity of the migrating planet in the systems with low mass donors is even more complicated by the oscillations of the accretion rate. Nevertheless, as one proceeds from the distant low-mass planet to the low-orbit giant planet, a similar trend can be found for these systems as well, see Fig.~\ref{fig:migration_different_donor_masses}, center and right plots.

Note that a relatively slow migration of low-mass planets takes place in the residual disk. Nevertheless, given the long lifetime of the residual disk, it shifts the planets a few times closer to the host star as compared to that at the AGB stage. Additionally, the type \RomanNumeralCaps{1} migration timescale decreases outwards due to the profiles of $\Sigma$ and $\aspectratio$ specific for residual disk, see Fig.~\ref{fig:migration_different_donor_masses}, left and center plots. This facilitates the relocation of distant planets toward the host star.


With regards to giant planets, the migration tracks show that as the planet opens (closes) the gap, its migration slows down (speeds up). This feature is confirmed analytically,
see Eq.~\eqref{eq:migration_velocities_add} above. The estimated ratio of 
migration rates is in accordance with the numerical calculations of migration of the gas giants, 
see Fig.~\ref{fig:migration_general_1} and \ref{fig:migration_different_donor_masses}, left plot.


\begin{figure}
    \centering
    \includegraphics[width=\columnwidth]{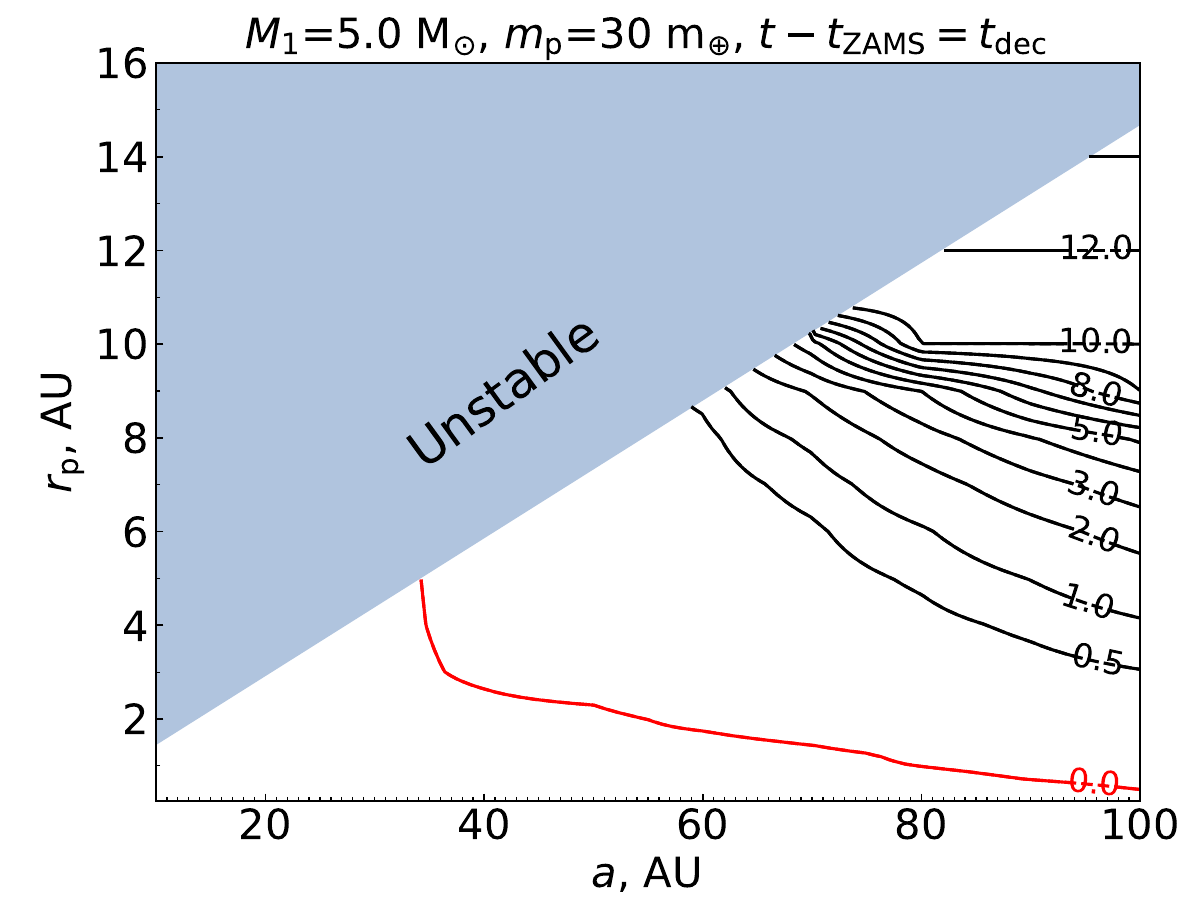}
    \caption{Size of the final planetary orbit as a function of its initial value, $r_{\mathrm{p}}$, and initial binary separation, $a$. Planet migrates all the time it is embedded in NS disk with $\alpha = 10^{-2}$. The curves show contours of the same size of the final planetary orbit marked by its value in AU. The masses are: $M_1 = 5.0$~M$_{\sun}$ and $m_{\mathrm{p}} = 30$~m$_{\oplus}$. The grey area at the upper left part of the figure shows the region of dynamically unstable planetary orbits calculated using Eq.~\eqref{eq:max_planet_orbit}. Migration contours are obtained by interpolation using the nodes taken with the step $\Delta a = 10$~AU and $\Delta r_{\mathrm{p}} = 1$~AU with additional values $r_{\mathrm{p}} = 0.25, 0.5, 0.75, 1.5$~AU closer to the host star. The curve marked as ``0.0'' (red color) shows the planets that reached the TM distance, $r_{\rm p}^{\rm fin}$, by the moment of the disk decay. }
    \label{fig:contour_NSD_different_times_1}
\end{figure}

In order to obtain a picture of an overall transformation of a planetary system caused by a planetary migration in a wind-fed disk, we plot contours of constant final orbits. First, they are shown on the plane of initial binary separations vs. initial planetary orbits, see Fig.~\ref{fig:contour_NSD_different_times_1}. We choose the case $m_{\rm p} = 30~ \mathrm{m}_\oplus$ as here the contribution of the type \RomanNumeralCaps{2} migration is still negligible for all combinations of $a$ and $~r_{\rm p}$. At the same time, such a planet already undergoes the substantial type \RomanNumeralCaps{1} migration. We find a prominent transformation of planetary orbits even for the widest binaries. For example, the planet at initial orbit $4$ AU migrates down to $1$ AU by the time of the disk decay in the system with the initial separation $100$ AU. Moreover, in systems with initial separations less than $\sim 35$ AU, the planets at all dynamically stable initial orbits finally reach the distance of TM. It is notable that as the initial orbit decreases, planets approach the host star progressively tighter. Thus, a large number of planets on the plane $(a,~r_{\rm p})$ end up within the distance $\lesssim 1$ AU from the host star. For example, this is the case for the planets at all dynamically stable initial orbits in the systems with initial separations less than $\sim 65$ AU.

\begin{figure}
    \centering
    \begin{minipage}{0.98\linewidth}
    \includegraphics[width=1.0\linewidth]{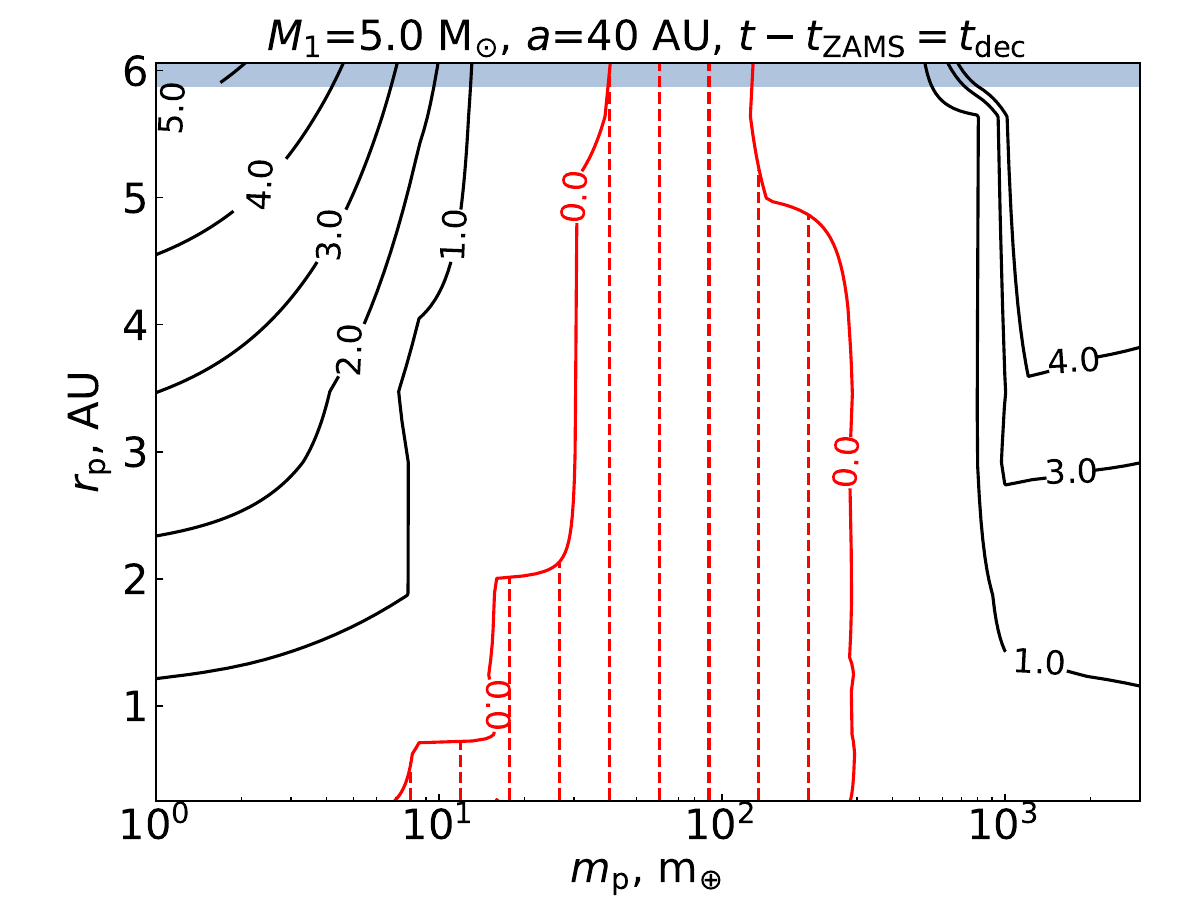}
    \end{minipage}    
    \caption{Size of the final planetary orbit as a function of the initial planetary orbit, $r_{\mathrm{p}}$, and planetary mass, $m_{\mathrm{p}}$. A planet migrates all the time it is embedded in an NSD with $\alpha = 10^{-2}$. The initial parameters are $M_1 = 5.0$~M$_{\sun}$ and $a = 40$~AU. The notations are the same as in Fig.~\ref{fig:contour_NSD_different_times_1}. The vertical dashed lines additionally highlight the planets reaching the region of tidal migration.}
    \label{fig:contour_different_planet_mass_3}
\end{figure}

Further, the contours of constant final orbits are shown on the plane of planetary masses, and initial orbits, see Fig.~\ref{fig:contour_different_planet_mass_3}. We choose the case of binary with intermediate initial separation, $a=40$~AU, and a massive donor with $M_1=5.0~\mathrm{M}_\sun$. The left part of the plot should be considered as an extension to the data in Fig.~\ref{fig:contour_NSD_different_times_1}. It shows the transformation of orbits belonging to the low-mass planets $1\text{--}40~\mathrm{m}_\oplus$ undergoing the type \RomanNumeralCaps{1} migration. It can be seen that the Earth-like planets shift by $\sim 10\%$ toward the host star. At the same time, all planets with masses $ \sim 10~\mathrm{m}_\oplus$ on the dynamically stable initial orbits relocate to orbits $\lesssim 2$~AU with a small number of them reaching the distance of TM in such a system, $r_{\rm p}^{\rm fin}$. Hereafter, the situation when a planet reaches $r_{\rm p}^{\rm fin}$ during the existence of the wind-fed disk is referred to as the complete migration. Planets with masses $ \sim 40~\mathrm{m}_\oplus$ are subject to complete migration from all dynamically stable initial orbits. We check numerically that such a great relocation of distant planets toward the star is also caused by the significant migration in a residual disk where the migration timescale decreases outwards, see the migration tracks above.

The range of planetary masses includes also the super-Jovian values $\lesssim 3000~\mathrm{m}_\oplus$. We demonstrate that planets undergoing complete migration are limited from above by the value $\sim 300~\mathrm{m}_\oplus$. This number just weakly depends on the initial planetary orbit, see the right red line in Fig.~\ref{fig:contour_different_planet_mass_3}. The migration tracks of the low-orbit super-Jupiters, $r_{\rm p} \lesssim 2$~AU and $m_{\rm p} \gtrsim 10^3~\mathrm{m}_\oplus$, show that their migration (being the type \RomanNumeralCaps{2}) is slowed down on the account of the small ratio of the disk characteristic mass enclosed within the low planetary orbit and the planetary mass. 

For planets of even smaller mass, two effects contribute to the enhancement of the overall migration: 
\begin{enumerate}
\item the type \RomanNumeralCaps{2} migration becomes less reduced due to the finite disk mass; 
\item the planet no longer opens the gap in the largest disk accumulated around $t=\tAGB$. This causes an intermediate period of the fast type \RomanNumeralCaps{1} migration. Such behavior can be seen from the contour describing the final orbit $1.0$~AU in Fig.~\ref{fig:contour_different_planet_mass_3}. 
\end{enumerate}
In the second case, the final orbit is almost independent of the initial orbit. This can be explained by a longer intermediate period of the fast type \RomanNumeralCaps{1} migration of the planet with the same mass located farther from the host star, thus, having a greater $\aspectratio$ and so a greater $m_{\rm p}^{\rm crit}$ in its vicinity. It is also notable that the sufficiently distant super-Jupiters with $m_{\rm p} \gtrsim 10^3~\mathrm{m}_\oplus$ migrate outwards. To see it, follow the contour describing the final orbits $3.0$ and $4.0$ AU in Fig.~\ref{fig:contour_different_planet_mass_3}. We consider this novel effect below in Sect.~\ref{sec:decretion_migration}. 

\subsubsection{The rolling migration}
\label{sec:rolling_migration}

In this Section, we discuss a special type of migration that can only take place in the NS disk. Let us consider the general case of a planet embedded in the disk with an aspect ratio variable over the radial coordinate as well as changing over time, $\aspectratio = \aspectratio(r,t)$. The rate of evolution of $\aspectratio$ at the planetary position can be written as
\begin{equation}
\label{eq:com_dot_delta}
\frac{\mathrm{d} \aspectratio}{\mathrm{d}t} = \dot \aspectratio + \aspectratio^\prime v_{\rm mig},
\end{equation}
where $ \dot \aspectratio$ and $\aspectratio^\prime$ denote the partial derivatives of $\aspectratio(r,t)$ over $t$ and $r$, respectively. $v_{\rm mig}$ is the migration rate of the planet having mass $m_{\rm p}$. We assume that the value of $v_{\rm mig}$ is equal to  $v_1$ in case of type~\RomanNumeralCaps{1} migration, and to $v_2$ for the type~\RomanNumeralCaps{2} migration. The migration rate is negative for the rolling migration. Hereafter, we consider the situation of the transition between the types of migration which means that $\aspectratio$ measured by the observer on the planet takes the critical value, $\aspectratio=\aspectratio_{\rm crit}$, related to $m_\mathrm{p}$ via Eq.~\eqref{q_cr_simple}. As discussed in Sect.~\ref{sec:migration_general}, we usually have $|v_2| < |v_1|$ in this situation, see Eq.~\eqref{eq:migration_velocities_add}. Therefore, if $\mathrm{d}\aspectratio/\mathrm{d}t >0~(<0)$, the gap is closed (opened) and the planet accelerates (decelerates) toward the host star changing to the migration type~\RomanNumeralCaps{1} (\RomanNumeralCaps{2}). As far as $\mathrm{d}\aspectratio/\mathrm{d}t$ saves its sign during such a transition, we find an ordinary single change of the migration type. Obviously, this is always the case for the stationary disk when $\dot \aspectratio = 0$. However, the NS disk adds a new option. We find that the sign of $\mathrm{d}\aspectratio/\mathrm{d}t$ changes while planet crosses the point where $\aspectratio=\aspectratio_{\rm crit}$, provided that 
\begin{eqnarray}
\label{roll_migr_conds}
& \dot \aspectratio > 0, \nonumber \\
& \aspectratio^\prime > 0, \\
& v_1 < v_{\rm roll} < v_2 < 0, \nonumber 
\end{eqnarray}
where 
\begin{equation}
\label{v_roll}
v_{\rm roll} \equiv -  \left . \frac{\dot\aspectratio}{\aspectratio^\prime}  \right |_{\aspectratio=\aspectratio_{\rm crit}}
\end{equation}
is the velocity of the constant disk aspect ratio. If the conditions~\eqref{roll_migr_conds} are satisfied, the planet starts migrating at an average rate equal to $v_{\rm roll}$. The detailed hydrodynamical consideration of gravitational disk-planet interaction in a more general case of NS disk is required in order to predict the exact behavior of the migrating planet in this situation. Since the timescales of the gap opening and closure along with the timescale of crossing the gap by the line $\aspectratio=\aspectratio_{\rm crit}$ are all of the same order, it is possible that the migration rate is subject to oscillations around $|v_{\rm roll}|$. We refer to this new type of planetary migration as the rolling migration, since it is determined by the propagation of line $\aspectratio=\aspectratio_{\rm crit}$ along the disk growing at the AGB stage. Let us also note that, generally, the other variants of the rolling migration are possible for other combinations of conditions entering Eq.~\eqref{roll_migr_conds}, see Table \ref{tab:roll_mig}.

\begin{table}
\caption{Conditions for rolling migration in a non-stationary disk}
\centering
\begin{tabular}{ccc}
$\dot\aspectratio$ & $\aspectratio^\prime$ & Migration Rates \\
\hline
positive & positive & $v_1 < v_{\rm roll} < v_2 < 0$ \\
positive & negative & $v_1 > v_{\rm roll} > v_2 > 0$ \\
negative & positive & $v_2 > v_{\rm roll} > v_1 > 0$ \\
negative & negative & $v_2 < v_{\rm roll} < v_1 < 0$ \\
\hline
\end{tabular}
\label{tab:roll_mig}
\end{table}

The conditions~\eqref{roll_migr_conds} are usually met in the NS disk considered in this work. Employing the numerical tests, we find that the rolling migration takes place for the giants $100 \text{--} 300~\mathrm{m}_\oplus$ orbiting the host star within $\sim 1$~AU in the binaries with high-mass donors and intermediate initial separations. In this case, the planet is embedded in the inner part of the sufficiently large disk growing at the AGB stage, which corresponds to the zone of evaporation of dust, see Fig.~\ref{fig:accretion_total}. Note that such planets undergo complete migration. Oscillations of the wind rate taking place in binaries with low-mass donors hinder the rolling migration. An example of a rather long period of the rolling migration is shown in Fig.~\ref{fig:migration_flipflop_and_outward}, left plot. It can be seen that, at first, the planet undergoes the type \RomanNumeralCaps{1} migration which leads to a gradual decrease of $m_{\rm p}^{\rm crit}$ due to the decrease of an aspect ratio around the planet approaching the inner part of a flaring disk. Further, the gap opens and the planet slows down changing to the type \RomanNumeralCaps{2} migration. This period of migration lasts for $\sim 400$~kyr. As soon as the system approaches $t=\tAGB$, the disk aspect ratio grows faster. Once the condition $|v_{\rm roll}| > |v_2|$ becomes satisfied, the period of the rolling migration starts, which lasts for $\sim 300$~kyr and ends with a return back to the type \RomanNumeralCaps{2} migration. Finally, the planet enters the innermost hot part of the disk puffed due to opacity enhanced by the partial ionization of the gas. This causes the new change to the fast type \RomanNumeralCaps{1} migration and the complete migration before the donor loses its envelope.

\subsubsection{Outward migration in the zone of decretion}
\label{sec:decretion_migration}

Here we show examples of the type \RomanNumeralCaps{2} migration in the zone of decretion. Details about this essentially NS variant of the disk evolution are presented in Sect.~\ref{sec:qsd_vs_nsd}. Since the critical mass of the planet opening the gap is large in the outer parts of the disk, we are obliged to consider a super-Jupiter with $m_{\rm p} = 2000~\mathrm{m}_\oplus$, see Fig.~\ref{fig:migration_flipflop_and_outward}, center and right plots. Therefore, we obtain a moderate outward migration reduced by the ratio of the characteristic disk mass and the planet mass. Note that it takes place for both high and low-mass donors. As soon as the accretion rate vanishes, outward migration is replaced by the usual inward migration in a residual disk, since it turns into a quasi-stationary state. If we took a smaller planet, it would undergo the fast type \RomanNumeralCaps{1} migration close to the peak of the accretion rate associated with the AGB stage. Since the type \RomanNumeralCaps{1} of migration is directed inwards in our simple model, it overwhelms a relatively weak outward migration due to the decretion of the disk matter. Nevertheless, an outward migration can be found in contours of constant final planetary orbits for the largest planets, see Fig.~\ref{fig:contour_different_planet_mass_3}.

There is a caveat that the disk structure in the zone of decretion is far from that of a simplified isothermal disk used to obtain the type~\RomanNumeralCaps{1} migration rate in \citet{tanaka2002}. An abnormal negative surface density gradient, which makes the net viscous torque positive despite the negative specific net viscous torque, may reverse the type~\RomanNumeralCaps{1} migration. If this is the case, we should expect a different shape of contours of the constant final planetary orbits for distant planets embedded in the zone of decretion: such planets may avoid the complete migration in the whole range of planet masses, see Fig.~\ref{fig:contour_different_planet_mass_3}. 

To get a feeling of the significance of possible outward type~\RomanNumeralCaps{1} migration, we show the characteristic timescale of decretion and the corresponding migration track of the planet embedded in the zone of decretion. This corresponds to the situation when the planet migrates at the same rate as the gas around it, see the upper panel of Fig.~\ref{fig:migration_flipflop_and_outward}, right plot. Clearly, the rate of migration significantly exceeds the one obtained for the outward migration in our simplified model and leads to fast escape of the planet to the zone of dynamically unstable orbits. The possible outward type \RomanNumeralCaps{1} migration of distant planets may change substantially the overall transformation of a planetary system caused by migration in NS wind-fed disk, as well as the picture of the complete migration of giant planets, see Sect.~\ref{sec:complete_migration} below. The gravitational interaction between a planet and an NS disk with the zone of decretion should be the subject of a future study.

The zone of decretion with sufficiently large surface density may probably lead to other effects
such as the formation of planets in density traps as discussed in~\citet{alessi2020}.

\subsubsection{Complete migration}
\label{sec:complete_migration}

In this Section, we consider the planets subject to complete migration (CM). We define the CM as follows: migration from the initial orbit to the innermost orbit close to the star where additional processes or the orbital evolution (e.g., the tides) are important.

\begin{figure}
    \centering
    \includegraphics[width=\columnwidth]{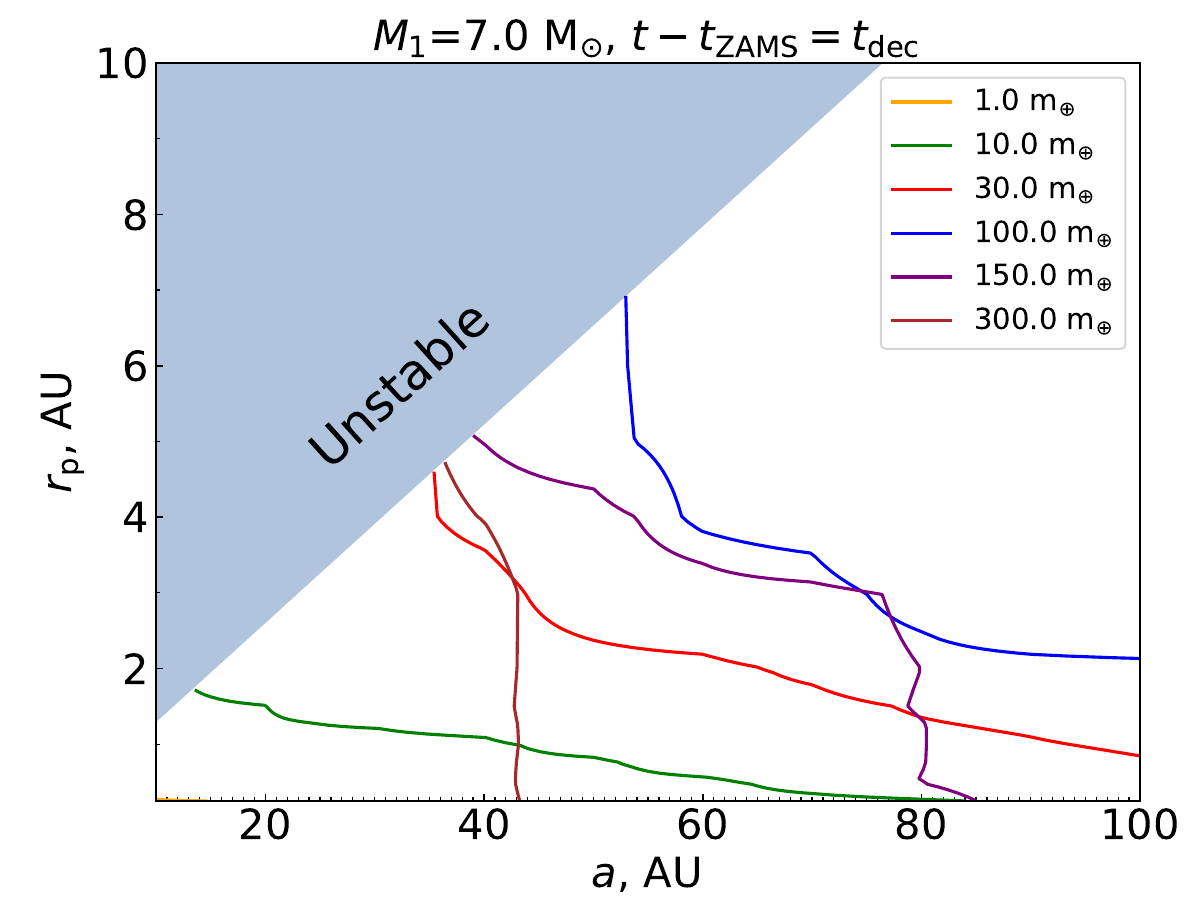}
    \caption{The contours show planets of a given mass that reached the tidal migration distance, $r_{\rm p}^{\rm fin}$, by the time of the disk decay. In other words, all planets of the same mass at the given contour and below will also reach the tidal migration distance. The yellow line for $m_{\mathrm{p}} = 1.0$\,m$_{\oplus}$ is invisible because it is below the plotted region. The donor initial mass is $M_1 = 7.0$~M$_{\sun}$. The notations are the same as in Fig.~\ref{fig:contour_NSD_different_times_1}.}
    \label{fig:contour_different_planet_mass_4}
\end{figure}

We represent this type of migration in the plane $(a, r_{\rm p})$ separately for each donor, see Fig.~\ref{fig:contour_different_planet_mass_4} and Fig.~\ref{fig:contours_different_donor_masses}. There are contours for planets of several given masses. The bottom left part of the plane $(a, r_{\rm p})$ concerning each contour introduces the CM of planets with the corresponding mass. We find these plots to be qualitatively similar to each other. Most importantly, there is always a limiting contour of CM which shows the largest initial orbit of the planet reaching $r_{\rm p}^{\rm fin}$ for the binary with a specified initial separation regardless of the planetary mass. Conversely, all planets preexisting in the binary on orbits above this limiting contour are not subject to CM in our definition. Thus, either they survive in an evolved binary or they are ejected beyond the Roche lobe of the host star due to a dynamical instability of their new orbits. 

The planetary mass representing the limiting contour of CM belongs approximately to the range $60 \text{--} 80~\mathrm{m}_\oplus$ for all donors. We check that all lower mass planets representing their contours of CM reach $r_{\rm p}^{\rm fin}$ undergoing the type \RomanNumeralCaps{1} migration. Therefore, the shape of the limiting contour is similar to that of CM for lower mass planets, see the contours for planets $10~\mathrm{m}_\oplus$ and $30~\mathrm{m}_\oplus$ in Fig.~\ref{fig:contour_different_planet_mass_4} and Fig.~\ref{fig:contours_different_donor_masses}. In the case of the high-mass donors, $5.0~\mathrm{M}_\sun$ and $7.0~\mathrm{M}_\sun$, the limiting contour of CM is close to that of CM of a planet with $m_{\rm p}=100~\mathrm{m}_\oplus$. For example, in the binaries with intermediate initial separation $a=40$ AU, the largest initial orbit of CM exceeds the largest dynamically stable orbit for all donors except for the lowest one, whereas it becomes smaller than $4$ AU or even $3$ AU in the binaries with initial $a=60$~AU, see Fig.~\ref{fig:contour_different_planet_mass_4} and Fig.~\ref{fig:contours_different_donor_masses}. The planets with $m_{\rm p} \lesssim 10~\mathrm{m}_\oplus$ undergo CM in close binaries starting with initial orbits $r_{\rm p} \lesssim 2$~AU. The intermediate-mass planets, $\sim 30~\mathrm{m}_\oplus$, undergo CM for the whole range of binary initial separations including the distant planets with $r_{\rm p} \gtrsim 2$~AU in sufficiently close binaries with $a \lesssim 40$~AU. CM of the lower mass planets becomes a little weaker as one switches from high-mass to low-mass donor. 

The existence of the limiting contour of CM is explained as follows. For larger planetary masses, the increasing rate of the type \RomanNumeralCaps{1} migration makes the planet reach $r_{\rm p}^{\rm fin}$ from a larger initial orbit. This is no longer the case as soon as episodes of the type \RomanNumeralCaps{2} migration appear. For relevant values of $r_\mathrm{p}$, $a$, and $m_{\rm p}$ the planet opens the gap for the first time in a residual disk. We check that the critical mass of the gap formation decreases down to the planet's mass as soon as the disk cools down in the absence of wind, and additionally, the planet reaches the inner part of the disk which has the least aspect ratio due to the evaporation of dust. Under these conditions, the further increase of the planetary mass leads to longer episodes of relatively slow type \RomanNumeralCaps{2} migration. Therefore, the largest initial orbit of CM stops growing for an increasing planetary mass in the given binary. Note that the final period of migration yielding the limiting contour of CM occurs always with a closed gap. We find that the latter takes place in the innermost hot region of the young residual disk puffed due to opacity enhanced by partial ionization of the gas. Therefore, the final period of migration in this case occurs at a comparatively high rate. As the planet becomes even more massive, it holds the gap opened all the time it stays in the residual disk. This slows down the migration and shifts the contour of CM back to more close binaries and lower initial planetary orbits on the plane $(a, r_{\rm p})$. This proceeds unevenly along the contour. Thus, we find that CM ceases in the binaries with the largest initial separations $a \gtrsim 80$~AU as the planet mass increases from $100~\mathrm{m}_\oplus$ up to $150~\mathrm{m}_\oplus$ regardless of the planet's initial location, see Figs.~\ref{fig:contour_different_planet_mass_4} and Fig.~\ref{fig:contours_different_donor_masses}, left plot for the case of high-mass donors. At the same time, CM ceases at initial separations $a \gtrsim 80$~AU already for the planetary mass $\sim 100~\mathrm{m}_\oplus$ and at $a \gtrsim 60$~AU for the planetary mass $\sim 150~\mathrm{m}_\oplus$, see Fig.~\ref{fig:contours_different_donor_masses}, center plot for the case of donor $3.0~\mathrm{M}_\sun$. 

The least massive donor $1.7~\mathrm{M}_\sun$ cannot provide CM of planets with $m_{\rm p}=~\sim100~\mathrm{m}_\oplus$ already at initial separations $a \gtrsim 70$~AU regardless of the planetary initial location, see Fig.~\ref{fig:contours_different_donor_masses}, right plot. Hence, a part of the contour describing CM for the largest initial binary separations is almost independent on $r_{\rm p}$ and shifts to the closer binaries as the mass of the giant further increases. 
 
The contour of CM for Jupiters, $m_{\rm p}=~\sim300~\mathrm{m}_\oplus$, becomes independent on $r_{\rm p}$ for all dynamically stable orbits regardless of the donor. It corresponds to binaries with initial separation slightly exceeding $40$, $35$, $30$, and $20$~AU from the largest to the smallest donor, respectively. Such a feature of CM contour is caused by the weak dependence of the final orbit of the giant on its initial orbit. This is also in accordance with contours of partial migration represented on the plane $(m_{\rm p}, r_{\rm p})$, see Fig.~\ref{fig:contour_different_planet_mass_3} above. In other words, distant giants catch up with close giants on the way to the host star. As discussed at the end of Sect.~\ref{sec:migration_general}, this is caused by the different ratio of the time periods of either opened or closed gap spent by the giant as migrating across the wind-fed disk from different initial orbits. Indeed, the further away the planet is from the host star initially, the earlier it switches to the type \RomanNumeralCaps{1} migration due to the growth of NS flaring disk while approaching $t=\tAGB$. Since the type \RomanNumeralCaps{1} migration under the conditions near the gap formation is always faster than the type \RomanNumeralCaps{2} migration, the total time of migration up to the distance of TM hardly increases with increasing initial orbit. We check that the structure of the NS disk in the vicinity of the planet undergoing CM does not facilitate the transition to the rolling migration. Thus, we usually find a single transition from the type \RomanNumeralCaps{2} to the \RomanNumeralCaps{1} migration in such a situation. Finally, the CM of Jupiters and super-Jupiters is additionally suppressed by the decrease of the type \RomanNumeralCaps{2} migration rate on account of the finite characteristic mass of the disk.

\section{Discussion}
\label{sec:discussion}


\subsection{Stellar wind properties}

\label{sec:wind_props}

In this study, we make simplifying assumptions on the stellar wind properties. This is motivated by the main goal, which is to consider the significant planetary migration in a relatively massive disk formed during the relatively short stages of an extensive mass loss by the primary. In the first place, we are interested in the AGB stage when the stellar wind is intensive and it has a low velocity. Typically, the corresponding value is taken as $\sim 10\text{--}20$~km\,s$^{-1}$. Such a value has been used in previous studies of accretion disks in binaries with evolved donors, e.g., \citet{perets2013, kulikova2019}. 

 In our model, we do not study the dependence of the disk structure on the stellar wind velocity.  Instead, we use the same value $20$~km\,s$^{-1}$ for the whole evolutionary track, which is unrealistic. Obviously, stars have much larger wind velocities at the Main sequence stage~\citep[e.g.,][]{cassinelli1979}. This should result in a far smaller Bondi radius and lower accretion rate. Thus, we significantly overestimate the accretion rate for a long-term formation of the disk before the RGB stage, as well as between the RGB and AGB stages of the low-mass donors. Still, we find that such an overestimation does not lead to any considerable planetary migration before the RGB stage of the primary. Additionally, the wind velocity might depend on the stellar parameters (mass, radius, temperature, metallicity), see, e.g.,~\citet{vink2000} and references therein. Analysis of these dependencies is out of the scope of this paper, which is to evaluate the upper range of the disk growth and planetary migration. More detailed calculations are a subject for future study.

Our computations of the NS disk show that a small and very light disk with a size less than $1$\,AU consisting entirely of R2 appears at an accretion rate much below the formal threshold value, $\dot M_{\mathrm{acc}}^{\mathrm{lim}}$. Within our simplified model of wind, this formally corresponds to the epochs far earlier than the RGB stage, or additionally, far later/earlier than the RGB/AGB stage in the case of the low mass donors, for the considered close binaries with $a \lesssim 100$\,AU. Of course, it is unlikely that such a disk will exist in the model with a variable wind velocity. However, we expect that such a disk may appear in wide binaries with $a > 100$\,AU, which provides a low accretion rate close to RGB  and AGB peaks of the low-velocity wind. If confirmed, the formation of a light wind-fed disk may cause moderate migration of giants in wide binaries.

There is also a caution that the appearance of such a disk can be prevented by counteraction of the wind from the accretor. The latter can be roughly assessed by comparing the wind ram pressure from both the donor and the accretor at the Bondi radius of the accretor, see Eq.~\eqref{eq:bondi_accretion_radius}. Given the spherical symmetry of winds, the corresponding estimate shows that counteraction of the solar-like accretor is not enough to prevent the disk formation at the wind rate of the donor above $\sim 10^{-10}$--$10^{-11}$~M$_\sun$\,yr$^{-1}$, which corresponds to the accretion rate well below $M_{\mathrm{acc}}^{\mathrm{lim}}$.



\subsection{Circularization radius versus Bondi radius and 
process of matter settling}

By circularization radius we mean a specific distance from the rotation axis of the accretion disk where the bulk of the captured matter gets into circular streamlines with a velocity close to the Keplerian rotation of the disk, and afterward, the matter cools down and settles into the disk. In realistic accretion disks the circularization radius can be much smaller than the radius of gravitational capture, $r_{\mathrm{a}}$, which is identified in our model with a radius under which all matter settles into the disk. Thus, we assume that $r_{\mathrm{a}}$ defines the efficiency of disk accretion $\eta$: $\dot M_{\mathrm{acc}} \sim \eta \dot M_{\mathrm{w}}$, see Eq.~\eqref{eq:bondi_accretion_rate}. However, the situation may differ.

An exact picture of the captured wind matter settling into the disk might be quite complicated 
\citep[see, e.g.,][]{huarte-espinosa2013}. Depending on the orbital, sound, and wind velocities, the structure of the flow inside the Bondi radius can vary. Additional deviations in comparison with a simplified Bondi–Hoyle–Lyttleton accretion (see, e.g., \cite{edgar2004} for a review and references to early studies) may be due to the properties of the outflow from the donor, especially in the case of AGB stars, see, e.g.,~\cite{elmellah2020} and references therein. The outflow can be spherically non-symmetric and the wind velocity may have a complicated profile depending on the detailed parameters of the donor (e.g., rotation and chemical composition). Additionally, the pulsations of the AGB star may influence the outflow as well as the accretion flow \citep{chen2017}. Some simplifications are necessary for the purposes of our study as we want to study the long-term evolution of an accretion disk. Hence, we follow the approach presented in~\citet{perets2013}\footnote{We note that eq. (1) in~\citet{perets2013} contains misprints: $\frac{j_\mathrm{a}}{j_2} = 1.2 \dots \left(\frac{M_1 + M_2}{2.5 \mathrm{~M}_{\sun}}\right)^{1} \dots$ should be replaced with $\frac{j_\mathrm{a}}{j_2} = 0.74 \dots \left(\frac{M_1 + M_2}{2.5 \mathrm{~M}_{\sun}}\right)^{1/2} \dots$} making no difference between the circularization radius and the Bondi radius.

\subsection{Photo-evaporation of a residual disk}

In this study, we follow the disk evolution after the donor forms a white dwarf (WD). The newborn WD is characterized by high effective temperature in the range from $\sim 10^{4.5}$~K for low mass donors up to $\sim 10^5$~K for high mass donors~\citep[see][for a recent review on white dwarfs physics]{blouin2024}. That leads to irradiation of the disk by high-energy photons. Such irradiation affects the outer parts of the disk via photo-evaporation. 

We closely follow~\citet{armitage2017} and estimate the photo-evaporation rate of the disk considering the irradiation of the outer edge of the disk by photons with energy sufficient to ionize and evaporate the disk matter. For simplicity, we assume that the disk consists of neutral hydrogen with threshold ionization energy 13.6\,eV. This value corresponds to the high-energy tail of the spectrum of WD. We estimate the photon illumination of the disk as
\begin{equation*}
    \Phi_{\mathrm{disk}} = \Phi_{\mathrm{total}} \frac{S_{\mathrm{disk}}}{S_{\mathrm{total}}} \approx \Phi_{\mathrm{total}} \frac{\verticalscale_{\mathrm{max}} r_{\mathrm{max}}}{\pi a^2},
\end{equation*}
where $\Phi_{\mathrm{total}}$ is the luminosity of WD in the band $\geq 13.6$\,eV, $S_{\mathrm{total}}$ is the surface of the sphere with the radius $a$ centered at WD, which is the final separation of the system after the AGB stage, $S_{\mathrm{disk}}$ is the cross-section of the disk edge as seen from WD, roughly estimated as the rectangle $2 \verticalscale_{\mathrm{max}} \times 2 r_{\mathrm{max}}$ corresponding to the largest radial and vertical sizes of the disk. 

We estimate the high-energy fraction of illumination by WD integrating the black-body energy distribution over the energies from 13.6\,eV to infinity. The result of integration gives the photon illumination in the range $10^{44} \text{--} 10^{46}$ photons~s$^{-1}$ for WD temperatures in the range $10^{4.5} \text{--} 10^5$~K. Finally, we obtain an estimate of the photo-evaporation rate of the disk by multiplying $\Phi_{\mathrm{disk}}$ by the hydrogen atom mass, $\mathrm{m}_{\mathrm{H}}$, as
\begin{multline}
\label{eq:evaporation_rate}
    \dot M_{\mathrm{evap}} = \mathrm{m}_{\mathrm{H}} \Phi_{\mathrm{disk}} \approx \\
    \approx 10^{-7} \frac{\mathrm{M}_\sun}{\mathrm{yr}} \left(\frac{\Phi_{\mathrm{total}}}{10^{46} \mathrm{~s}^{-1}}\right) \left(\frac{\verticalscale_{\mathrm{max}}}{1 \mathrm{~AU}}\right) \left(\frac{r_{\mathrm{max}}}{10 \mathrm{~AU}}\right) \left(\frac{a}{60\mathrm{~AU}}\right)^{-2}.
\end{multline}
The estimate~\eqref{eq:evaporation_rate} gives us an upper limit on the evaporation rate of the accretion disk. 

Photo-evaporation at rate $\lesssim 10^{-7}\,\mathrm{M}_\sun\,\mathrm{yr}^{-1}$ 
depletes the disk much faster than its viscous evolution. Indeed, the timescale of photo-evaporation is estimated as $t_{\mathrm{evap}} \sim M_{\mathrm{disk}} {\dot M_{\mathrm{evap}}}^{-1} \gtrsim 10^4 \mathrm{~yrs}$ for typical disk mass $\sim 10^{-3}$\,$\mathrm{M}_\sun$ after AGB, which is about two orders of magnitude below the viscous timescale approaching $t_\nu \sim 10^6$ yrs. However, it is known that thermal wind from the disk surface produced by external high-energy radiation stalls under the radius of gravitational capture of matter heated in the vicinity of the host star, see~\citet{hollenbach1994}. The temperature of the corresponding ionized atmosphere above the disk surface attains $10^4$~K, which yields gravitational capture of the heated matter up to $r\sim 9$ AU around a Solar-like star. Here, we obtain that the largest residual disk only slightly exceeds this value, see Table \ref{tab:values_30}. Thus, we conclude that photo-evaporation of a residual disk has little effect on its evolution and the corresponding planetary migration.

\subsection{Lower disk viscosity}
\label{sec:discussion_disk_turbulence}


In this work, we have assumed that $\alpha=0.01$. 
However, current observations do not exclude the much lower values of $\alpha$. 
Such uncertainty obliges us to track how the results can change.
Here we briefly report additional calculations of the accretion disk evolution 
and the corresponding planetary migration for $\alpha = 0.001$. 

The lower viscosity leads to a substantial decrease of the radial velocity in a disk and the respective increase of its viscous timescale. As far as the disk evolution stays in a quasi-stationary regime, we expect the respective increase of surface density provided that the accretion rate remains the same. Note that according to the analytical stationary solutions specified by the opacity typical in protoplanetary disks, the temperature increases only slightly in this case. We find that the NSD solution is in accordance with these conclusions at the stages of donor evolution with slowly varying wind. However, as the wind rate rapidly grows towards RGB and AGB peaks, the surface density increases more slowly than it might be expected from the comparison with the corresponding NSD solution obtained for $\alpha=0.01$. The same applies to the disk mass. We attribute this to the non-stationary lag behind the corresponding QSD solution, which becomes even greater than that in the case of $\alpha=0.01$. Thus, the lower viscosity disk attains only twice as the mass of the disk described in Sect.~\ref{sec:accretion_general} by $t=\tAGB$ for the most massive donor, while its size does not exceed $\sim 15$~AU. We find that at that moment the zone of decretion expands inside the disk considerably farther than that in the case of $\alpha=0.01$, while the abnormal amplitude of positive radial velocity decreases considerably less than it might be expected from the difference between the viscosities. Note that the residual disk with $\alpha=0.001$ lives for more than $10$ Myr, which is by order of magnitude longer as compared with the case of $\alpha=0.01$.

Due to the larger disk volume density the type~\RomanNumeralCaps{1} migration 
becomes faster. 
In contrast, the type~\RomanNumeralCaps{2} migration 
becomes slower due to the slower accretion. The transition between the migration types 
occurs for less massive planets according to Eq.~\eqref{q_cr_simple}. 
Thus, the distant sub-Jupiters at several AU and low-orbit Neptunes at one AU become 
the gap-opening planets, which is not the case for $\alpha=0.01$.
We expect planetary migration to become extensive also due to the long-lived residual disk.

Let us introduce the comparative results with regards to the complete migration (CM here below) for the most massive donor.
We find that the limiting contour of CM shifts to the lower planetary mass, 
$40$--$50\, \mathrm{m}_\oplus$. At the same time, CM covers the much larger part of the 
plane $(a,\, r_{\rm p})$ due to the prominent type~\RomanNumeralCaps{1} migration.
For example, planets $m_{\rm p} = 10$~m$_{\oplus}$ 
are subject to CM from any initial dynamically stable orbit 
provided that the initial separation of the binary is $a \leq 60$~AU, whereas 
in systems with $a \leq 100$~AU 
the same is the case for all initial orbits $r_{\rm p} < 6$~AU.
Moreover, the Earth-mass planets $m_{\rm p} = 1$~m$_{\oplus}$
undergo CM from any initial dynamically stable orbit
for initial binary separations $a \leq 30$ AU. In contrast, the giants migrate much slower than that in the disk with $\alpha=0.01$. So, planets $m_{\rm p} = 300$~m$_{\oplus}$ 
undergo CM from any initial dynamically stable orbit for initial binary separation as small as $a \leq 20$~AU. 

\subsection{Advanced migration}
\label{sec:discussion_migration}





During the last decades, the description of migration has been developed far beyond the basic variant introduced above, see the review by~\citet{paardekooper2023}. Foremost, it was shown that the unsaturated corotation torque arising in real disks with thermal and viscous diffusion slows down type \RomanNumeralCaps{1} migration of super-Earths, see, e.g.,~\citet{paardekooper2006} and \citet{kley2008}. The generalized migration rates derived by \citet{paardekooper2010}, \citet{paardekooper2011} and by \citet{jimenez2017} later on have been used to upgrade, for example, simulations of planetary populations \citep{dittkrist2014} and planet formation models \citep{guilera2019}. It has become clear that the generalized type \RomanNumeralCaps{1} migration is quite sensitive to the disk structure. For example, it may trap the planet in the vicinity of the opacity transitions, see, e.g., \citet{kretke2012}. 
The original type \RomanNumeralCaps{2} migration built on the assumption of an empty gap has been also
refined. Numerical studies of the giants interacting with the surrounding accreting matter showed 
that the gap is never swept out completely. The remaining gap-crossing flow allows the planet to 
migrate independently on the accreting gas so that the actual type 
\RomanNumeralCaps{2} migration rate may exceed its original value by a few times, see, e.g.,~\citet{durmann2015}. An advanced analytical model of type \RomanNumeralCaps{2} migration
suggests that gravitational torque on the planet is determined by the surface density at the 
bottom of the gap, see \citet{kanagawa2018}. The dependence of the latter on the viscosity and
aspect ratio of the disk as well as the mass of the planet determines the gradual 
transition between the two types of migration. 
We relegate the application of the advanced migration theory to the migration of planets pre-existing 
in evolved wide binaries for future work.

\section{Conclusions}
\label{sec:conclusions}

We study planetary migration in a wind-fed accretion disk around a secondary component of a binary system. For that, we construct the detailed model of the NS disk within the paradigm of the standard disk accretion. For the first time, the energy balance across the disk includes both irradiation by the host star and by the donor. We find the donor heating to be significant at the outer parts of the disk for the whole range of binary separations, as well as for all considered donor masses. This is the case due to the combination of the small enough disk size and aspect ratio along with the proximity of the donor and its large size at RGB and AGB stages. At the same time, the disk energy balance is generalized onto the optically thin medium. The latter makes it possible to take into account the formation of very light disks, of the order of $\sim 10^{-7}\,\mathrm{M}_\odot$ and even smaller, which takes place in the sufficiently wide binaries and/or for the low mass donors, see Table \ref{tab:values_30}, the details in Section \ref{sec:accretion_other_donors} and discussion in Section \ref{sec:wind_props}. 

We show that the disk exists up to 5\,Myr before the envelope loss of the most massive donor in binaries with intermediate initial separation $\sim 30$\,AU. The disk accumulates mass up to $0.01\,\mathrm{M}_\odot$ reaching the size up to $18$\,AU by the AGB peak from the wind of the most massive donor in closer binaries with the initial separation $\sim 20$\,AU. The further decrease of initial separation leads to the tidal truncation of the disk. The disk is non-stationary because it is supplied by a variable source of matter. This variability is provided by two reasons. First, the rapid evolution of a donor at RGB and AGB stages. Second, the accompanying change of binary separation and mass ratio. 

We find that the disk behaves approximately quasi-stationary only sufficiently far from RGB and AGB peaks of the donor wind. At the same time, it becomes essentially non-stationary close to the maxima of the wind rate, which is expressed in the appearance of an extended zone of decretion in its outer parts beyond $\sim 2$\,AU. The absolute value of the radial velocity of decreting matter substantially exceeds the standard quasi-stationary value. The zone of decretion exists up to several hundred kyr before the AGB peak of the wind rate. The full NS disk model allows us to determine the residual disk left after the envelope loss by the donor. It depletes for $\sim 1$\mbox{ Myr} in the quasi-stationary regime.

We find the new physical features that planetary migration acquires in the NS disk. In a growing disk, a planet undergoing the type \RomanNumeralCaps{2} migration can be accelerated up to the rate, which is intermediate between the rates of the type \RomanNumeralCaps{2} and the faster type \RomanNumeralCaps{1} migrations. It looks like the planet is trapped by the line of the critical value of the disk aspect ratio determining the gap formation, which propagates towards the host star through a flaring disk. That is why we refer to this new type of migration as the rolling migration. We suppose that rolling migration is activated by the partially closed gap, though the corresponding torque on the planet should differ from that in the advanced type \RomanNumeralCaps{2} migration developed by \citet{kanagawa2018} and others in the case of a stationary disk with constant aspect ratio. That is the subject of future research to give a comprehensive hydrodynamic description of rolling migration. In the zone of decretion, the planet that opens the gap should migrate outwards. Moreover, the corresponding abnormal distribution of matter in the zone of decretion should presumably provide an outward type \RomanNumeralCaps{1} migration. 
Additionally,
there is a general issue of the transition between the two types of migration 
in the regions of disk with substantial radial and/or time derivatives of an aspect ratio and the surface density while considering them jointly with the radial motion of the migrating planet.
This may be important to address, since the characteristic time of variation of these quantities 
in the vicinity of the migrating planet can be comparable to the characteristic time of gap 
opening/closure.
However, these conjectures should be verified through the solution of the corresponding hydrodynamical problems, which are far beyond the scope of this study.


We find that planets with masses up to $\sim 60\,\mathrm{m}_\oplus$ mostly are subject to the type \RomanNumeralCaps{1} migration regardless of the donor mass. Already the Neptune-like planets from initial orbits at a few AU significantly approach the host star until the disk decays -- even in the systems with initial $a\sim 100$\,AU. Moreover, the Neptune-like planets from all dynamically stable orbits reach the tidal migration distance in the close systems with initial $a \lesssim 20$\,AU for all donors. We also check that migration in the residual disk substantially contributes to that. 

As planetary mass increases up to $\sim 100\,\mathrm{m}_\oplus$, the most distant planets reaching the tidal migration distance begin to open the gap in the innermost parts of the decaying residual disk. The corresponding episode of the relatively slow type \RomanNumeralCaps{2} migration inhibits the further increase of the largest initial orbit of the planet subject to complete migration. Therefore, for every system with a specified initial separation, there is a limiting initial planetary orbit beyond which no one planet approaches the tidal migration distance. Even heavier planets undergo the fast type \RomanNumeralCaps{1} migration in the vicinity of the AGB peak only, while the slow type \RomanNumeralCaps{2} migration takes place the rest of the time. The limiting initial orbit of the planet subject to complete migration equals the largest dynamically stable planetary orbit in the systems with the initial separation $\sim 30$\,AU and $\sim 50$\,AU for the low and the high mass donors, respectively. It decreases up to $\sim 2$\,AU in the wider systems with the initial separation $\sim 80$ to $ 100$\,AU for all donors.  



\begin{acknowledgements} 
We thank the anonymous referee for helpful and insightful suggestions which allowed us to improve the paper. The authors thank Galina~V.~Lipunova and Nikolai~I.~Shakura for the insightful discussions. The authors thank Alexander S. Andryushin for providing the stellar evolution tracks calculated with MESA. This work was partially supported by the German research foundation, \emph{Deut\-sche For\-schungs\-ge\-mein\-schaft, DFG\/}.
VVZ acknowledges the support of Roscosmos. This research makes use of the open-source \textsc{Python} libraries \textsc{numpy}, 
\textsc{scipy}, 
\textsc{astropy} 
and \textsc{matplotlib}.
\end{acknowledgements}

\section*{Data availability}
The data underlying this article will be shared on reasonable request to the corresponding author.




\bibliographystyle{aa} 
\bibliography{binary_planets} 



\begin{appendix}


\onecolumn 

\section{Additional figures}
\label{app:figures}

Here we present our additional Figures used in the corresponding Sections of the main text.

\noindent\begin{minipage}{\textwidth}
    \centering
    \includegraphics[width=0.33\textwidth]{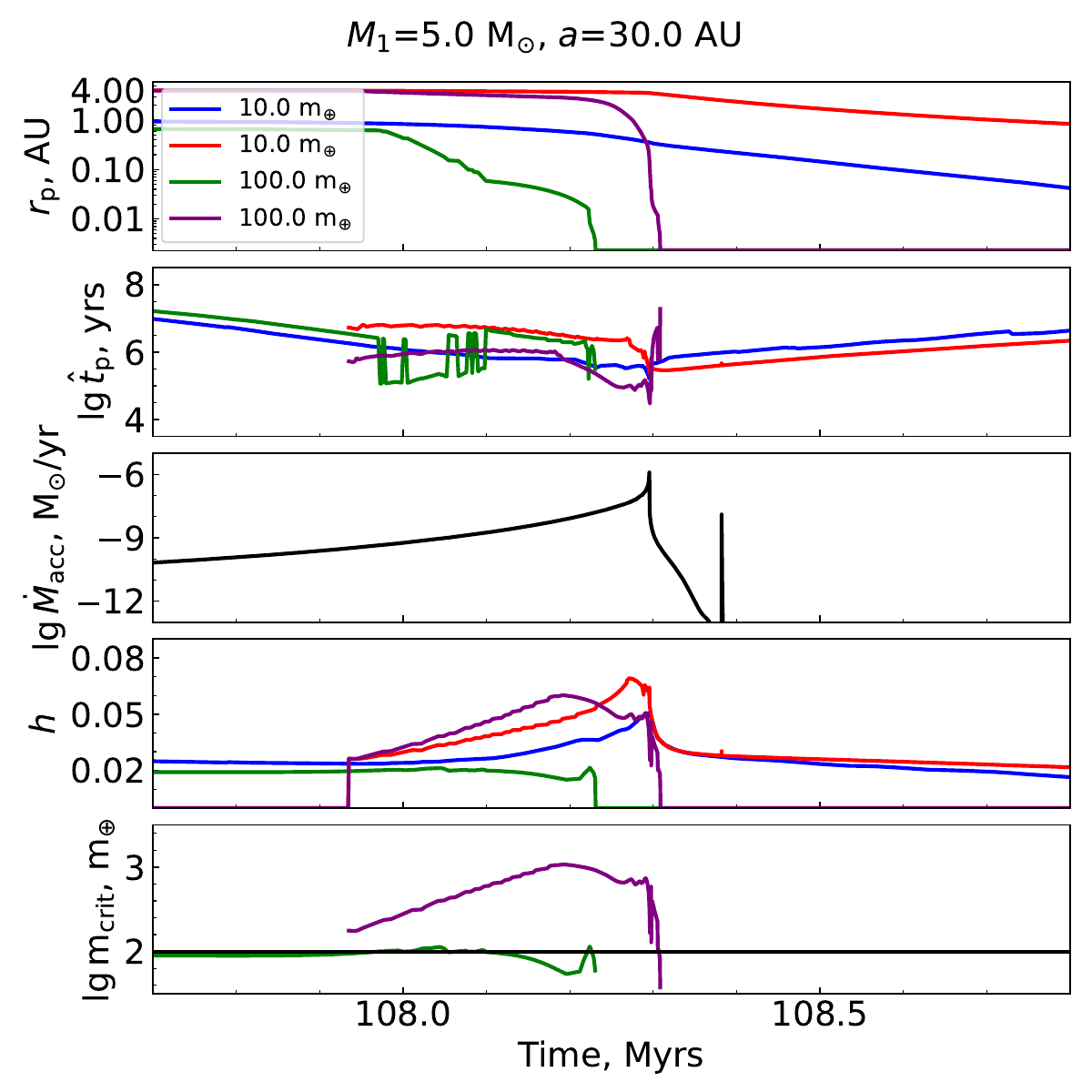}
    \includegraphics[width=0.33\textwidth]{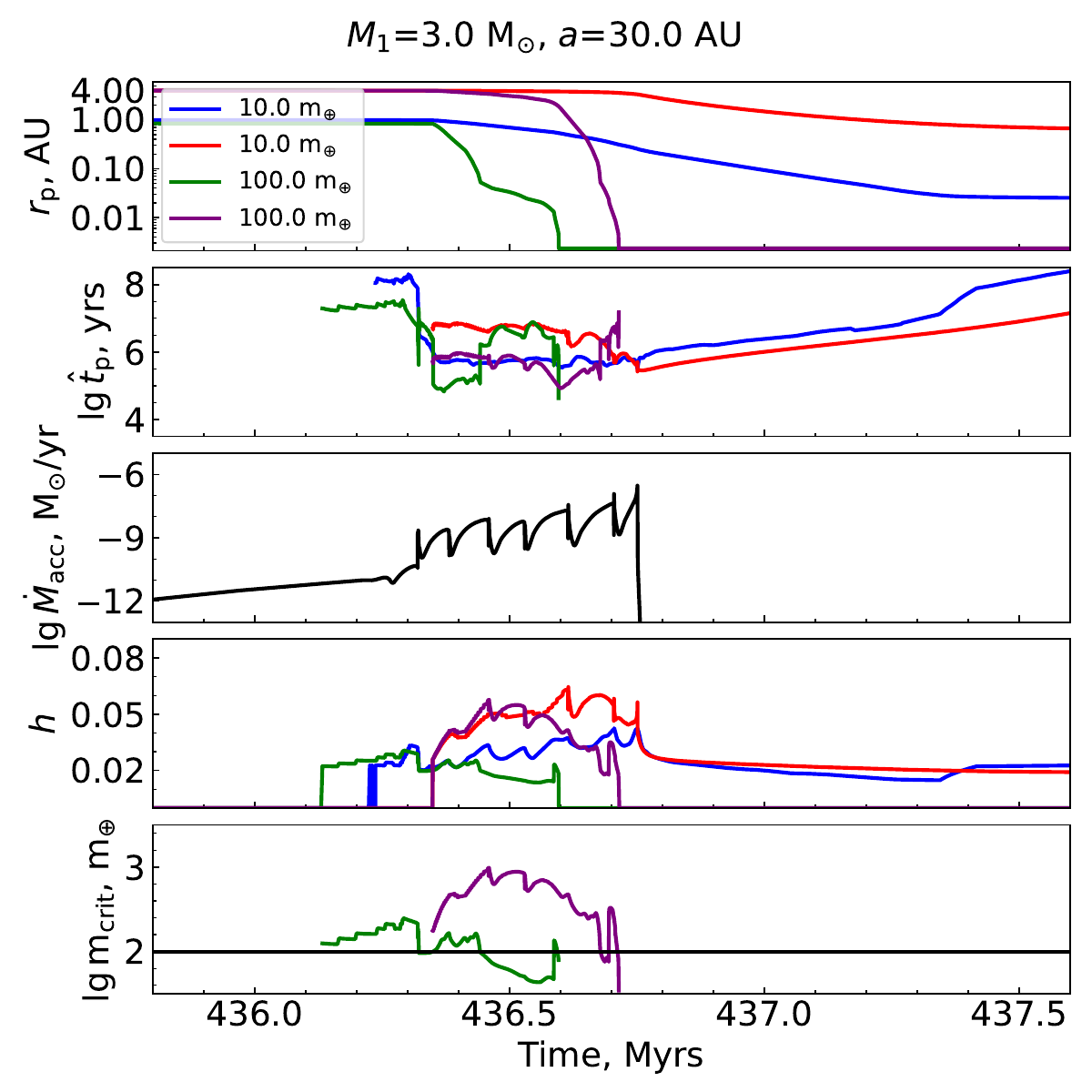}
    \includegraphics[width=0.33\textwidth]{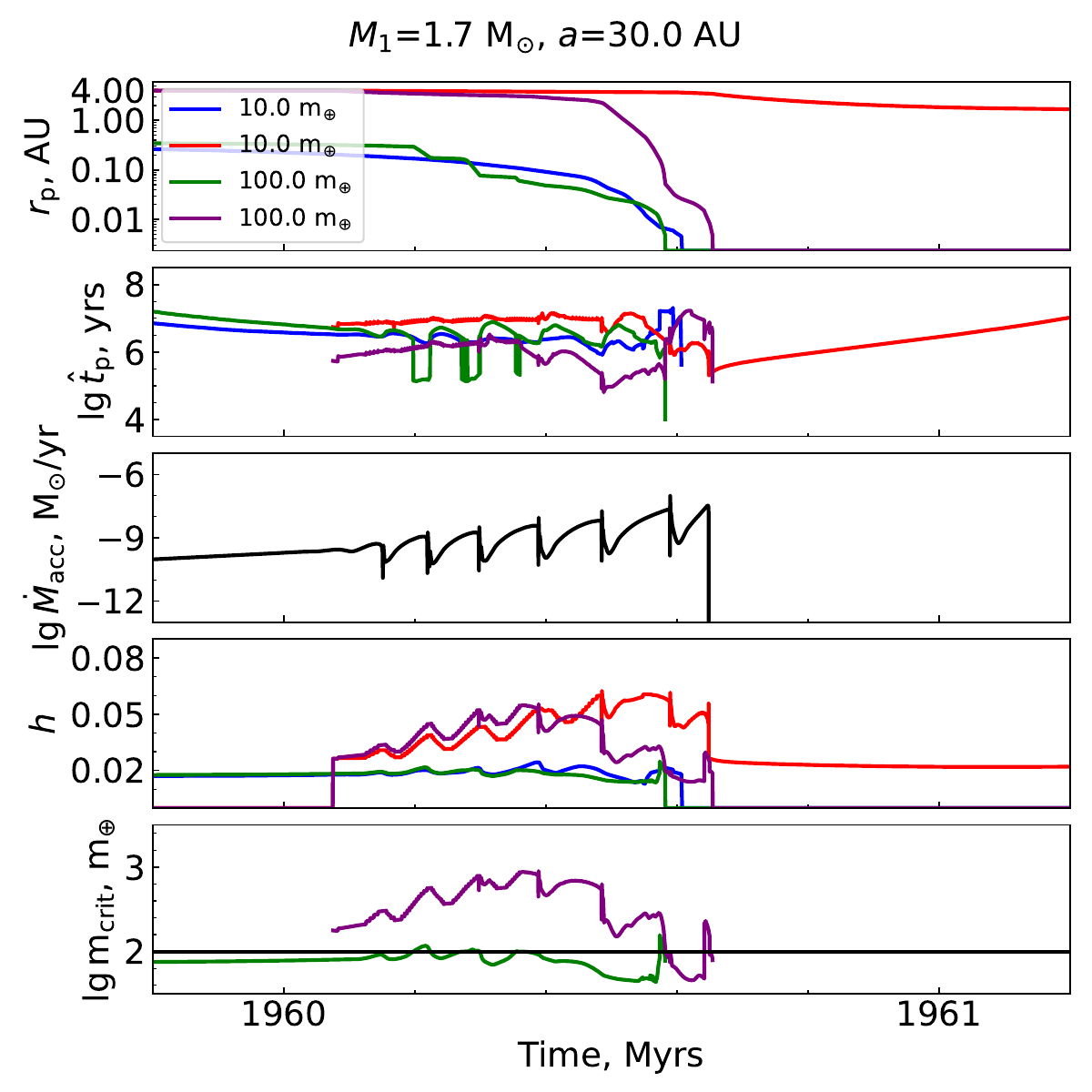}
    \captionof{figure}{The same as Fig.~\ref{fig:migration_general_1} but for the systems with $M_1 = 5.0$~M$_{\sun}$ (left plot), $M_1 = 3.0$~M$_{\sun}$ (center plot), $M_1 = 1.7$~M$_{\sun}$ (right plot). 
    }
    \label{fig:migration_different_donor_masses}
\end{minipage}

\vspace{1\baselineskip}

\noindent\begin{minipage}{\textwidth}
    \centering
    \includegraphics[width=0.33\textwidth]{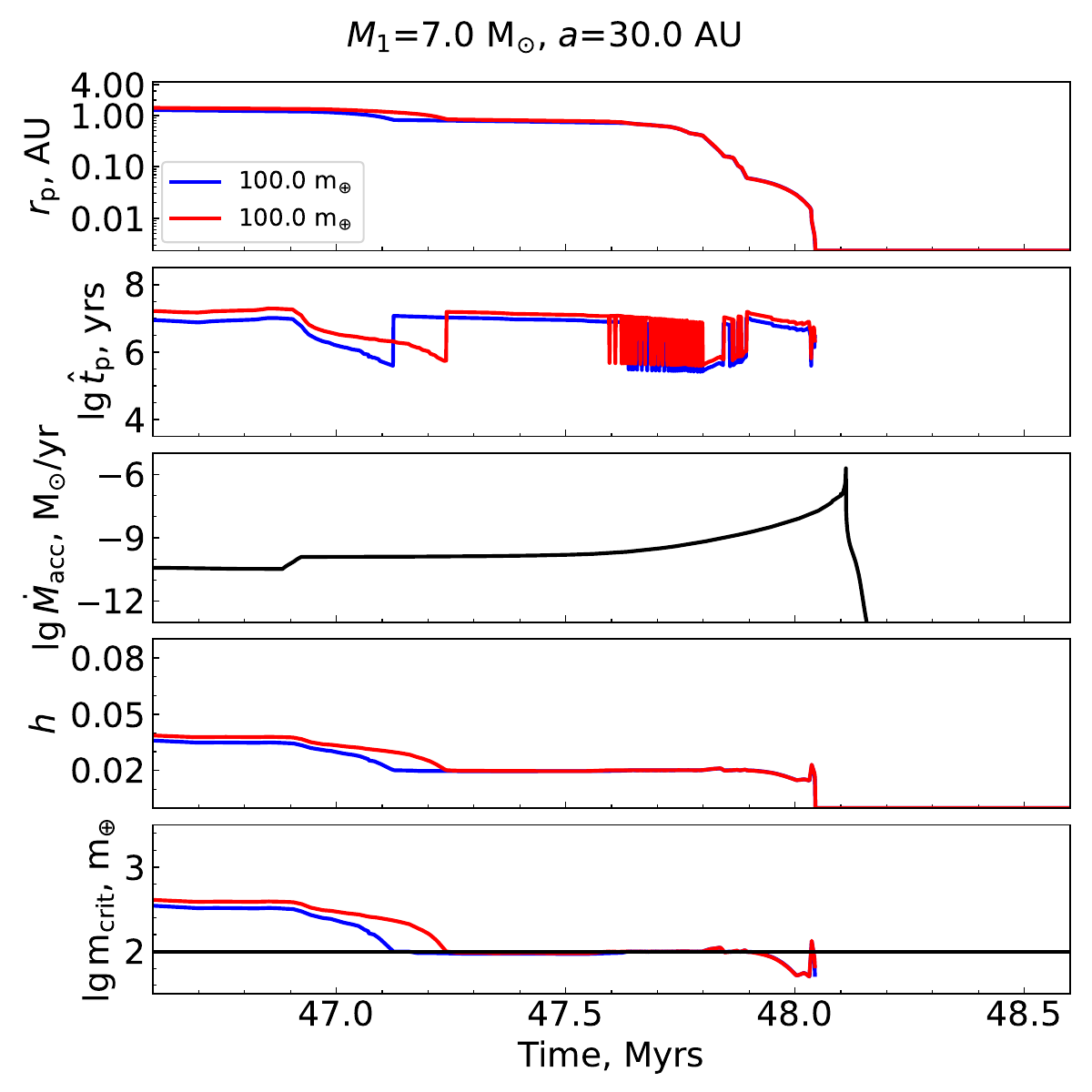}
    \includegraphics[width=0.33\textwidth]{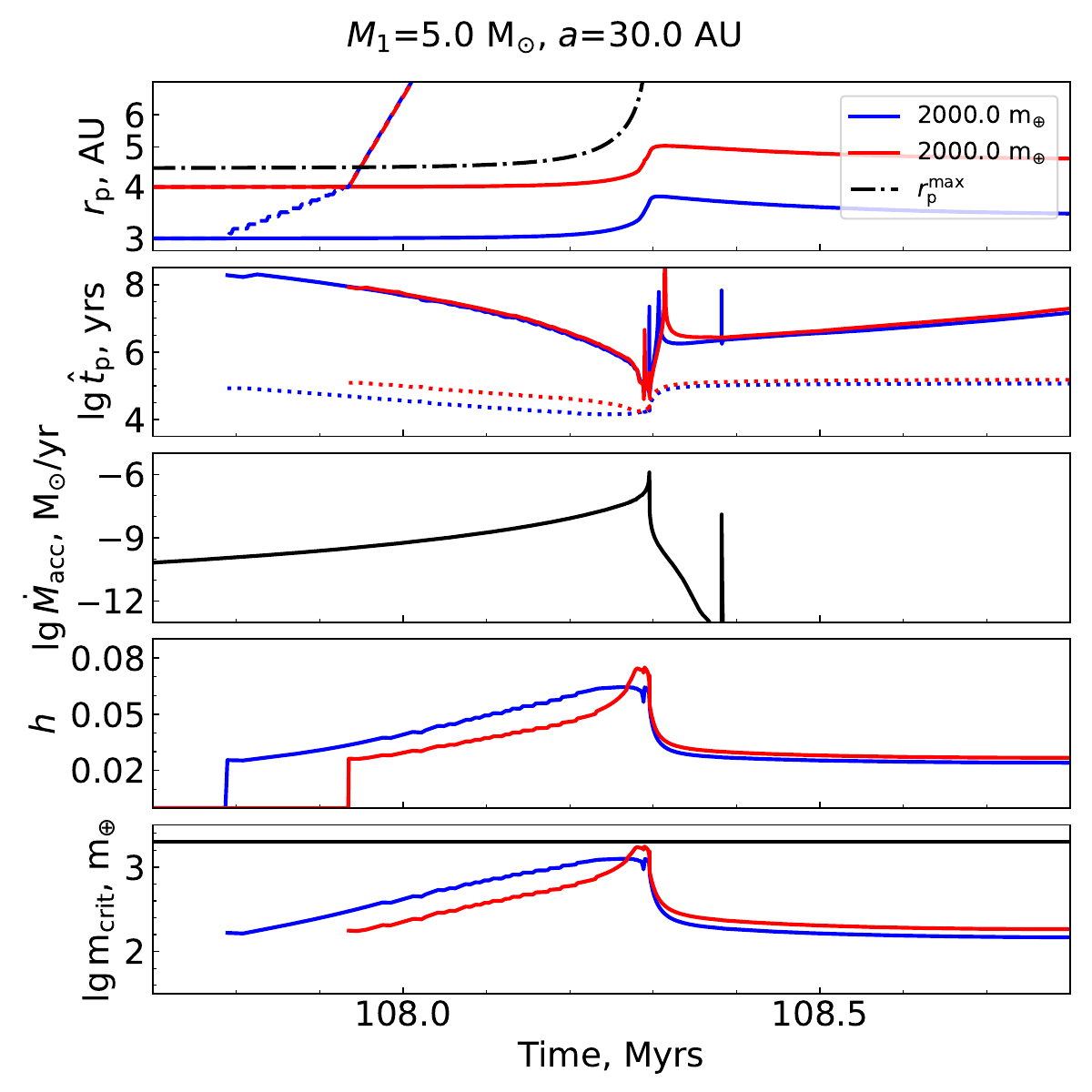}
    \includegraphics[width=0.33\textwidth]{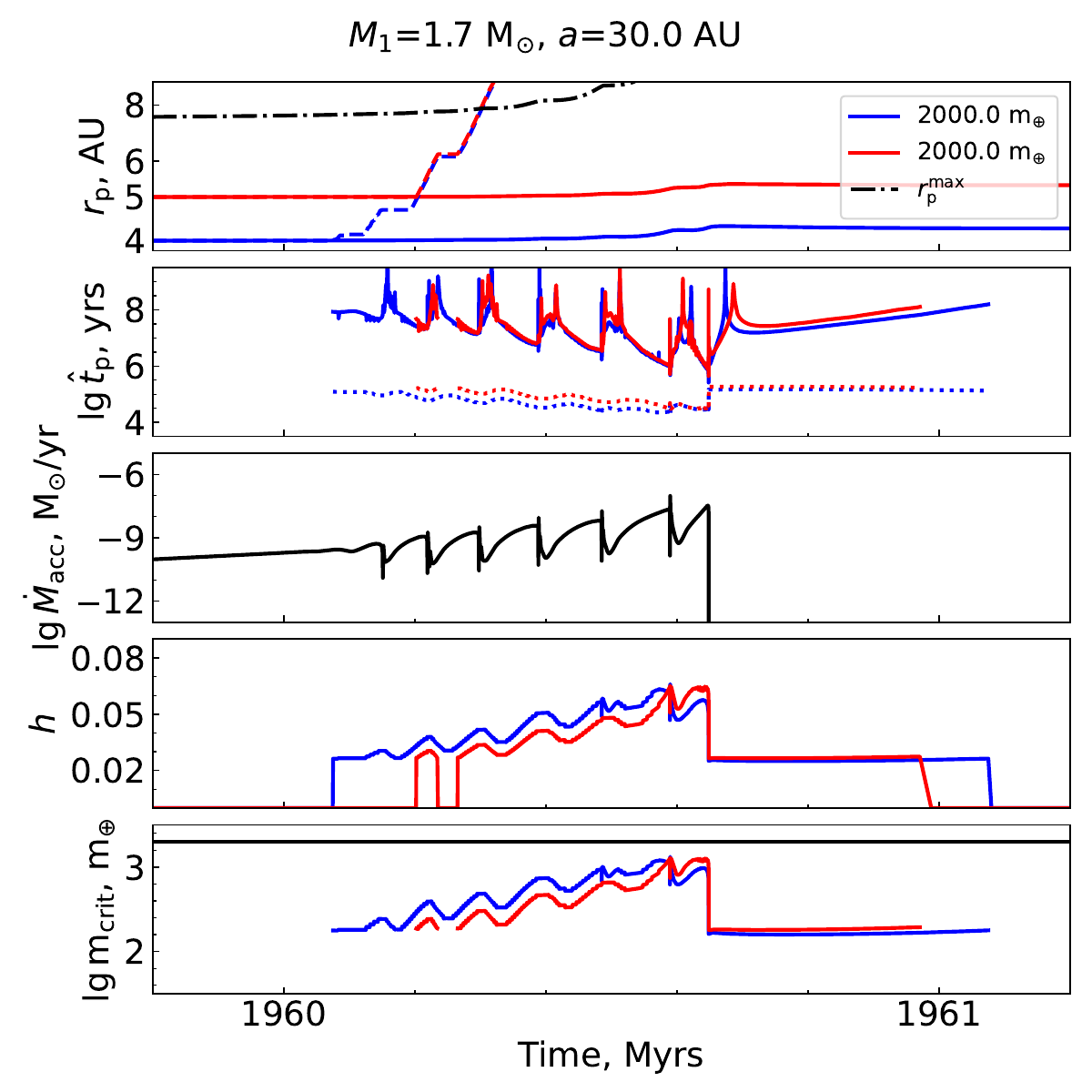}
    \captionof{figure}{The same as Fig.~\ref{fig:migration_general_1} but the left plot is for the planet with $m_{\mathrm{p}} = 100$~m$_{\oplus}$ and  initial orbits $r_{\mathrm{p}} = 2$~AU and $3$~AU in the system with initial $M_1 = 7.0$~M$_{\sun}$ and $a = 30$~AU. The center plot is for the planet with $m_{\mathrm{p}} = 2000$~m$_{\oplus}$ and initial orbits $r_{\mathrm{p}} = 3$~AU and $4$~AU in the system with initial $M_1 = 5.0$~M$_{\sun}$ and $a = 30$~AU. Additionally, the dashed curve shows the migration track determined by the rate of decretion. The dotted and the dot-dashed curves show, respectively, the timescale of decretion and the location of the largest dynamically stable planetary orbit. The right plot is as the center plot but for $M_1 = 1.7$~M$_{\sun}$ and for planets with initial orbits $r_{\mathrm{p}} = 4$~AU and $5$~AU.}
    \label{fig:migration_flipflop_and_outward}
\end{minipage}

\vspace{1\baselineskip}

\noindent\begin{minipage}{\textwidth}
    \centering
    \includegraphics[width=0.33\textwidth]{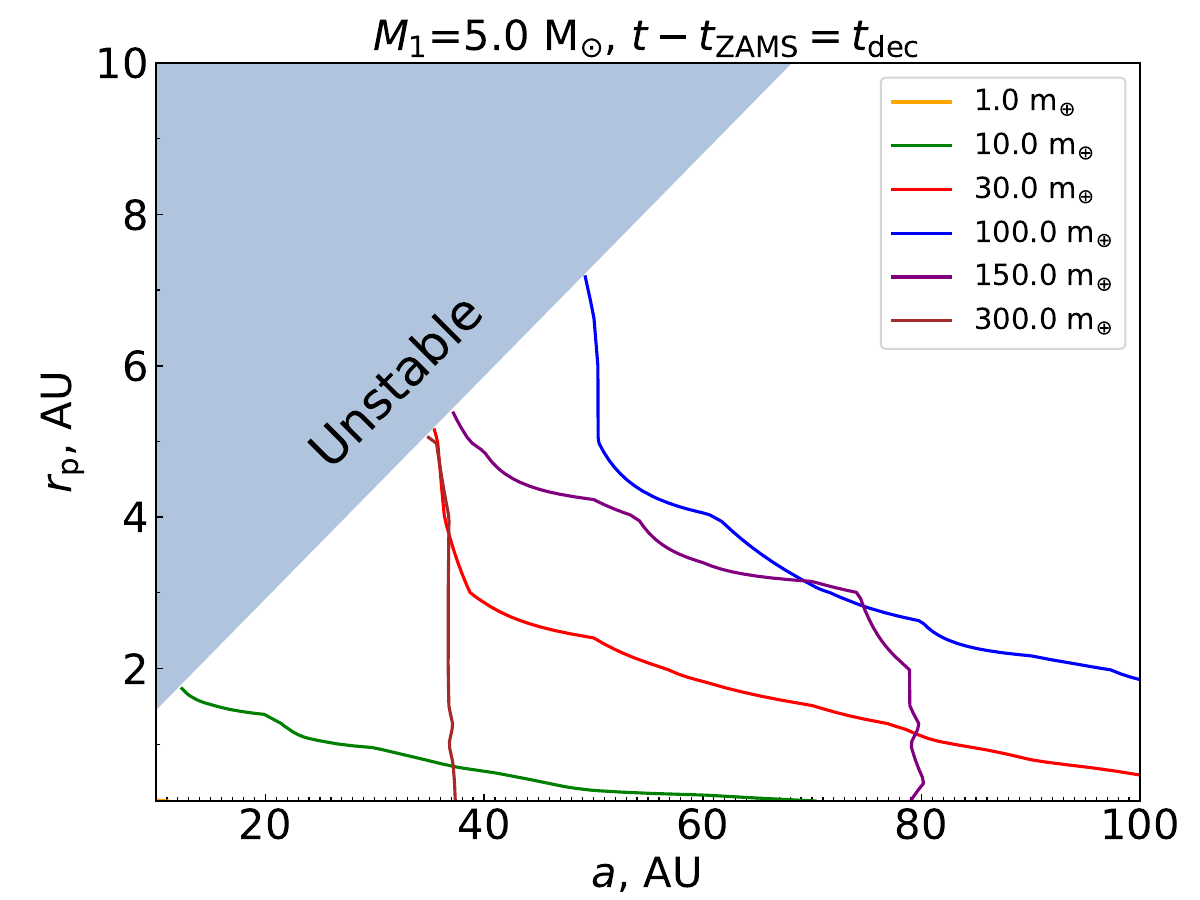}
    \includegraphics[width=0.33\textwidth]{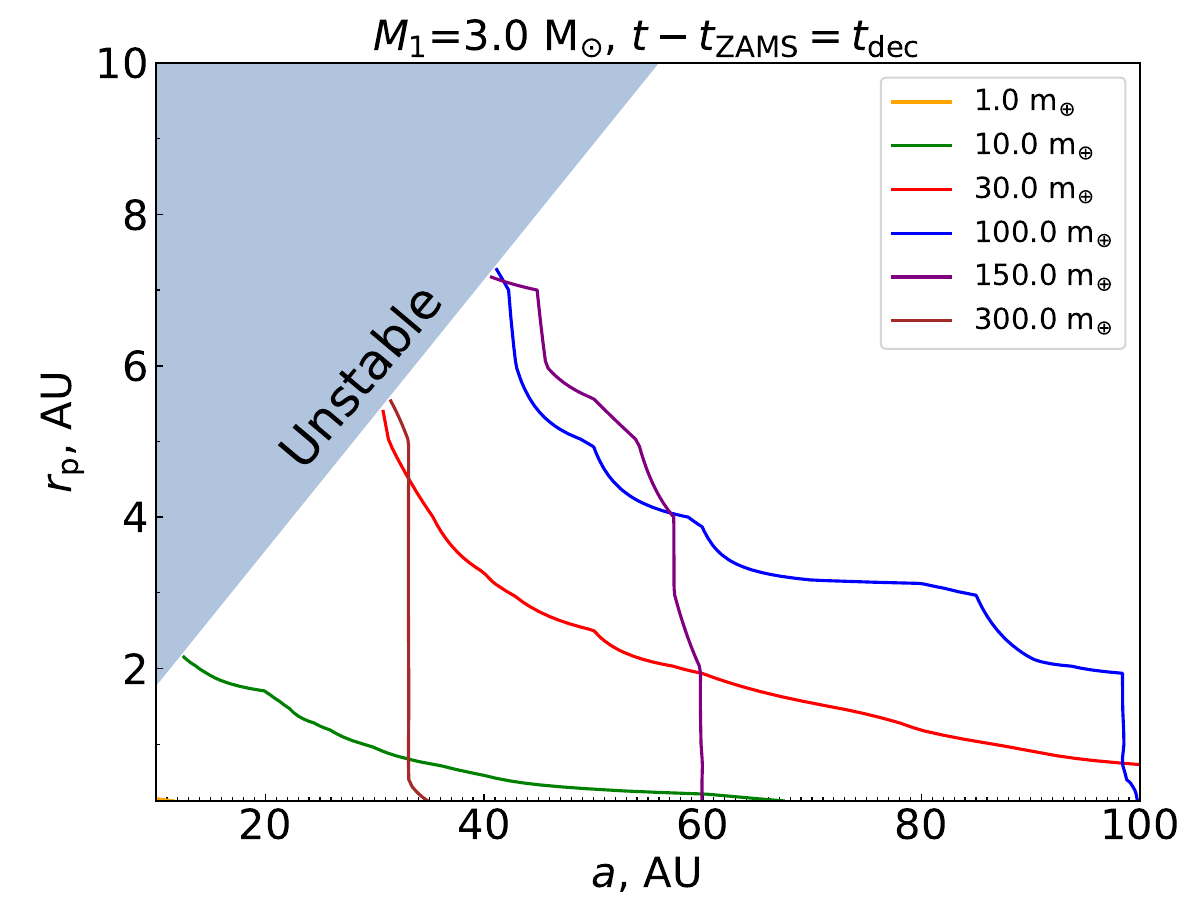}
    \includegraphics[width=0.33\textwidth]{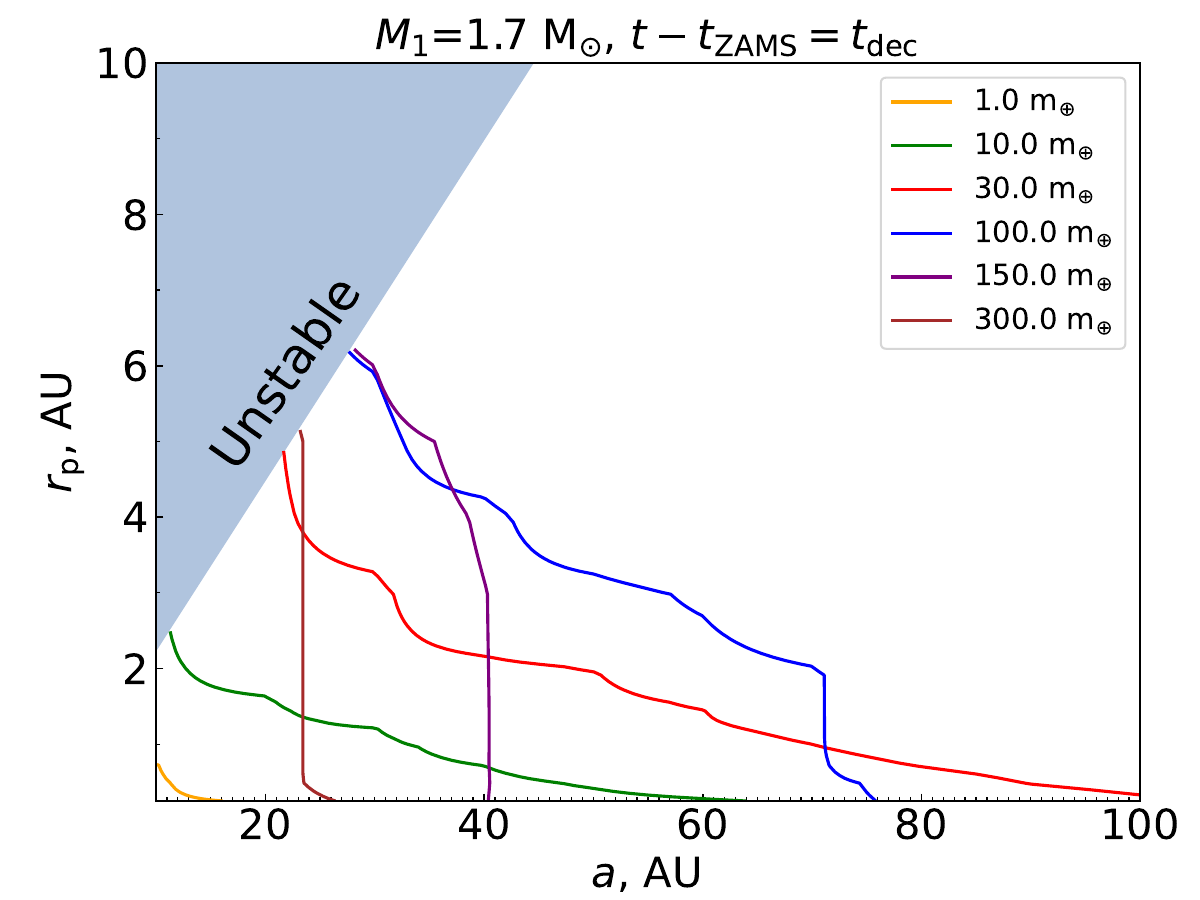}
    \captionof{figure}{The same as Fig.~\ref{fig:contour_different_planet_mass_4} but for $M_1 = 5.0$~M$_{\sun}$ (left plot), $M_1 = 3.0$~M$_{\sun}$ (center plot), $M_1 = 1.7$~M$_{\sun}$ (right plot). The yellow line for $m_{\mathrm{p}} = 1.0$\,m$_{\oplus}$ is invisible (except for the right plot) because it is below the plotted region.}
    \label{fig:contours_different_donor_masses}
\end{minipage}

\twocolumn

\section{Derivation of energy balance equation}
\label{app:temperature}

Following~\citet{nakamoto1994}, 
the local energy balance for a geometrically thin accretion disk reads:
\begin{equation}
\label{eq:app_energy_balance}
\dot E_{\mathrm{v}} = 
2 F_\mathrm{z}, 
\end{equation}
where $\dot E_{\mathrm{v}}$ is the viscous energy dissipation rate per unit surface area, 
while $F_\mathrm{z}$ is net energy flux from 
either side of the disk. 

The viscous energy dissipation rate is, explicitly:
\begin{equation}
\label{eq:app_viscous_rate}
\dot E_{\mathrm{v}} = \nu \Sigma \left( r \frac{\partial \Omega}{\partial r}\right)^{2} = \frac94 \nu \Sigma \Omega_{\mathrm{K}}^2.
\end{equation}
The net energy flux $F_\mathrm{z}$ is the sum of components representing various physical 
processes:
\begin{equation}
\label{eq:app_fluxes}
F_\mathrm{z} 
= F_{\mathrm{d}} - F_{{\mathrm{I}}, {\mathrm{A}}} - F_{{\mathrm{I}}, {\mathrm{D}}} - F_{\mathrm{w}},
\end{equation}
where $F_{\mathrm{d}}$ is the radiation flux from the disk surface, $F_{{\mathrm{I}}, {\mathrm{A}}}$ is the disk surface irradiation flux from the host star, $F_{{\mathrm{I}}, {\mathrm{D}}}$ is the disk surface irradiation flux from the primary, and $F_{\mathrm{w}}$ is the heating flux due to the kinetic energy dissipation of the settling wind material.

Below we consider each component of Eq.~\eqref{eq:app_fluxes}
in detail.


\subsection{Radiation flux from the disk surface} 


\subsubsection{Optically thick regime, $\tau_\mathrm{z, R} > 1$}

Introducing the effective temperature of the disk surface, $T_{\mathrm{d}}$, we have: 
\begin{equation}
\label{eq:app_disk_flux_thick}
F_{\mathrm{d}} = \sigma_{\mathrm{B}} T_{\mathrm{d}}^4,
\end{equation}
where $\sigma_{\mathrm{B}}$ is the Stefan-Boltzmann constant. 
The surface temperature of the disk is generally not equal to the midplane disk temperature, $T_{\mathrm{c}}$. 
The diffusive approximation of radiation transfer yields the corresponding relation between $T_{\mathrm{d}}$ and $T_{\mathrm{c}}$, see, e.g.,~\citet{ruden1986}:
\begin{equation}
\label{eq:app_Td_Tc_thick}
\sigma_{\mathrm{B}} T_{\mathrm{d}}^4 = \sigma_{\mathrm{B}} T_{\mathrm{c}}^4 - \frac{3}{8} \tau_{\mathrm{z}, \mathrm{R}} \dot E_{\mathrm{v}},
\end{equation}
where $\tau_{\mathrm{z }, \mathrm{R}} = \kappa_{\mathrm{R}} \Sigma / 2$ is the Planck mean optical thickness in $z$-direction.

\subsubsection{Optically thin regime, $\tau_\mathrm{z, R} < 1$}

Following~\citet{nakamoto1994}, we obtain (see Appendix~\ref{app:thin_temperature} for the details):
\begin{equation}
\label{Fd_thin}
F_{\mathrm{d}} = 4 \tau_{\mathrm{z }, \mathrm{P}} \sigma_{\mathrm{B}} T_{\mathrm{d}}^4, 
\end{equation}
where $\tau_{\mathrm{z }, \mathrm{P}} = \kappa_{\mathrm{P}} \Sigma / 2$ is the Planck mean optical thickness in $z$-direction
estimated by the Plank mean opacity $\kappa_{\mathrm{P}}$ 
taken at the disk midplane.
In this case, 
we assume the disk to be isothermal in vertical direction:
\begin{equation}
\label{eq:app_Td_Tc_thin}
T_{\mathrm{d}} = T_{\mathrm{c}}.
\end{equation} 

\subsection{Irradiation flux from the host star 
}


In this case, the disk is heated by the radiation absorbed at the surface layers of the disk. The corresponding energy flux can
be obtained similarly to~\citet{chiang1997}. Assuming that the disk is conical we obtain:
\begin{equation}
\label{eq:app_accretor_flux_thick}
F_{{\mathrm{I}}, {\mathrm{A}}} = \frac{2}{3 \pi} \left( \frac{R_2}{r}\right)^{3} \sigma_{\mathrm{B}} T_2^4, 
\end{equation}
where $R_2$ is the host star radius and $T_2$ is its effective temperature.

\subsection{Irradiation flux from the primary}


Under certain conditions, the disk surface can be additionally irradiated by a donor. It is possible to obtain analytically the corresponding irradiation flux for a conical disk, see App.~\ref{app:donor} for detailed calculations. Here we give the final result:
\begin{multline}
\label{app_flux_donor_thick}
F_{{\mathrm{I}}, {\mathrm{D}}} = {I_1 \cos\aspectratio} \frac{\sqrt{R_1^2 - \verticalscale^2}}{a} \times \\
\times \left[ \frac{\left(R_1 - \verticalscale\right)^2}{a^2} - \frac{\left(\verticalscale_{\tau} - \verticalscale\right)^2}{r_{\tau}^2} \left( 1 + \frac{r^2}{2 r_{\tau}^2} + \frac{r}{a} \frac{r}{r_{\tau}}\right) + \right.\\
\left. + \tan\aspectratio \left( \frac{r}{r_{\tau}} \frac{\verticalscale_{\tau} - \verticalscale}{r_{\tau}} - \frac{r}{a} \frac{R_1 - \verticalscale}{a} \right) \right].
\end{multline}
where $R_1$ is the radius of the donor, $T_1$ is its effective temperature, and $r_{\tau}$ is the distance from the host star where the disk scale height attains a maximum, $\verticalscale_{\tau}$. In all cases $r_{\tau} = r_{\mathrm{max}}$ due to conical geometry.

\subsection{Heating flux due to the kinetic energy dissipation of the settling wind material}

This contribution to the energy balance is taken following~\citet{perets2013}:
\begin{equation}
\label{eq:app_kinetic_flux}
F_{\mathrm{w}} = \frac{G M_2 \dot M_{\mathrm{acc}}}{2 \pi r^2 r_{\mathrm{a}}} = \frac{G M_2 \dot M_{\mathrm{w}} r_{\mathrm{a}}}{8 \pi a^2 r^2} \theta (r_{\mathrm{a}} - r), \quad \forall \tau_{\mathrm{z}, \mathrm{R}},
\end{equation}
where all the values are defined in Sect.~\ref{s:binary}. 

\subsection{Final equation}

Putting Eq.\eqref{eq:app_viscous_rate} along with Eqs.~\eqref{eq:app_disk_flux_thick}\,--\,\eqref{eq:app_kinetic_flux} 
together into Eq.~\eqref{eq:app_energy_balance}
we obtain energy balance equations for the two cases:
\begin{multline}
\label{energyEq_fin_thick}
T_{\mathrm{c}}^4 = \frac{ 9 \nu \Sigma \Omega_{\mathrm{K}}^2}{8 \sigma_{\mathrm{B}}} \left( 1 + \frac{3 \tau_{\mathrm{z }, \mathrm{R}}}{4}\right) + \frac{F_{{\mathrm{I}}, {\mathrm{A}}}}{\sigma_{\mathrm{B}}} + \frac{F_{{\mathrm{I}}, {\mathrm{D}}}}{\sigma_{\mathrm{B}}} + \frac{G M_2 \dot M_{\mathrm{w}} r_{\mathrm{a}}}{8 \sigma_{\mathrm{B}} \pi a^2 r^2} \theta (r_{\mathrm{a}} - r), 
\end{multline}
for the optically thick case, and:
\begin{multline}
\label{energyEq_fin_thin}
T_{\mathrm{c}}^4 = \frac{ 9 \nu \Sigma \Omega_{\mathrm{K}}^2}{32 \sigma_{\mathrm{B}} \tau_{\mathrm{z }, \mathrm{P}}} + \frac{F_{{\mathrm{I}}, {\mathrm{A}}}}{4 \sigma_{\mathrm{B}} \tau_{\mathrm{z }, \mathrm{P}}} + \frac{F_{{\mathrm{I}}, {\mathrm{D}}}}{4 \sigma_{\mathrm{B}} \tau_{\mathrm{z }, \mathrm{P}}} + \frac{G M_2 \dot M_{\mathrm{w}} r_{\mathrm{a}}}{32 \sigma_{\mathrm{B}} \tau_{\mathrm{z}, \mathrm{P}} \pi a^2 r^2} \theta (r_{\mathrm{a}} - r),
\end{multline}
for the optically thin case.

Combining Eqs.~\eqref{energyEq_fin_thick}\,--\,\eqref{energyEq_fin_thin} we obtain the energy balance equation which is applicable for any value of the optical thickness, $\tau_{z, \mathrm{R}}$:
\begin{equation}
\label{energyEq_fin}
T_{\mathrm{c}}^4 = \frac{ 9 \nu \Sigma \Omega_{\mathrm{K}}^2}{8 \sigma_{\mathrm{B}}} \left( 1 + \frac{3 \tau_{\mathrm{z }, \mathrm{R}}}{4} + \frac{1}{4 \tau_{\mathrm{z }, \mathrm{P}}} \right) + \left( T^4_{{\mathrm{I}}, {\mathrm{A}}} + T^4_{{\mathrm{I}}, {\mathrm{D}}} + T^4_{\mathrm{w}} \right) \left( 1 + \frac{1}{4 \tau_{\mathrm{z}, \mathrm{P}}} \right),
\end{equation}
where the terms $T^4_{{\mathrm{I}}, {\mathrm{A}}}$, $T^4_{{\mathrm{I}}, {\mathrm{D}}}$, $T^4_{\mathrm{w}}$  are related to Eqs.~\eqref{eq:app_accretor_flux_thick}\,--\,\eqref{eq:app_kinetic_flux} according to the Stefan-Boltzmann law. 

\section{Optically thin disk irradiation by the host star}
\label{app:thin_temperature}

Following~\citet{nakamoto1994} we assume that an optically thin disk is locally represented by a plane-parallel layer of finite thickness exposed to the ambient radiation. It is useful to define the optical thickness of a disk per unit frequency as:
\begin{equation}
\tau_{\nu} (z) = \int\limits_{-\verticalscale}^z \kappa_{\nu} \rho (z) \mathrm{d } z,
\end{equation}
where $\kappa_{\nu}$ is opacity at frequency $\nu$.

The radiation transfer equation for the intensity of radiation  with frequency $\nu$ yields the solution:
\begin{equation}
\label{eq:transfer_equation}
I_{\nu}(\tau_{\nu} (z), \alpha) = I_{\nu}(0, \theta) e^{-\tau_{\nu} (z) /\cos\alpha} + \left( 1 - e^{-\tau_{\nu} (z) /\cos\alpha} \right)B_{\nu} (T_{\mathrm{c}}),
\end{equation}
where $B_{\nu} (T_{\mathrm{c}})$ is the Planck function taken for the midplane temperature of disk, $T_{\mathrm{c}}$, and $\alpha$ is the angle between 
the direction of propagating radiation and the normal to the disk surface. Note that the disk is assumed to be isothermal here. Since radiation from the host star is assumed to be a blackbody, the total frequency-integrated intensity at $z = -\verticalscale$, where $\tau_{\nu} (z) \equiv 0$ for the incoming radiation, along the line of sight, intersecting the star photosphere simplifies and becomes:
\begin{equation}
I (0, \alpha) = \int\limits_0^{\infty} B_{\nu} (T_2) d \nu = \frac{\sigma_{\mathrm{B}}}{\pi} T_2^4,
\end{equation}
where $T_2$ is the effective temperature of the host star.

The first term of Eq.~\eqref{eq:transfer_equation} becomes negligible during the following propagation of radiation through the disk, since in this case $\tau_{\nu} (z) / {\cos\alpha} \gg 1$ because the source of radiation is located at $\alpha \approx \pi/2$. On the contrary, for the second term the extraction by $\tau_{\nu} (z) / {\cos\alpha} \ll 1$ can be used since the disk itself irradiates mostly at $\alpha \approx 0$. Hence, Eq.~\eqref{eq:transfer_equation} becomes:
\begin{equation}
I_{\nu}(\tau_{\nu} (z), \alpha) \approx \frac{\tau_{\nu}(z)}{\cos\alpha} B_{\nu} (T_{\mathrm{c}}).
\end{equation}
The total frequency-integrated intensity at $z = +\verticalscale$ reads:
\begin{equation}
\label{A_I}
I(\verticalscale, \alpha) = \int\limits_0^{\infty} I_{\nu}(\tau_{\nu}, \alpha) \mathrm{d }\nu
= \int\limits_0^{\infty}\frac{\tau_{\nu}}{\cos\alpha} B_{\nu} (T_{\mathrm{c}}) \mathrm{d }\nu = \frac{\tau_{z, \mathrm{P}}}{\cos\alpha} \frac{\sigma_{\mathrm{B}}}{\pi} T_{\mathrm{c}}^4,
\end{equation}
where $\tau_{\mathrm{z }, \mathrm{P}}$ is the Planck mean opacity given by:
\begin{equation}
\tau_{\mathrm{z }, \mathrm{P}} = \int\limits_0^{\infty} \tau_{\nu} B_{\nu} \mathrm{d }\nu \Biggm/ \int\limits_0^{\infty} B_{\nu} \mathrm{d }\nu.
\end{equation}

Generally, the radiation flux is related to the intensity in the following way:
\begin{equation}
\label{RubEq}
F = \int I \cos\alpha \mathrm{d} \Omega,
\end{equation}
where $\mathrm{d} \Omega = \sin\theta \mathrm{d}\theta \mathrm{d}\phi$ is an element of the solid angle 
($\theta$ and $\phi$ are polar and azimuthal angles of the
spherical coordinates, which we select in this case with $z$-direction to the center of the star of the linked Cartesian system), $\alpha$ is the angle between the direction of propagating radiation and the normal to the disk surface. In the considered case, the angle $\alpha$ is defined as follows: $\cos\alpha = \sin\theta \cos\phi$. Then, for the upward flux, we obtain:
\begin{multline}
\label{Fplus}
F^+(\verticalscale) = \int \mathrm{d} \phi \int I(\verticalscale, \alpha) \cos\alpha \sin\theta \mathrm{d}\theta = \\
= \left(\phi_2 - \phi_1 \right) \int\limits_{\theta_2}^{\theta_1} \mathrm{d} (\cos\theta) \frac{\tau_{\mathrm{z}, \mathrm{P}}}{\cos\alpha} \frac{\sigma_{\mathrm{B}} T_{\mathrm{c}}^4}{\pi} \cos\alpha = \\
= \frac{\phi_2 - \phi_1}{\pi} \tau_{\mathrm{z }, \mathrm{P}} \sigma_{\mathrm{B}} T_{\mathrm{c}}^4 \left(\cos\theta_1 - \cos\theta_2\right) = 2 \tau_{\mathrm{z }, \mathrm{P}} \sigma_{\mathrm{B}} T_{\mathrm{c}}^4,
\end{multline}
where $\phi_1 = -\pi/2$, $\phi_2 = \pi/2$, $\theta_1 = 0$ and $\theta_2 = \pi$ sets the range of angles when the disk radiates from the surface. Due to the symmetry of the disk, the downward intensity is: 
\begin{equation}
I^- (\verticalscale, \alpha) = - I^+ (0, \alpha) = - \frac{\sigma_{\mathrm{B}} T^4_2}{\pi},
\end{equation}
while the corresponding flux is:
\begin{multline}
\label{Fminus}
F^- (\verticalscale) = \int\limits_{\phi_1^{\prime}}^{\phi_2^{\prime}} \mathrm{d}\phi \int\limits_{\theta_1^{\prime}}^{\theta_2^{\prime}} \mathrm{d}\theta I^-(\verticalscale, \alpha) \sin^2\theta \sin\phi = \\
= - \frac{\sigma_{\mathrm{B}} T^4_2}{\pi} \left(\phi_2^{\prime} - \phi_1^{\prime}\right) \left. \frac{\theta - \sin\theta\cos\theta}{2}\right|_{\theta_1^{\prime}}^{\theta_2^{\prime}} \approx - \frac{2}{3\pi} \left(\frac{R_2}{r}\right)^3 \sigma_{\mathrm{B}} T_2^4,
\end{multline}
where the integration limits  $\phi_1^{\prime} = - \pi/2$, $\phi_2^{\prime} = \pi/2$, $\theta_1^{\prime} = 0$, $\theta_2^{\prime} \approx R_2/ r$ set the range of angles when the star is visible from the disk surface, as assumed that $R_2 \ll r$.

The net radiation flux from the upper surface of the disk is
\begin{multline}
\label{thin_flux}
F(\verticalscale) = F^+(\verticalscale) + F^-(\verticalscale) = \\
= 2 \tau_{\mathrm{z }, \mathrm{P}} \sigma_{\mathrm{B}} T_{\mathrm{c}}^4 - \frac{2}{3 \pi}\left(\frac{R_2}{r}\right)^{3} \sigma_{\mathrm{B}} T_2^4 = F_{\mathrm{d}} - F_{{\mathrm{I}}, {\mathrm{A}}},
\end{multline}
where with $F_{\mathrm{d}}$ we denote the flux of radiation escaping from the optically thin disk surface,
and by $F_{{\mathrm{I}}, {\mathrm{A}}}$~--~the irradiation flux from the central star. Note that $F_{\mathrm{d}}$ differs from $F_{\mathrm{d}}$ introduced in equation~\eqref{Fd_thin} by a factor of two. This formal difference is due to the definition of the optical thickness here corresponding to the thickness of disk rather than its semi-thickness. Due to the similarity of the evaluation, the equation, same to Eq.~\eqref{thin_flux} can be obtained for the donor irradiation $F_{{\mathrm{I}}, {\mathrm{D}}}$.

\section{Optically thick disk irradiation by the primary star}
\label{app:donor}

\subsection{Flat disk approximation}
\label{app:donor_flat}

At first, we assume that a disk is flat with $\verticalscale=const$ and negligible thickness,
$\verticalscale \ll R_1$. Let us consider the irradiation of a disk at the point with cylindrical coordinates ($r, \varphi, z=0$), where we introduce the spherical coordinate system $r^\prime, \theta, \phi$. 
The polar angle $\theta$ is measured with respect to the direction of the center of the donor. Further, we define the angle $\alpha$ between the normal to the disk surface and the direction toward the emitting point of the donor, similar to the angle in Eq.~\eqref{RubEq}. The flux from the donor is also determined by Eq.~\eqref{RubEq}. Let the angle $\phi$ be measured with respect to normal to the disk surface. Then $\alpha$ is explicitly
\begin{equation}
\label{cosalpha}
 \cos\alpha = \sin\theta \cos\phi.
\end{equation}
The limits of integration in Eq.~\eqref{RubEq} are specified now as follows (Note that only half of the stellar surface is visible from the disk surface):
\begin{equation}
\label{angles}
\phi \in \left(-\frac{\pi}{2}, +\frac{\pi}{2}\right), \quad \theta \in \left(0, \theta_{\mathrm{max}} \approx \frac{R_1}{R (\varphi)} \right),
\end{equation}
where $R(\varphi)$ is the distance from the given point of the disk to the center of the donor:
\begin{equation}
\label{Rphi}
R (\varphi) = \sqrt{a^2 + r^2 - 2 a r \cos\varphi},
\end{equation}
where $\varphi$ is an azimuthal angle in the disk plane and $\varphi = 0$ is selected as the direction to the center of the donor. Hereafter it is assumed that $R_1 \ll R(\varphi)$.

Let us calculate the flux at the specified point of the disk, 
taking into account that the intensity of radiation coming from the donor, $I_1$, is constant over the solid angle: 
\begin{multline}
\label{eq:C3}
F (\varphi) = \int\limits_{-\frac{\pi}{2}}^{\frac{\pi}{2}}\mathrm{d} \phi \int\limits_{0}^{\theta_{\mathrm{max}}} I_1 \cos\phi \sin\theta \sin\theta \mathrm{d} \theta = \\
= I_1 \left(\theta_{\mathrm{max}} - \sin\theta_{\mathrm{max}} \cos\theta_{\mathrm{max}}\right) \approx \\
\approx I_1 \left( \frac{R_1}{R} - \left(\frac{R_1}{R} - \frac{R_1^3}{3 R^3} \right) \sqrt{1 - \frac{R_1^2}{R^2}}\right) \approx \\
\approx I_1 \left(\frac{R_1}{R} + \frac{R_1^3}{3 R^3} - \frac{R_1}{R} + \frac{1}{2} \frac{R_1^3}{R^3}\right) = \frac{2 I_1}{3} \frac{R_1^3}{R^3 (\varphi)}.
\end{multline}

The intensity is determined in a usual way through the effective temperature of the donor:
\begin{equation}
I_1 = \frac{\sigma_{\mathrm{B}}}{\pi} T_1^4.
\end{equation}

Since 
the energy balance of a standard accretion disk is being established on the timescale exceeding the Keplerian time, we should average $F(\varphi)$ over the azimuthal angle: 
\begin{multline} 
F = \frac{1}{2 \pi} \int\limits_{0}^{2\pi} F (\varphi) \mathrm{d}\varphi = \frac{2 I_1 R_1^3}{3 \pi} \int\limits_{0}^{\pi} \frac{\mathrm{d}\varphi}{\left( a^2 + r^2 - 2 a r \cos\varphi\right)^{\frac32}} = \\ = \frac{2 I_1 R_1^3}{3 \pi (a - r)(a + r)^2}  E\left( \frac{\pi}{2} \right| \left. - \frac{4 a r}{(a - r)^2} \right), 
\end{multline}
where $E \left(\varphi \right| \left. m \right)$ is a complete elliptic integral of the second kind.  Since the ratio $r/a$ is small, the expansion of the elliptic integral over the parameter $m \ll 1$ is valid:
\begin{multline*}
E\left( \pi / 2  \right| \left. m \right) = \frac{\pi}{2} \left(1 - \frac{m}{4} - \frac{3 m^2}{64} - \dots \right) = \\
= \frac{\pi}{2} \left(1 + \frac{a r}{(a - r)^2} - \frac{3 a^2 r^2}{4 (a - r)^4} - \dots \right).
\end{multline*}
Using this expansion we obtain an averaged flux of irradiation from the donor in the case of a razor-thin disk:
\begin{multline}
F \approx \frac{2 I_1 R_1^3}{3 \pi (a - r)(a + r)^2} \frac{\pi}{2} \left(1 + \frac{a r}{(a - r)^2}\right) = \\ = \frac{I_1 R_1^3}{3 a^3} \frac{1 + r/a + \left( r/a\right)^2}{\left( 1 - r/a\right)^3 \left( 1 + r/ a\right)^2} \approx \frac{I_1 R_1^3}{3 a^3} \left(1 + 2 \frac{r}{a} + 5 \left( \frac{r}{a}\right)^{2}\right).
\end{multline}

\subsection{Conical disk approximation}
\label{app:donor_conical}

We assume that the outer boundary of the disk is defined by the radial optical thickness equal to unity, $r_{\tau} = r (\tau_{\mathrm{r }, \mathrm{R}} = 1)$. At $r < r_{\tau}$ the disk becomes optically thick in $r$-direction but remains optically thin in $z$-direction, which allows to use the energy equation in the optically thin case derived in App.~\ref{app:thin_temperature}. 

Now we consider a more general case of a conical disk with constant aspect ratio $\aspectratio = \verticalscale / r = const \ll 1$. We also take into account shading by the outer parts of the disk which are closer to the donor. 

\begin{figure}
    \centering
    \includegraphics[width=\columnwidth]{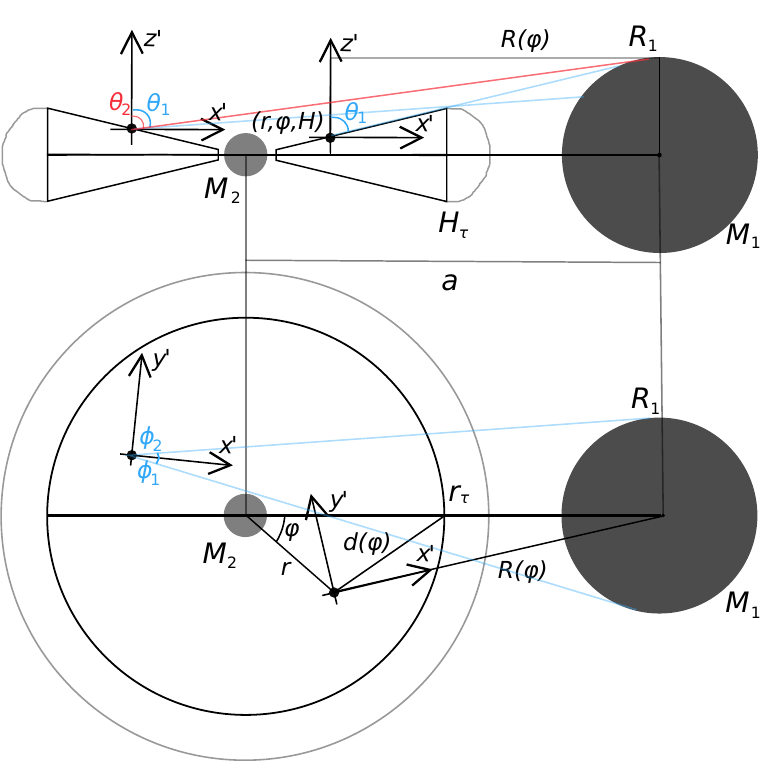}
    \caption{Not to scale: basic geometry for calculations in a conical disk approximation. 
    Edge-on (top) and face-on (bottom) views.
    Two arbitrary points at the conical disk surface are marked. Two corresponding coordinate systems $(x, y, z)$ for each point are shown. They define the chosen spherical coordinate systems $(r^\prime, \phi, \theta)$ centered on each of these points. The distance $R (\varphi)$ is defined by Eq.~\eqref{Rphi}. 
    The maximum conical disk height, $\verticalscale_{\tau}$, corresponds to the shading radius $r_{\tau}$ at $\tau_{\mathrm{r}, \mathrm{R}} = 1$. 
    $M_1$ and $M_2$ are the donor and the accretor masses. 
    The binary separation is $a$. }
    \label{fig:donor_appendix}
\end{figure}

Hereafter, let us denote the coordinate system O centered on secondary: ($r, \varphi, z$). Similarly to Appendix~\ref{app:donor_flat}, 
let us consider an arbitrary point of a disk photosphere ($r, \varphi, \verticalscale$) which is above its midplane, $|z|>0$. Since the disk is conical, the normal to the disk surface taken at this point differs from the vertical direction by an angle $\aspectratio$.

In order to obtain the irradiation flux, we introduce the new coordinate system O$^\prime$: ($x^\prime, y^\prime, z^\prime$), centered on the selected arbitrary point of the disk ($r, \varphi, \verticalscale$).~$x^\prime$-axis of the system is directed to the vertical of the donor and intersects it at point with $z$-coordinate $z = \verticalscale$,~$z^\prime$-axis is coplanar with $z$-axis and $y^\prime$ is selected so to define the right triplet of vectors, see Fig.~\ref{fig:donor_appendix}. We also define the corresponding spherical system ($r^\prime, \phi, \theta$), associated with O$^\prime$.

The irradiation flux is obtained analytically using Eq.~\eqref{RubEq}. With this choice of the spherical coordinate system O$^\prime$, we obtain angles of integration ($\theta, \phi$) to be calculated as follows:
\begin{multline}
\label{eq:integration_angles}
\phi_2 (\varphi) \approx \frac{\sqrt{R_1^2 - \verticalscale^2}}{R (\varphi)}, \quad \phi_1 (\varphi) \approx -\frac{\sqrt{R_1^2 - \verticalscale^2}}{R (\varphi)}, \\
\theta_2 (\varphi) \approx \frac{\pi}{2} - \frac{R_1 - \verticalscale}{R (\varphi)}, \quad \theta_1 (\varphi) \approx \frac{\pi}{2} - \frac{\verticalscale_{\tau} - \verticalscale}{d (\varphi)},
\end{multline}
where $R (\varphi)$ is defined as in Eq.~\eqref{Rphi} and $d (\varphi) = \sqrt{r_{\tau}^2 + r^2 - 2 r r_{\tau} \cos \varphi}$ is a distance between given disk point ($r, \varphi$) and closest to the donor disk point ($r_{\tau}, \varphi=0$).

Note that all terms in Eq.~\eqref{eq:integration_angles} are small, but can be comparable to each other: $R_1/R \ll 1$, $\verticalscale/R \ll 1$, $\verticalscale_{\tau}/r_{\tau} \ll 1$, etc.

The angle $\alpha$ for Eq.~\eqref{RubEq} is defined in O$^\prime$ as follows:
\begin{equation}
\label{eq:alpha_conical}
\cos\alpha = \cos\aspectratio\cos\theta - \sin\aspectratio \sin\theta \cos (\varphi + \psi - \phi),
\end{equation}
where $\psi$ is rotation angle of the system O$^\prime$ with respect to the system O, $\sin\psi = r\sin\varphi / R(\varphi)$.

With given expressions, the irradiation flux is calculated as:
\begin{multline}
\label{eq:flux_phi}
\frac{F(\varphi)}{I_1 \cos\aspectratio} = \int\limits_{\phi_1}^{\phi_2}\mathrm{d} \phi \int\limits_{\theta_1}^{\theta_2} \left(\cos\theta - \tan\aspectratio \sin\theta \cos \left(\varphi + \psi - \phi\right)\right) \sin\theta  \mathrm{d}\theta = \\
= \left(\phi_2 - \phi_1\right) \frac{\cos^2\theta_2 - \cos^2\theta_1}{2} - \tan\aspectratio  \cos\left(\varphi + \psi\right) \times \\
\times \sin\frac{\phi_2 - \phi_1}{2} \left( \theta_2 - \theta_1 \right) \left( 1 - \cos\left(\theta_1 + \theta_2\right) \right) \approx \\
\approx \frac{\phi_2 - \phi_1}{2} \left[ \cos^2\theta_2 - \cos^2\theta_1 - \tan\aspectratio \cos\left(\varphi + \frac{r \sin\varphi}{R(\varphi)}\right) \times \right.\\
\biggl. \times \left( \theta_2 - \theta_1 \right) \left( 1 + \cos\left(\pi - \theta_1 - \theta_2\right) \right) \biggr],
\end{multline}
hereinafter we use small values $r/R$ and $\pi/2 - \theta_{1, 2}$ in order to simplify the terms in Eq.~\eqref{eq:flux_phi}:
\begin{equation*}
\cos(\pi - \theta_1 - \theta_2) \approx 1, \quad \cos\left(\varphi + \frac{r \sin\varphi}{R(\varphi)}\right) \approx \cos\varphi - \frac{r}{R} \sin^2 \varphi.
\end{equation*}

After that, the flux is:
\begin{multline}
\label{eq:flux_phi_simplified}
\frac{F(\varphi)}{I_1 \cos\aspectratio} \approx \frac{\phi_2 - \phi_1}{2} \left[ \left(\pi/2 - \theta_2\right)^2 - \left(\pi/2 - \theta_1\right)^2 - \right.\\
\Bigr. - 2 \tan\aspectratio \left( \cos\varphi - \frac{r}{R} \sin^2 \varphi \right) \left( \theta_2 - \theta_1 \right) \Bigr] \approx \\
\approx \frac{\sqrt{R_1^2 - \verticalscale^2}}{R (\varphi)} \left[ \frac{(R_1 - \verticalscale)^2}{R^2 (\varphi)} - \frac{(\verticalscale_{\tau} - \verticalscale)^2}{d^2 (\varphi)} - \right.\\
\left. - 2 \tan\aspectratio \left( \cos\varphi - \frac{r}{R} \sin^2 \varphi \right) \left( \frac{R_1 - \verticalscale}{R (\varphi)} - \frac{\verticalscale_{\tau} - \verticalscale}{d (\varphi)} \right) \right].
\end{multline}

Then we should calculate the averaged flux, but since Eq.~\eqref{eq:flux_phi_simplified} contains radicals and cannot be integrated with elementary functions, we integrate this equation numerically by $\varphi$ in the limits $(0, 2\pi)$. We note that Eq.~\eqref{eq:flux_phi_simplified} is obtained under the assumption that the donor size is larger than the maximum disk vertical scale height $R_1 > \verticalscale_{\tau}$. Otherwise, the radiation from the donor is completely absorbed in the outer edge of the disk. Therefore, in the case of $R_1 \leq \verticalscale_{\tau}$ one should use the zero value of the flux in calculations, $F = 0$.

The azimuthal averaging is physically justified only when the dynamical timescale of the disk at a given radius is smaller than the heating/cooling timescale. Otherwise, the disk may acquire a non-axisymmetric structure due to the donor heating. We additionally check that the heating and cooling timescales are larger than the dynamical timescale in the optically thick region of the disk, which has a size within $\sim 10$ AU in our calculations.

\section{Numerical scheme}
\label{sec_app:numerical_method}

\subsection{Calculation of the accretion disk evolution}
\label{sec_app:accretion_numerical}

The set of Eqs.~\eqref{eq:torque},~\eqref{eq:diffusion_equation},~and~\eqref{eq:energy_equation} is advanced numerically with an implicit and unconditionally stable numerical scheme using finite difference approximations of derivatives at the nodes of the introduced numerical grid ($t_k, \angularmomentum_{k, n}$) including the imposed boundary and initial conditions,
\begin{multline}
\label{eq:diffusion_equation_numerical}
\frac{\Sigma_{k + 1, n} - \Sigma_{k, n}}{t_{k + 1} - t_k} = \frac{(G M_2)^2}{4 \pi \angularmomentum_{k, n}^3} \frac{F_{k + 1, n + 1} - 2 F_{k + 1, n} + F_{k + 1, n - 1}}{\Delta_k^2} + \\
+ \left(\frac{\partial \Sigma}{\partial t}\right)_{\mathrm{ext}, k + 1, n},
\end{multline}
where $\Delta_k = \angularmomentum_{k, n} - \angularmomentum_{k, n-1}$ is a time-dependent step of the spatial grid, defined to be uniform in \angularmomentum terms, 
\begin{multline}
\label{eq:energy_equation_numerical}
0 = \left(T_{\mathrm{c}}\right)_{k, n}^4 - \frac{ 9 \nu_{k, n}\left(T_{\mathrm{c}}\right) \Sigma_{k, n} \left( G M_2\right)^4}{8 \sigma_{\mathrm{B}} \angularmomentum_{k, n}^6} \times \\
\times \left( 1 + \frac{3 \kappa_{\mathrm{R}, k, n}\left(T_{\mathrm{c}}\right) \Sigma_{k, n}}{8} + \frac{1}{2 \kappa_{\mathrm{P}, k, n}\left(T_{\mathrm{c}}\right) \Sigma_{k, n}} \right) - \\ - T^4_{\mathrm{I}, \mathrm{A}, k, n} - T^4_{\mathrm{I},\mathrm{D}, k, n} - T^4_{\mathrm{w}, k, n} \left(1 + \frac{1}{2 \kappa_{\mathrm{R}, k, n}\left(T_{\mathrm{c}}\right) \Sigma_{k, n}} \right),
\end{multline}
\begin{equation}
\label{eq:torque_numerical}
0 = F_{k, n} - 3 \pi \angularmomentum_{k, n} \nu_{k, n}\left(T_{\mathrm{c}}\right) \Sigma_{k, n}.
\end{equation}

The set of Eqs.~\eqref{eq:diffusion_equation_numerical}\,--\,\eqref{eq:torque_numerical} allows us to determine the evolution of NSD. 
In the case of QSD, Eq.~\eqref{eq:diffusion_equation_numerical} is replaced by a numerical analog of~Eq.~\eqref{eq:nu_sigma}:
\begin{equation}
\label{eq:nu_sigma_numerical}
F_{k, n} = 3 \pi \angularmomentum_{k, n} \nu_{k, n} \Sigma_{k, n} = \dot M_{\mathrm{acc}} (t_k) \angularmomentum_{k, n} f ( \angularmomentum_{k, n}, \angularmomentum_{\mathrm{a}, k, n} (t_k)),
\end{equation}
and the set of Eqs.~\eqref{eq:energy_equation_numerical}\,--\,\eqref{eq:nu_sigma_numerical} allows us to model the evolution of QSD.

For both QSD and NSD, Eq.~\eqref{eq:energy_equation_numerical} is 
solved along with Eqs.~\eqref{eq:diffusion_equation_numerical},~\eqref{eq:torque_numerical} or~\eqref{eq:torque_numerical},~\eqref{eq:nu_sigma_numerical} using the Brent algorithm~\citep{brent1973} for solving the algebraic equations.

The optical thickness is also calculated numerically. Equations~\eqref{eq:optical_thickness_vertical} and~\eqref{eq:optical_thickness_radial} are approximated as follows:
\begin{multline}
\label{eq:thickness_numerical}
\tau_{\mathrm{z}, \mathrm{R}, k, n} = \frac{\kappa_{\mathrm{R}} (r_{k, n}, t_k) \Sigma_{k, n}}{2}, \\
\tau_{\mathrm{r}, \mathrm{R}, k, n} = \sum\limits_{i = n}^{N} \tau_{\mathrm{z}, \mathrm{R}, k, i} \frac{r_{k, i} - r_{k, i-1}}{\verticalscale_{k, i}}.
\end{multline}

\subsection{Calculation of the migration}
\label{sec_app:migration}

We calculate planetary migration using the numerical approximations of  Eqs.~\eqref{eq:migration_type1},~\eqref{eq:migration_type2} and checking the transition between the types of migration with the help of Eq.~\eqref{eq:critical_mass_planet}. For equations describing the migration, we use precalculated values of $\left(\Sigma\right)_{k, n}, \left(T_{\mathrm{c}}\right)_{k, n}, \left(F\right)_{k, n}$. 
The difference equations for migration are the following:
\begin{multline}
\label{eq:migration_numerical}
\frac{r_{\mathrm{p}, k + 1} - r_{\mathrm{p}, k}}{t_{k + 1} - t_k} = f \left( r_{\mathrm{p}, k}, \left\{\Sigma, T_{\mathrm{c}}, F\right\} \left( t_k, r_{\mathrm{p}} (t_k) \right), \dots \right),
\end{multline}
where $t_k$ is the time slice which may differ from that in the previous Sect.~\ref{sec_app:accretion_numerical}. The quantities of the disk are taken at the location of a planet. This additionally requires their interpolation in space and time. Therefore, for the calculation of migration, we interpolate the relevant pre-calculated quantities of the disk evolution between the inner disk edge, $r_{\rm in}$, and the largest dynamically stable circular planetary orbit, $r_{\rm p}^{\rm max}$, given by Eq.~\eqref{eq:max_planet_orbit}, using $N = 1000$ uniform spacial grid cells.

Equation~\eqref{eq:migration_numerical} corresponds to an explicit numerical scheme that requires the time step to be small enough for the numerical stability of the solution.  
During the simulations, we check that the time step is $\lesssim 10^3$ years (the corresponding number of grid points is $K \gtrsim 10^3$ per a million year) is sufficient for the solution to be numerically stable. Therefore, we interpolate the values and the time grid to $K = 50000$ uniformly distributed cells added to the base MESA time grid in the range from the moment of disk formation to the moment of the disk decay rounded up to a Myr.

We calculate planet migration for planets with masses $m_{\rm p} = 1.0$, $3.0$, $6.0$, $10.0$, $30.0$, $60.0$, $100.0$, $300.0$, $500.0$, $750.0$, $1000.0$, $2000.0$, $3000.0$~m$_{\oplus}$ for all selected combinations of parameters of disk evolution $a$, $M_1$.

\subsection{Migration contours}
\label{sec_app:contours}

Due to a large number of free parameters, we introduce the method used to visualize the results of our calculations of planetary migration. We would like to plot curves of 
the constant final size of the planetary orbit, which we refer to as the migration 
contours hereafter. 
Since there are many free parameters of the problem (initial mass of donor, initial binary separation, planetary mass, initial size of planetary orbit, etc.), it is worthwhile to set some of them to particular values. We specify the initial mass of the donor and the planetary mass and obtain
the migration contours via bi-linear interpolation on the plane of parameters of the initial binary separation and initial size of planetary orbit, i.e., $a$ and $r_{\mathrm{p}}$ for the specified $m_{\rm p}$, or $r_{\mathrm{p}}$ and $m_{\rm p}$ for the specified $a$:
\begin{multline}
\label{eq:interpolation}
f \left(x, y\right) = \frac{1}{\left(x_{i + 1} - x_i\right) \left(y_{j + 1} - y_j\right)} \left[ f \left(x_i, y_j\right) \left(x_{i + 1} - x\right) \left(y_{j + 1} - y\right) + \right.\\ 
+ f \left(x_{i + 1}, y_j\right) \left(x - x_i\right) \left(y_{j + 1} - y\right) + f \left(x_i, y_{j + 1}\right) \left(x_{i + 1} - x\right) \left(y - y_j\right) + \\
\left. + f \left(x_{i + 1}, y_{j + 1}\right) \left(x - x_i\right) \left(y - y_j\right) \right],
\end{multline}
where $f$ is an interpolated value (e.g., size of planetary orbit after overall migration), $x$ and $y$ are values in the range of the plane of interpolation (initial $a$ and $r_{\mathrm{p}}$), $x_i$ and $y_j$ are parameters of numerical experiments, and $f \left(x_i, y_j\right)$ are values of the interpolated function at the corresponding points.

We note that in Figs.~\ref{fig:contour_NSD_different_times_1}~--~\ref{fig:contour_different_planet_mass_4},~\ref{fig:contours_different_donor_masses} boundary artifacts may occur. These artifacts are due to the transition from a four-point interpolation to a three-point extrapolation.

\section{The zone of decretion: an additional test}
\label{sec_app:decretion}

\begin{figure}
    \centering
    \includegraphics[width=\columnwidth]{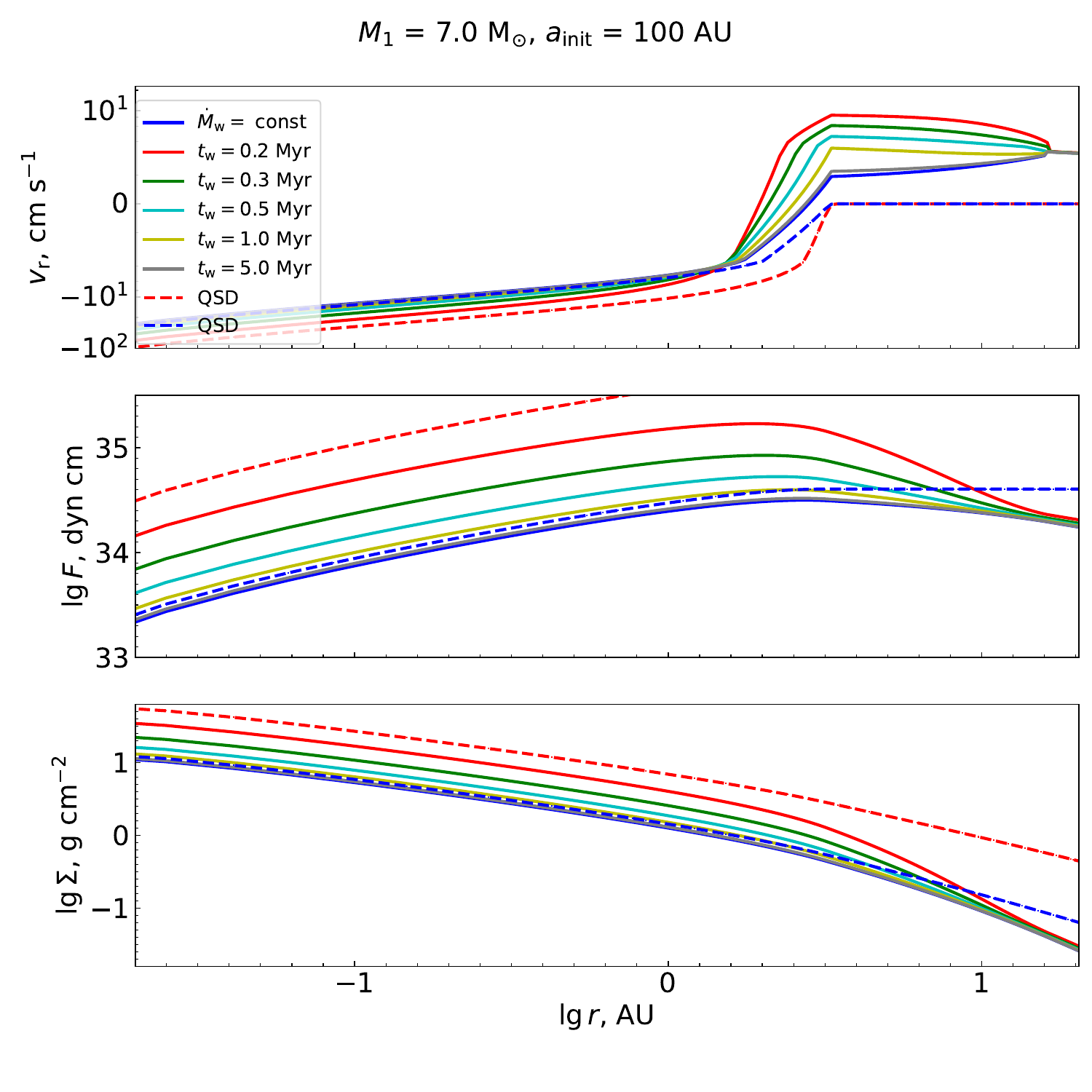}
    \caption{Radial velocity, viscous torque, and surface density radial profiles for the simplified NS disk model as described in App.~\ref{sec_app:decretion}. The profiles are taken at the time moment $0.5$~Myr after the start of the wind exponential growth. Before that, the disk has been accumulated from the wind with the constant rate $\dot M_{\rm w} = 5 \times 10^{-8}$~M$_\sun$~yr$^{-1}$ during 5 Myr. The solid curves with different colors correspond to different wind variation timescales, $t_{\rm w} \equiv \dot M_{\rm w} / \ddot M_{\rm w}$. The black, yellow, cyan, green, and red curves represent the cases of $t_{\rm w} = 5.0, 1.0, 0.5, 0.3, 0.2$~Myr, respectively. 
    The solid blue curve shows the solution for the constant wind rate. The dashed blue curve shows the QSD solution obtained for $\dot M_{\rm w} = 5 \times 10^{-8}$~M$_\sun$~yr$^{-1}$. The dashed red curve shows the QSD solution obtained for
    the instant value of the wind rate attained after 0.5 Myr of its exponential growth with 
    $t_{\rm w}=0.2$~Myr.
    The initial donor mass $M_1 = 7.0$~M$_{\sun}$, the binary separation is constant $a = 100$~AU. In all cases, the disk satisfies the floating outer boundary condition~\eqref{eq:boundary_fl_F}.}
    \label{fig:decretion_timescales}
\end{figure}

In order to verify the validity of an enhanced decretion which we find in NSD profiles at the AGB stage of the donors, we perform an additional numerical test employing a simplified accretion disk model. This test confirms that enhanced decretion is not a numerical artifact, but physically justified essentially NS disk feature.
 
We consider simplified NSD assuming that the disk temperature is determined solely by the viscous dissipation in Eq.~\eqref{eq:energy_equation}. The opacity of disk is taken to be constant, $\kappa_{\mathrm {R}} = 3.0$ cm$^2$g$^{-1}$, while $\kappa_{\mathrm{P}}=7.17$ cm$^2$g$^{-1}$. Binary separation is set to the constant initial value. 

At the preliminary stage of the test, the wind rate is constant, $\dot M_{\rm w} = 5 \times 10^{-8}$~M$_\sun$~yr$^{-1}$, for $t_0 = 5$ Myr, which is plausible in astrophysical situations. During this time the disk accumulates matter and approaches the corresponding quasi-stationary solution. We check that $t_{\nu}$ evaluated locally along the disk attains $\sim 5$ Myr in the vicinity of the outer boundary. At the main stage of the test, the wind rate grows exponentially, $\dot M_{\rm w} = 5 \times 10^{-8}\, e^{(t-t_0) / t_{\rm w}}$~M$_\sun$~yr$^{-1}$, for the other 5 Myr. Therefore, we consider how the disk transforms under the action of the source of matter with the constant characteristic variation timescale, provided that the latter differs from the viscous timescale of the disk evaluated at the outer boundary. The profiles of radial velocity, viscous torque, and surface density are shown in Fig.~\ref{fig:decretion_timescales} $0.5$~Myr after the wind starts growing. The different curves demonstrate varying $t_{\rm w}$. 

We find that while $t_{\rm w} > t_0$, the disk stays close to the quasi-stationary solution obtained for $\dot M_{\rm w} = 5 \times 10^{-8}$~M$_\sun$~yr$^{-1}$, see the curves corresponding to the constant wind and wind with $t_{\rm w} = 5 \times 10^6$\,yr along with the QSD solution for a constant wind in Fig.~\ref{fig:decretion_timescales}. The accretion takes place almost up to the Bondi radius, where QSD radial velocity attains zero. At the same time, the decretion persists beyond the Bondi radius. Note that this is not a consequence of the floating boundary condition, Eq.~\eqref{eq:boundary_fl_F}, at the time-span $\sim 5$~Myr. We find that the corresponding solution obtained with another boundary condition, Eq.~\eqref{eq:boundary}, imposed at $r=r_\mathrm{out}$ regardless of the location of $r_\tau < r_{\rm out}$ exhibits quite similar profile of decretion up to $r\sim 10$~AU. The two solutions deviate from each other at $r_\tau \gtrsim 10$~AU tending to different values of $v_{\rm r}$ at their own boundaries. Yet, we additionally check that these two solutions deviate from each other already beyond the Bondi radius as they are advanced further up to $\sim 10$~Myr and more. Of course, the decretion eventually ceases in the latter solution, whereas it is sustained in the former solution by the condition $v_r>0$ at the floating boundary.

The situation changes as soon as $t_{\rm w} < t_0$. At the early time of the wind growth, the NSD profile of radial velocity closely follows the corresponding QSD profile determined by the instant value of the growing wind rate only in the inner part of the disk, $r \lesssim 1$~AU. This is clearly seen by the red solid and dashed curves on the top panel in Fig.~\ref{fig:decretion_timescales}. Note that at $r\lesssim 1$~AU the negative radial velocity describes the accretion which accelerates over time along with the growing wind rate. In contrast, the NSD radial velocity takes rapidly growing positive values on both sides of the Bondi radius. The typical velocity of an enhanced decretion significantly exceeds the quasi-stationary value imposed at the outer boundary according to Eq.~\eqref{eq:boundary_fl}. It attains almost $10$~cm~s$^{-1}$ ($\approx 2 \times 10^{-5}$~AU~yr$^{-1}$) at several AU in the case of $t_{\rm w}=0.2$~Myr. Also, the zone of enhanced decretion expands inside the disk, see a number of solid curves in Fig.~\ref{fig:decretion_timescales}. This is in accordance with the large deviation of NSD profiles of viscous torque and surface density from their QSD counterparts at $r\gtrsim 1$~AU, see the red solid and dashed curves on the middle and the bottom panels in Fig.~\ref{fig:decretion_timescales}. As the viscous torque gets larger, its inverse radial dependence, $\mathrm{d} F / \mathrm{d} r < 0$, which defines the zone of decretion, gets steeper. We check that the inner edge of the zone of decretion is approximately defined by the equality of $t_{\rm w}$ with $t_\nu$ locally evaluated in a disk at the end of the preliminary stage of this test. Note that $t_{\rm w}$ can take much smaller values in astrophysical situations as shown in Fig.~\ref{fig:viscous_times}. Therefore, the zone of enhanced decretion may expand even further inside the disk. Accordingly, it may propagate inside the disk longer than $\sim 0.5$~Myr found in this test. We also check that the solution obtained with another boundary condition, Eq.~\eqref{eq:boundary}, imposed at $r=r_\mathrm{out}$ in the case $t_{\rm w} < t_0$ recovers exactly the zone of enhanced decretion described above.

As $\sim 0.5$~Myr has passed since the wind rate starts growing, the zone of decretion for the case $t_{\rm w}=0.2$~Myr attains its maximum size and amplitude as measured by the profile of radial velocity. After that, it begins to reduce back outwards. However, the positive radial velocity beyond the Bondi radius retains its large value, which is required to take an increased amount of angular momentum from the grown disk. The viscous timescale of the new disk evaluated at the Bondi radius becomes less than the constant $t_{\rm w}$ again. The new disk continues growing as it approaches the new quasi-stationary state determined by the instant growing value of the wind rate (not shown in Fig.~\ref{fig:decretion_timescales}). We note that such a secular wind growth seems to be unfeasible in astrophysical situations. Instead, we should expect the abrupt change of wind to the attenuation long before the disk overtakes the new quasi-stationary state. It is more likely that the disk does it already at the stage of the closed evolution after the wind ceases, see main text.

At last, we check that our test numerical solutions do not depend on the value of spurious initial viscous torque, i.e., we vary the amplitude of the sine in $F_{\rm init}$ defined in Sect.~\ref{sec:numerical_method} from $10^{20}$~dyn\,cm to zero. As long as this amplitude is sufficiently small and the dependence $F_{\rm init} (\angularmomentum)$  satisfies the boundary conditions, the exact choice of the initial condition does not affect the following evolution. For instance, the amplitude of $10^{20}$~dyn\,cm is more than $12$ orders of magnitude lower than the smallest values of viscous torque presented in Fig.~\ref{fig:decretion_timescales}.

\end{appendix}



\end{document}